\documentclass[twoside,twocolumn,english,pra,twocolumn,aps,superscriptaddress,showpacs,longbibliography]{revtex4-1}

\usepackage[T1]{fontenc}
\usepackage[latin9]{inputenc}
\usepackage{color}
\usepackage{babel}
\usepackage{amsmath}
\usepackage{amsthm}
\usepackage{amssymb}
\usepackage{graphicx}
\usepackage[unicode=true,pdfusetitle,
 bookmarks=true,bookmarksnumbered=false,bookmarksopen=false,
 breaklinks=false,pdfborder={0 0 0},pdfborderstyle={},backref=false,colorlinks=true]
 {hyperref}

\makeatletter
\theoremstyle{plain}

\usepackage{tabularx}

\usepackage{makecell}
\usepackage{fontenc}

\hypersetup{urlcolor=blue}

\usepackage{cmap}
\usepackage[T1]{fontenc}

\input glyphtounicode
\makeatother

\begin{document}
\title{Relation between quantum illumination and quantum parameter estimation}
\author{Wei Zhong }
\email{zhongwei1118@gmail.com}

\affiliation{Institute of Quantum Information and Technology, Nanjing University
of Posts and Telecommunications, Nanjing 210003, China}
\author{Dong-Qing Wang}
\affiliation{Institute of Quantum Information and Technology, Nanjing University
of Posts and Telecommunications, Nanjing 210003, China}
\author{Wen-Yi Zhu}
\affiliation{Institute of Quantum Information and Technology, Nanjing University
of Posts and Telecommunications, Nanjing 210003, China}
\author{Lan Zhou}
\affiliation{School of Science, Nanjing University of Posts and Telecommunications,
Nanjing 210003, China}
\author{Ming-Ming Du}
\affiliation{College of Electronic and Optical Engineering, Nanjing University
of Posts and Telecommunications, Nanjing 210003, China}
\author{Yu-Bo Sheng}
\email{shengyb@njupt.edu.cn}

\affiliation{College of Electronic and Optical Engineering, Nanjing University
of Posts and Telecommunications, Nanjing 210003, China}
\begin{abstract}
Quantum illumination leverages entangled lights to detect the presence
of low-reflectivity objects within a thermal environment. In a related
vein, quantum parameter estimation utilizes nonclassical probes to
precisely determine unknown system parameters. Although both fields
have been studied extensively, their performances have traditionally
been assessed using different figures of merit: signal-to-noise ratio
for QI and quantum Fisher information for parameter estimation. In
this paper, we reveal the intrinsic connection between these two measures
in the context of target detection, thereby providing explicit operational
criteria for identifying optimal measurements. We further apply this
relationship to various target detection protocols that employ exotic
non-Gaussian states derived from coherent states and two-mode squeezed
vacuum states.
\end{abstract}
\maketitle

\section{Introduction}

Quantum illumination (QI) is a promising protocol that surpasses the
performance of optimal classical illumination (CI) in target detection
\citep{Lloyd2008science,Pirandola2018NPreview,Shapiro2020Review,Sorelli2022Review}.
In QI, a signal beam, which is entangled with an idler (or ancilla
\citep{Lloyd2008science}) beam, is sent to illuminate a low-reflectivity
target embedded in a bright background. A joint measurement on the
reflected signal and idler then determines the presence or absence
of the target (see Fig.~1). Although QI was first proposed by Lloyd
\citep{Lloyd2008science}, it exhibits the counterintuitive feature
that its advantage persists even when the final state of the reflected
signal and idler is not entangled \citep{Sacchi2005PRA,Sacchi2005PRA-report}.
This observation challenges the common notion that entangled states
are only beneficial in noiseless conditions, given their inherent
fragility in noisy environments.

Inspired by Lloyd\textquoteright s pioneering work \citep{Lloyd2008science},
both theoretical \citep{Tan2008PRL,Shapiro2009NJP,Guha2009PRA,Barzanjeh2015PRL,Zhuang2017PRL,Zhuang2017JOSAB,Yung2020npj,Lee2021PRA,Jo2021PRR,GallegoTorrome2024review,Kronowetter2024PRApp,wei2024quantumilluminationadvantagequantum}
and experimental \citep{Lopaeva2013PRL,Zhang2015PRL-QI,England2019PRA,Xu2021PRL}
studies have advanced the field considerably. Among the various QI
protocols, the one proposed by Tan \emph{et al.} \citep{Tan2008PRL}
is especially notable, as it employs the two-mode squeezed vacuum
(TMSV) state to achieve up to a $6$-dB improvement in the error probability
exponent over coherent states with the same signal intensity. Recently,
Bradshaw \emph{et al.} rigorously demonstrated that the coherent state
is the optimal continuous-variable probe for CI in the limit of zero
object reflectivity \citep{Bradshaw2021PRA}. Correspondingly, the
TMSV state was shown to be optimal for QI \citep{Bradshaw2021PRA},
a finding further supported by comparisons with displaced TMSV states
\citep{Kim2023PRR}. However, realizing the full $6$-dB gain experimentally
remains challenging due to the complexity of the required joint measurements,
which often involve a quantum Schur transform that necessitates a
quantum computer unavailable currently \citep{Calsamiglia2008PRA}.
In contrast, local measurement strategies based on local operations
and classical communication have demonstrated a more experimentally
feasible 3-dB advantage \citep{Guha2009PRA}, albeit at half the gain
of global measurements \citep{Tan2008PRL,Zhuang2017PRL}.

More recently, Sanz \emph{et al.} recast the QI problem as a quantum
parameter estimation task aimed at estimating the target\textquoteright s
reflectivity, showing that TMSV states offer a 3-dB advantage in quantum
Fisher information (QFI) over coherent states \citep{Sanz2017PRL}.
This finding aligns with the advantage observed via shot-to-noise
ratio (SNR) \citep{Guha2009PRA} and suggests an underlying connection
between SNR and QFI---a connection that has often been overlooked
as these quantities were traditionally treated as separate performance
metrics for target detection and sensitivity, respectively \citep{Guha2009PRA,Lopaeva2013PRL,Barzanjeh2015PRL,Sanz2017PRL,Barzanjeh2020SciAdv,shi2023fulfillingentanglementsoptimaladvantage,Lee2021PRA}.
In this paper, we address this gap by elucidating the intrinsic relationship
between SNR and QFI in the context of target detection. This connection
not only provides a unified framework for understanding performance
limits but also offers practical guidance for selecting optimal measurement
strategies that approach the ultimate error probability bounds under
local operations.

Furthermore, we apply our established relationship to various target
detection protocols employing generalized coherent states, which are
obtained by transforming coherent states through nonlinear light-matter
interactions \citep{Titulaer1966PR,BialynickaBirula1968PR,Stoler1971PRD,Yurke1986PRL,Uria2023PRR,Walschaers2021PRXQuantum,LewisSwan2020PRL},
and de-Gaussified TMSV states, which are generated via photon-addition
and photon-subtraction operations. The choice of these states is motivated
by recent findings. Uria \emph{et al.} \citep{Uria2023PRR} demonstrated
that generalized coherent states can achieve Heisenberg-limited sensitivity
for small field displacements, and earlier works \citep{Zhang2014PRA-QI,Fan2018PRA}
have explored photon-subtraction and photon-addition in QI. However,
these studies were limited to single-photon processes, leaving open
the question of the performance of TMSV states modified by multi-photon
addition and subtraction. In addition to assessing the ultimate performance
achievable with these candidate states, our approach also identifies
the optimal measurements for each protocol based on our established
connection between SNR and QFI.

The paper is organized as follows. In Sec.~\ref{sec:Quantum-Illumination-Formalism}
we briefly introduce the theoretical framework for QI and outline
the two key performance measures, SNR and QFI. In Sec.~\ref{sec:Quantum-Fisher-information},
we provide analytical derivations of the QFI for target detection
protocols employing generalized coherent states, and multi-photon-added
TMSV (MPA-TMSV) or multi-photon-subtracted TMSV (MPS-TMSV) states,
and identify the optimal measurements that saturate the performance
bounds. Section~\ref{sec:Discussion} discusses the roles of photon
addition and subtraction in QI. Finally, our conclusions are presented
in Sec.~\ref{sec:Conclution}.

\section{Quantum illumination formalism \label{sec:Quantum-Illumination-Formalism}}

\subsection{Quantum illumination \label{subsec:quantum-illumination}}

\begin{figure}
\begin{centering}
\includegraphics[scale=0.2]{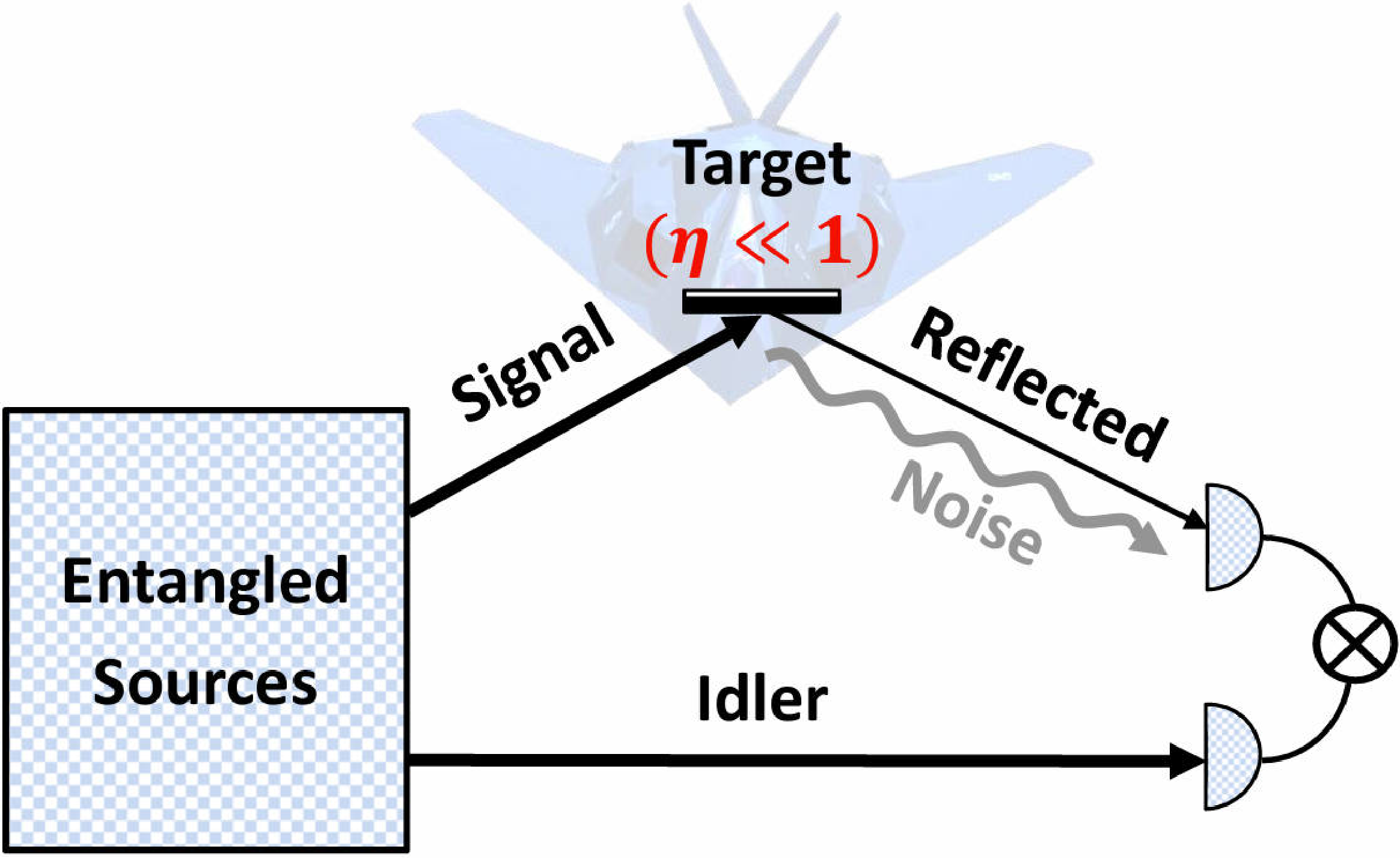}
\par\end{centering}
\centering{}\caption{(Color online) Schematic representation of the concept of quantum
illumination. \label{fig:QI}}
\end{figure}

Quantum illumination aims to detect a low-reflectivity object embedded
in a thermal bath (see Fig.~\ref{fig:QI}). A typical QI protocol
involves generating a two-mode entangled light source, where one mode,
termed as the signal, is directed towards the target region, while
the other mode, termed as the idler, is retained. A joint measurement
on the reflected signal and idler then determines the presence of
the object.

To formalize the QI problem, we assume that the probe light is initially
prepared in an arbitrary two-mode entangled pure state $\rho_{SI}\!=\!\vert\psi\rangle_{SI}\langle\psi\vert$,
where the subscripts $S$ and $I$ denote the signal and idler modes,
respectively. The advantage of QI lies in the photon correlations
between these modes, hence $\vert\psi\rangle_{SI}$ must be nonseparable.
In contrast, CI is realized by discarding the idler mode, so that
the probe is described by a single signal-mode state, e.g., $\rho_{S}\!=\!\vert\psi\rangle_{S}\langle\psi\vert$
\citep{Xu2021PRL}. The ambient environment is modeled as a thermal
state,
\begin{equation}
\rho_{B}=\sum_{n=0}^{\infty}\varrho_{n}\vert n\rangle_{B}\langle n\vert,\label{eq:thermalstate}
\end{equation}
with $\varrho_{n}\!=\!N_{B}^{n}/\!\left(N_{B}+1\right)^{n+1}$ representing
the thermal distribution with an average photon number $N_{B}$. Thus,
the overall initial state is given by the tensor product $\rho_{SI}\!\otimes\!\rho_{B}$.
The low-reflectivity object is modeled as a beam splitter represented
by the unitary operator
\begin{eqnarray}
U\!\left(\eta\right) & = & \exp\!\left[\eta\!\left(a_{S}^{\dagger}b-a_{S}b^{\dagger}\right)\right],\label{eq:beam-splitter}
\end{eqnarray}
where $a_{S}$ and $b$ are the annihilation operators for the signal
and environment modes, respectively \citep{Yurke1986PRA,Sanz2017PRL}.
Under the assumption of low reflectivity ($\eta\!\ll\!1$), the object's
reflectivity is given by $R\!=\!\sin^{2}\eta\!\sim\!\eta^{2}$. When
the initial state $\rho_{SI}\!\otimes\!\rho_{B}$ evolves under $U\!\left(\eta\right)$
and the signal mode is traced out, the final state is
\begin{eqnarray}
\rho_{RI}^{\left(1\right)} & \equiv\rho_{RI}\left(\eta\right)= & {\rm Tr}_{S}\!\left[U\!\left(\eta\right)\rho_{SI}\!\otimes\!\rho_{B}\,U{}^{\dagger}\!\left(\eta\right)\right].\label{eq:eta-density-matrix}
\end{eqnarray}
This state, defined under the hypothesis $H_{1}$ (object present),
is the reduced density matrix for the reflected and idler modes. In
the absence of the object $\left(\eta=0\right)$, the state becomes

\begin{equation}
\rho_{RI}^{\left(0\right)}\equiv\rho_{RI}\left(\eta\right)\vert_{\eta=0}=\rho_{I}\otimes\rho_{B},
\end{equation}
with $\rho_{I}\!=\!{\rm Tr}_{S}\!\left(\rho_{SI}\right)$. This corresponds
to the hypothesis $H_{0}$ (object absent). For brevity, we denote
$\rho_{0}\!\equiv\!\rho_{RI}^{\left(0\right)}$ and $\rho_{1}\!\equiv\!\rho_{RI}^{\left(1\right)}$.

An important observation from this formulation is that determining
whether the object is present reduces to a state discrimination problem
between $\rho_{0}$ and $\rho_{1}$ \citep{Audenaert2007PRL}. Since
the difference between $\rho_{0}$ and $\rho_{1}$ is infinitesimal
(due to small $\eta$), distinguishing them can be recast as estimating
the parameter $\eta$, that is, a QI problem can also be mapped onto
a parameter estimation problem \citep{Braunstein1994PRL}. Consequently,
two figures of merit have been employed for the study of QI: SNR \citep{Guha2009PRA,Jo2021PRR}
and QFI \citep{Sanz2017PRL}. Although a close connection between
these two measures is intuitively expected, such a relationship has
not yet been established. In what follows, we reveal this intrinsic
connection. We also remark that although QI maps onto a reflectivity
estimation problem, which is closely related to transmission estimation
\citep{Woodworth2020PRA,Woodworth2022EPJ,Dowran2021QST,Monras2007PRL,Adesso2009PRA,Invernizzi2011PRA},
the two differ subtly: in QI, the environment is assumed to be in
a thermal state, whereas in transmission estimation it is typically
taken to be in a vacuum state. This subtle difference leads to fundamentally
different consequences. As we will show, certain quantum states that
are advantageous for transmission estimation may prove ineffective
for QI. 

\subsection{Signal-to-noise ratio \label{subsec:Signal-to-noise-ratio}}

Discriminating between $\rho_{0}$ and $\rho_{1}$ constitutes a binary
hypothesis testing problem. Here, $\rho_{0}$ corresponds to the null
hypothesis $H_{0}$ (object absent) and $\rho_{1}$ to the alternative
hypothesis $H_{1}$ (object present). We assume equal prior probability,
$\pi_{0}\!=\!\pi_{1}\!=\!1/2$, reflecting no prior knowledge of the
object's presence. 

An effective discrimination protocol relies on the repeated measurements
with identical copies. When $M$ copies are available, the states
become $\rho_{0}^{\otimes M}$ and $\rho_{1}^{\otimes M}$. To ensure
reliable discrimination, the number of samples is assumed to be sufficiently
large, i.e., $M\!\gg\!1$. In realistic scenarios, such as continuous-wave
spontaneous parametric down-conversion, where a $T$-second pulse
produces $M\!=\!WT\!\gg\!1$ signal-idler mode pairs with $W$ the
phase-matching bandwidth, the large-copy assumption is justified.

A collective measurement on these $M$ copies can be described by
a collective operator $\tilde{\mathcal{O}}\!=\!\sum_{i=1}^{M}\!\mathcal{O}_{i}$,
where $\mathcal{O}_{i}$ acts on the $i$th copy. Under the hypothesis
$H_{x}\left(x\!=\!0,1\right)$, the expectation value and variance
of $\mathcal{O}$ are defined as
\begin{eqnarray}
\mu_{x}=\left\langle \mathcal{O}\right\rangle _{\!\rho_{x}} & \text{ and } & \sigma_{\!x}^{2}=\left\langle \mathcal{O}^{2}\right\rangle _{\!\rho_{x}}\!\!-\left\langle \mathcal{O}\right\rangle _{\!\rho_{x}}^{\!2}.
\end{eqnarray}
For $M\!\gg\!1$, by the central limit theorem the collective measurement
outcome $\lambda$ is Gaussian with mean $M\mu_{x}$ and variance
$M\sigma_{x}^{2}$ under hypothesis $H_{x}$. The decision rule is
defined by a threshold $\lambda_{{\rm th}}$: if $\lambda\!<\!\lambda_{{\rm th}}$,
$H_{0}$ is declared; otherwise, $H_{1}$ is declared.

The performance of this decision-making process is quantified by the
total error probability \citep{Fuchs1999IEEE,Trees2013Book}
\begin{eqnarray}
P_{\!{\rm err}} & = & \frac{1}{2}\!\left[p\!\left(1\vert0\right)+p\!\left(0\vert1\right)\right],\label{eq:classicalEP}
\end{eqnarray}
where $p\!\left(1\vert0\right)$ is the false-alarm probability (declaring
the target present when it is absent), and $p\!\left(0\vert1\right)$
is the miss probability (declaring the target absent when it is present).
By minimizing $P_{\!{\rm err}}$ with respect to $\lambda_{{\rm th}}$,
and by setting $\lambda_{{\rm th}}\!=\!M\!\left(\sigma_{0}\mu_{1}+\sigma_{1}\mu_{0}\right)\!/\!\left(\sigma_{0}+\sigma_{1}\right),$
one obtains the minimum error probability \citep{Guha2009PRA,Jo2021PRR}
\begin{align}
P_{\!{\rm err}}\!\left(R\right) & =\frac{1}{2}{\rm erfc}\bigg(\!\sqrt{\frac{M}{2}}R\!\bigg)<\frac{1}{4}\exp\!\big(\!\!-\frac{MR^{2}}{2}\big)\!,\label{eq:upperbound-EP}
\end{align}
where the SNR is defined as
\begin{eqnarray}
R & = & \frac{\mu_{1}-\mu_{0}}{\sigma_{1}+\sigma_{0}}\equiv{\rm SNR},\label{eq:SNR}
\end{eqnarray}
and the inequality follows from ${\rm erfc}\!\left(x\right)\!<\!e^{-x^{2}}\!\!/2$
\citep{Trees2013Book}. Equation~\eqref{eq:upperbound-EP} above
indicates that a higher SNR corresponds to a lower error probability.
Thus, optimizing the strategy for distinguishing between $\rho_{0}$
and $\rho_{1}$ reduces to finding the optimal measurement $\mathcal{O}$
that maximizes the SNR. We address the conditions for the optimality
in the following subsection.

\subsection{Quantum Fisher information \label{subsec:Fisher-information} }

As discussed previously, the QI problem can be recast as the estimation
of the reflectivity amplitude $\eta$ of the object. The performance
of this estimation is characterized by the mean-square deviation of
an unbiased estimator $\hat{\eta}$. Denoting the $\eta$-dependent
state by $\rho_{\eta}$ (with $\rho_{\eta\neq0}\!\equiv\!\rho_{1}$
and $\rho_{\eta=0}\!\equiv\!\rho_{0}$), and given a measurement $\mathcal{O}$
performed on $\rho_{\eta}$, the sensitivity in estimating $\eta$
is limited by the quantum Cram\'er-Rao bound \citep{Helstrom1976Book,Holevo1982Book,Braunstein1994PRL,Zhong2014JPA}
\begin{equation}
\Delta^{2}\hat{\eta}\equiv\frac{1}{M}\frac{\left\langle \mathcal{O}^{2}\right\rangle _{\!\rho_{\eta}}\!\!-\left\langle \mathcal{O}\right\rangle _{\!\rho_{\eta}}^{\!2}}{\big\vert\partial_{\eta}\!\left\langle \mathcal{O}\right\rangle _{\!\rho_{\eta}}\!\big\vert^{2}}\geq\frac{1}{M\mathcal{F}},\label{eq:error-propagation}
\end{equation}
where $\partial_{\eta}f\!\equiv\!\partial f/\partial\eta$ and $\mathcal{F}$
is the QFI defined as
\begin{eqnarray}
\mathcal{F} & = & {\rm Tr}\!\left(\rho_{\eta}L^{2}\right),\label{eq:QFI}
\end{eqnarray}
with $L$ being the symmetric logarithmic derivative operator, which
satisfies 
\begin{eqnarray}
\frac{\partial\rho_{\eta}}{\partial\eta} & = & \frac{1}{2}\!\left(\rho_{\eta}L+L\rho_{\eta}\right).\label{eq:SLD}
\end{eqnarray}
Using the Robertson-Schr\"odinger uncertainty relation, it has been
shown that the necessary and sufficient condition for the measurement
$\mathcal{O}$ to saturate the bound given by Eq\emph{.~}\eqref{eq:error-propagation}
is \citep{Zhong2014JPA}
\begin{eqnarray}
\big(\mathcal{O}-\left\langle \mathcal{O}\right\rangle _{\!\rho_{\eta}}\!\big)\sqrt{\rho_{\eta}} & = & \alpha L\sqrt{\rho_{\eta}},\,\forall\alpha\in\mathbb{R\!}\setminus\!\left\{ 0\right\} .\label{eq:optimal-measurement-condition}
\end{eqnarray}

A crucial step in solving the problem stated in the previous section
is to establish the relationship between the SNR defined by Eq.~\eqref{eq:SNR}
and the sensitivity defined by Eq.~\eqref{eq:error-propagation}.
In the limit of small $\eta$, we express the measurement operator
in the interaction picture as $\mathcal{O}_{\eta}\!\equiv\!U^{\dagger}\!\left(\eta\right)\mathcal{O}U\!\left(\eta\right)$.
Using the cyclic property of the trace, the expectation value of $\mathcal{O}$
with respect to the detected state $\rho_{\eta}$ can be equivalently
expressed as the expectation value of $\mathcal{O_{\eta}}$ with respect
to the initial state $\rho_{SI}\!\otimes\!\rho_{B}$, i.e., $\left\langle \mathcal{O}\right\rangle _{\!\rho_{\eta}}\!\!=\!\left\langle \mathcal{O}_{\eta}\right\rangle _{\!S\!I\!B}$.
For small $\eta$, we approximate $\mathcal{O}_{\eta}$ by expanding
to the first order in $\eta$ as $\mathcal{O}_{\eta}\!\approx\!\mathcal{O}\!+\!\eta\!\left[\mathcal{O},\!\left(a_{S}^{\dagger}b\!-\!a_{S}b^{\dagger}\right)\right]\!+\!o\!\left(\eta^{2}\right)$,
which yields 
\begin{eqnarray}
\left\langle \mathcal{O}_{\eta}\right\rangle _{\!S\!I\!B} & \approx & \left\langle \mathcal{O}\right\rangle _{\!S\!I\!B}+\eta\!\left\langle \left[\mathcal{O},\left(a_{S}^{\dagger}b-a_{S}b^{\dagger}\right)\right]\right\rangle _{\!S\!I\!B}.
\end{eqnarray}
Recalling that $\rho_{\eta=0}\!\equiv\!\rho_{0}$ and $\rho_{\eta\neq0}\!\equiv\!\rho_{1}$,
and using the identity $\left\langle \mathcal{O}\right\rangle _{\!\rho_{\eta}}\!\!=\!\left\langle \mathcal{O}_{\eta}\right\rangle _{\!S\!I\!B}$,
we deduce in the asymptotic limit $\eta\!\rightarrow\!0$,
\begin{eqnarray}
\mu_{1}-\mu_{0} & = & \eta\,\partial_{\eta}\!\left\langle \mathcal{O}\right\rangle _{\!\rho_{\eta}},\label{eq:mean-difference}\\
\sigma_{1}^{2} & = & \big(\!\left\langle \mathcal{O}^{2}\right\rangle _{\!\rho_{\eta}}\!\!-\left\langle \mathcal{O}\right\rangle _{\!\rho_{\eta}}^{\!2}\!\big)\!\big\vert_{\eta\rightarrow0}=\sigma_{0}^{2}.\label{eq:variance-equality}
\end{eqnarray}
Thus, the variances under both hypotheses become identical as $\eta\rightarrow0$.
This equality has been used in the derivation of Eq.~\eqref{eq:upperbound-EP}.
Submitting these into the definition of SNR given by Eq.~\eqref{eq:SNR}
yields
\begin{eqnarray}
\bigg(\!\frac{R}{\eta}\!\bigg)^{\!2} & \!= & \frac{1}{4}\frac{\big(\partial_{\eta}\!\left\langle \mathcal{O}\right\rangle _{\!\rho_{\eta}}\big)^{\!2}}{\left\langle \mathcal{O}^{2}\right\rangle _{\!\rho_{\eta}}\!\!-\left\langle \mathcal{O}\right\rangle _{\!\rho_{\eta}}^{\!2}}\Bigg\vert_{\eta\rightarrow0}\!\!\!=\frac{1}{4M\Delta^{2}\hat{\eta}}.\label{eq:SNR-EP}
\end{eqnarray}
It is remarkable that in the low reflectivity limit the SNR and estimation
sensitivity are intrinsically linked, as the SNR is entirely determined
by the inverse of the error-propagation formula that characterizes
estimation sensitivity \citep{Zhong2014JPA}. This equivalence relationship
is a key result of our paper, as these two concepts have previously
been treated separately in the literature \citep{Guha2009PRA,Lopaeva2013PRL,Barzanjeh2015PRL,Sanz2017PRL,Barzanjeh2020SciAdv,shi2023fulfillingentanglementsoptimaladvantage,Lee2021PRA}.
Moreover, the established connection unifies the two approaches of
state discrimination and parameter estimation in the context of QI.
Strictly speaking, this equivalence is established for all target
detection protocols, regardless of whether they are formulated within
the framework CI or QI. Hence, the results presented in what follows
are applicable to both CI and QI. 

Combining Eqs.~\eqref{eq:upperbound-EP}, \eqref{eq:error-propagation}
and \eqref{eq:SNR-EP}, we obtain a tight upper bound for the minimum
error probability
\begin{eqnarray}
P_{\!{\rm err}}\!\left(\mathcal{F}\right) & = & \frac{1}{4}\exp\!\bigg(\!\!-\frac{\eta^{2}M\mathcal{F}}{8}\bigg).\label{eq:Pe-bound}
\end{eqnarray}
This bound coincides with that first derived by Sanz \emph{et al.}
\citep{Sanz2017PRL} and later recovered by Noh \emph{et al.} \citep{Noh2022JOSAB}.
Unlike these previous works, our derivation explicitly uses the equivalence
established by Eq.~\eqref{eq:SNR-EP}, thereby endowing an operational
meaning to the relationship between SNR and QFI, as clarified below.
It is important to note that Eq.~\eqref{eq:Pe-bound} is derived
under the assumption of local measurements. A tighter bound may be
achieved with global measurement strategies, as suggested by the quantum
Chernoff bound \citep{Audenaert2007PRL,Calsamiglia2008PRA,Tan2008PRL,Pirandola2008PRA}.
However, implementing such measurements would be a nontrivial task
in experiments, despite some efforts in this direction \citep{Zhuang2017PRL}.
This aspect is beyond the scope of our paper.

\subsection{Optimal measurements \label{subsec:classical-Fisher-information} }

The two error probability bounds given by Eqs.~\eqref{eq:upperbound-EP}
and \eqref{eq:Pe-bound} have clear physical interpretations. The
bound expressed in terms of SNR, $P_{\!{\rm err}}\!\left(R\right)$,
represents the practically achievable limit for a given measurement
$\mathcal{O}$, while the bound expressed in terms of QFI, $P_{\!{\rm err}}\!\left(\mathcal{F}\right)$,
represents the ultimate limit independent of the specific measurement
employed. These bounds satisfy the inequality
\begin{equation}
P_{\!{\rm err}}\!\left(\mathcal{F}\right)\leq P_{\!{\rm err}}\!\left(R\right),\label{eq:QFI_SNR}
\end{equation}
with equality achieved when 
\begin{equation}
\frac{R}{\eta}=\frac{\sqrt{\mathcal{F}}}{2}\,,\label{eq:optimal_measurement_condition}
\end{equation}
which provides a clear operational criterion for identifying the optimal
measurement. In particular, when the condition in Eq.~\eqref{eq:optimal_measurement_condition}
is satisfied, it indicates that the chosen measurement is optimal
for target detection. This condition is more tractable than that given
in Eq.~\eqref{eq:optimal-measurement-condition}. Notably, determining
the optimal measurement according to Eq.~\eqref{eq:optimal-measurement-condition}
is a challenging task, and the resulting measurement may not be practically
implementable with current technology. Therefore, in our subsequent
analysis, we adopt the condition given by Eq.~\eqref{eq:optimal_measurement_condition}
.

In addition, as a classical analog of QFI, one may consider the classical
Fisher information, defined as
\begin{equation}
F=\sum_{\lambda}\frac{\left[\partial_{\eta}p\left(\lambda\vert\eta\right)\right]^{2}}{p\left(\lambda\vert\eta\right)},\label{eq:CFI}
\end{equation}
where $p\left(\lambda\vert\eta\right)\!=\!{\rm Tr}\left(\rho_{\eta}\Pi_{\lambda}\right)$
is the conditional probability of outcome $\lambda$ from the projective
measurement $\Pi_{\lambda}\!\equiv\!\vert\lambda\rangle\langle\lambda\vert$
corresponding to $\mathcal{O}$. Using the classical Cram\'er-Rao
inequality $\Delta^{2}\hat{\eta}\!\geq\!1/MF$ \citep{Kholevo1974TPA,Uys2007PRA,Braunstein1994PRL},
one can derive an upper bound analogous to Eq.~\eqref{eq:Pe-bound}
as
\begin{eqnarray}
P_{\!{\rm err}}\!\left(F\right) & = & \frac{1}{4}\exp\!\bigg(\!\!-\frac{\eta^{2}MF}{8}\bigg).\label{eq:Pe-bound-CFI}
\end{eqnarray}
Similar to $P_{{\rm err}}\!\left(R\right)$, this bound is also dependent
on the choice of the specific projective measurement $\left\{ \Pi_{\lambda}\right\} $.
It is evident that $P_{{\rm err}}\!\left(F\right)$ is tighter than
or equal to $P_{{\rm err}}\!\left(R\right)$ because of the inequality
\citep{Kholevo1974TPA,Uys2007PRA,Braunstein1994PRL}
\begin{equation}
\frac{\big(\partial_{\eta}\!\left\langle \mathcal{O}\right\rangle _{\!\rho_{\eta}}\big)^{\!2}}{\left\langle \mathcal{O}^{2}\right\rangle _{\!\rho_{\eta}}\!\!-\left\langle \mathcal{O}\right\rangle _{\!\rho_{\eta}}^{\!2}}\leq F,\label{eq:CCRB}
\end{equation}
yet it is looser than or equal to $P_{\!{\rm err}}\!\left(\mathcal{F}\right)$
because of $F\!\leq\!\sup_{\left\{ \Pi_{\lambda}\right\} }\!F\!\equiv\!\mathcal{F}$.
Thus, we have the hierarchy of error probabilities
\begin{equation}
P_{\!{\rm err}}\!\left(\mathcal{F}\right)\leq P_{\!{\rm err}}\!\left(F\right)\leq P_{\!{\rm err}}\!\left(R\right).
\end{equation}
This hierarchy generalizes the inequality in Eq.~\eqref{eq:QFI_SNR},
and encapsulates the successive optimizations needed to saturate the
ultimate error probability bound $P_{\!{\rm err}}\!\left(\mathcal{F}\right)$.
The first equality is achieved when $F\!=\!\mathcal{F}$, which requires
optimization over all projective measurements $\left\{ \Pi_{\lambda}\right\} $,
and the second equality holds when $R/\eta\!=\!\sqrt{F}\!/2$, necessitating
an optimal estimator, such as the maximum likelihood estimator \citep{Kholevo1974TPA,Uys2007PRA,Braunstein1994PRL}.

\section{Quantum illumination with non-Gaussian states \label{sec:Quantum-Fisher-information}}

In this section, we apply the formalism established above to evaluate
the performance of various target detection protocols and to identify
the corresponding optimal measurements for saturating the ultimate
bound set by Eq.~\eqref{eq:Pe-bound}. Two classes of probe states
are considered: generalized coherent states and de-Gaussified TMSV
states via multi-photon addition or subtraction.

\subsection{Generalized coherent states \label{subsec:classical illumination} }

Generalized coherent states form a family of states that fully characterize
the evolution of a coherent state under nonlinear light-matter interactions.
They are defined by
\begin{equation}
\vert\alpha_{\chi,\epsilon}\rangle=e^{-i\chi\hat{N}^{\epsilon}}\vert\alpha\rangle,
\end{equation}
where $\chi\!=\!gt$ (with $g$ the coupling strength of Hamiltonian
$\hat{N}^{\epsilon}$) and $\hat{N}\!=\!a^{\dagger}a$ is the photon
number operator \citep{Titulaer1966PR,BialynickaBirula1968PR,Stoler1971PRD,Yurke1986PRL}.
For instance, $\vert\alpha_{0,\epsilon}\!\left(t\right)\rangle\!=\!\vert\alpha\rangle$,
$\vert\alpha_{\theta,1}\!\left(t\right)\rangle\!=\!\vert\alpha e^{-i\theta}\rangle$,
and $\vert\alpha_{\chi,2}\!\left(t\right)\rangle$ represent Kerr
states. Generalized coherent states exhibit non-classical features,
as they represent a superposition of different probability distribution
components in the phase-space representation and may display negative
values in their Wigner function \citep{Uria2023PRR}. However, their
photon statistics remain Poissonian, and they do not exhibit quadrature
squeezing \citep{Agarwal1992PRA}. Recently, Uria \emph{et al.} demonstrated
that these states are beneficial for precision measurement as they
can achieve Heisenberg-limited sensitivity in detecting small field
displacements \citep{Uria2023PRR}.

To evaluate their role in target detection, we compute the QFI for
generalized coherent states. Since these states are single-mode, the
protocols employing them fall within the framework of CI, where only
the signal mode is used to illuminate the target. As shown in \citep{Sanz2017PRL},
the QFI for CI with an arbitrary probe state $\vert\psi\rangle$ is
given by 
\begin{equation}
\mathcal{F}_{{\rm CI}}=\frac{4\left|\langle\psi\vert a_{S}\vert\psi\rangle\right|^{2}}{2N_{B}+1}.\label{eq:QFI_Classical}
\end{equation}
A detailed derivation is also provided in Appendix A. It is evident
that any states for which $\langle\psi\vert a_{S}\vert\psi\rangle=0$
are ineffective for target detection, examples being single-mode squeezed
vacuum states \citep{Andersen2016PS} and Fock states \citep{Uria2020PRL}.
This result conflicts with their performance in transmission estimation
tasks, where such states are optimal \citep{Monras2007PRL,Adesso2009PRA}. 

For generalized coherent states $\vert\alpha_{\chi,\epsilon}\rangle$
with a mean photon number $N_{S}\!=\!\left|\alpha\right|^{2}$, one
finds
\begin{equation}
\langle\alpha_{\chi,\epsilon}\vert a_{S}\vert\alpha_{\chi,\epsilon}\rangle=\alpha\sum_{n=0}^{\infty}C_{n}^{2}e^{i\chi\left[n^{\epsilon}-\left(n+1\right)^{\epsilon}\right]},
\end{equation}
as a consequence of the identity $a_{S}f(\hat{N})\!=\!f(\hat{N}\!+\!1)a_{S}$
\citep{Uria2023PRR,Klimov2009Book}. By employing the Schwarz--Cauchy
inequality, one obtains $\left|\langle\psi\vert a_{S}\vert\psi\rangle\right|^{2}\!\leq\!N_{S}$.
Submitting this into Eq.~\eqref{eq:QFI_Classical} yields
\begin{eqnarray}
\mathcal{F}_{\alpha_{\chi,\epsilon}} & \leq & \mathcal{F}_{\alpha}=\frac{4N_{S}}{1+2N_{B}},\label{eq:classicalQFI}
\end{eqnarray}
where $\mathcal{F}_{\alpha}$ is the QFI for coherent states \citep{Sanz2017PRL}.
Unfortunately, this implies that generalized coherent states do not
offer an advantage over coherent states in target detection, thereby
reaffirming the conclusion presented in \citep{Bradshaw2021PRA}.

We now show that the error probability bound for CI given by Eq.~\eqref{eq:QFI_Classical}
is capably saturated by a quadrature operator. Consider the general
quadrature operator $\mathcal{O}\!=\!a_{R}e^{-i\phi}+a_{R}^{\dagger}e^{i\phi}$.
In deriving the SNR according to Eqs.~\eqref{eq:mean-difference}
and \eqref{eq:variance-equality}, one must calculate the first and
second moments of $\mathcal{O}$ on $\rho_{\eta}$. Using the Bogoliubov
transformation $a_{R}\!=\!\eta a_{S}+\sqrt{1-\eta^{2}}b$ (corresponding
to the action of $U\!\left(\eta\right)$), and approximating to first
order in $\eta$, while noting that $\left\langle b\right\rangle _{\!\rho_{B}}\!=\!\left\langle b^{2}\right\rangle _{\!\rho_{B}}\!=\!0$
, we obtain
\begin{align}
\left\langle \mathcal{O}\right\rangle _{\!\rho_{\eta}} & =\eta\left\langle a_{S}e^{-i\phi}+a_{S}^{\dagger}e^{i\phi}\right\rangle _{\!\vert\psi\rangle},\\
\left\langle \mathcal{O}^{2}\right\rangle _{\!\rho_{\eta}} & =\left\langle b^{\dagger}b+bb^{\dagger}\right\rangle _{\!\rho_{B}}.
\end{align}
It then follows that 
\begin{align}
\mu_{1}-\mu_{0} & =2\eta\left|\langle\psi\vert a_{S}\vert\psi\rangle\right|\cos\left(\phi-\theta\right),\\
\sigma_{1}^{2} & =\sigma_{0}^{2}=1+2N_{B},
\end{align}
where $\theta$ denotes the argument of the complex $\langle\psi\vert a_{S}\vert\psi\rangle$.
The equality $R/\eta\!=\!\sqrt{\mathcal{F}_{{\rm CI}}}/2$ is thus
achieved when the condition $\cos\left(\phi-\theta\right)\!=\!1$
is satisfied. This proves that the general quadrature operator $\mathcal{O}\!=\!a_{R}e^{-i\phi}+a_{R}^{\dagger}e^{i\phi}$
with the phase chosen such that $\cos\left(\phi-\theta\right)\!=\!1$
is optimal for CI.

\subsection{De-Gaussified TMSV states \label{subsec:quantum-illumination-with-deGaussifying-TMSV} }

The de-Gaussified TMSV states of interest are classified into two
groups: MPA-TMSV states and MPS-TMSV states. An MPA-TMSV (or MPS-TMSV)
state is obtained by adding (or subtracting) an equal number $\kappa$
of photons to each mode of the TMSV state, and is denoted by $\vert{\rm TMSV};+\kappa\rangle_{SI}$
(or $\vert{\rm TMSV};-\kappa\rangle_{SI}$). The multi-photon addition
(or subtraction) process can be viewed as a series of successive single-photon
additions (or subtractions) that can be experimentally implemented
using a weakly parametric amplifier (or a weakly reflecting beam splitter)
\citep{Navarrete-Benlloch2012PRA,Barnett2018PRA}. However, since
the single-photon addition (or subtraction) process is probabilistic,
the success probability for the multi-photon process decreases exponentially
with $\kappa$, limiting its practical applicability. Nevertheless,
numerous studies have demonstrated the utility of photon addition
and subtraction in areas such as entanglement enhancement \citep{Dakna1997PRA,Dakna1999PRA,Wenger2004PRL,Fiurafmmodeheckslsesiek2005PRA,Kitagawa2006PRA,Biswas2007PRA,Ourjoumtsev2007PRL,Kim2008JPB,Navarrete-Benlloch2012PRA},
entanglement distillation \citep{Duan2000PRL,Eisert2004AP,Wenger2004PRL,Takahashi2010nphoton},
quantum teleportation \citep{Opatrny2000PRA,Cochrane2002PRA,Olivares2003PRA,DellAnno2007PRA,Yang2009PRA,DellAnno2010PRA},
and quantum phase estimation \citep{Birrittella2014JOSAB,Braun2014PRA,Wang2019OC,Zhong2020SC}.
In the following, we assume that the MPA-TMSV and MPS-TMSV states
are available for our analysis.

The TMSV state is a linear superposition of Holland-Burnett states
\citep{Holland1993PRL} and is expressed by
\begin{eqnarray}
\vert{\rm TMSV}\rangle_{SI} & = & \sum_{n=0}^{\infty}C_{n}\vert n\rangle_{S}\vert n\rangle_{I},\label{eq:TMSV}
\end{eqnarray}
with $C_{n}\!=\!\sqrt{1-z^{2}}z^{n}$, where $z\!=\!\tanh r$ and
$r$ is the squeezing parameter. Both the signal and idler modes have
a mean photon number $N_{S}\!=\!N_{I}\!=\!\sinh^{2}r$. The MPA-TMSV
and MPS-TMSV states are defined as 
\begin{eqnarray}
\vert{\rm TMSV};+\kappa\rangle_{SI} & = & \frac{a_{S}^{\dagger\kappa}a_{I}^{\dagger\kappa}}{\sqrt{\mathcal{N}_{\kappa,\kappa}^{+}}}\vert{\rm TMSV}\rangle_{SI},\\
\vert{\rm TMSV};-\kappa\rangle_{SI} & = & \frac{a_{S}^{\kappa}a_{I}^{\kappa}}{\sqrt{\mathcal{N}_{\kappa,\kappa}^{-}}}\vert{\rm TMSV}\rangle_{SI},
\end{eqnarray}
where $\mathcal{N}_{\kappa,\kappa}^{\pm}$ are the corresponding normalization
factors. Using Eq.~\eqref{eq:TMSV}, these states can be rewritten
in the compact form
\begin{eqnarray}
\vert{\rm TMSV};\pm\kappa\rangle_{SI} & = & \sum_{n=n_{\pm}}^{\infty}\!\!C_{n}^{\left(\pm\kappa\right)}\vert n\pm\kappa\rangle_{S}\vert n\pm\kappa\rangle_{I},\label{eq:PA=000026PS-TMSV}
\end{eqnarray}
with $n_{+}\!=\!0$ and $n_{-}\!=\!\kappa$, and $+$ and $-$ signs
denote the photon-added and photon-subtracted cases, respectively.
Here the expansion coefficients $C_{n}^{\left(\pm\kappa\right)}$
are real and given by \citep{Navarrete-Benlloch2012PRA}
\begin{eqnarray}
C_{n}^{\left(+\kappa\right)} & = & \frac{z^{n}\binom{n+\kappa}{\kappa}}{\sqrt{_{2}F_{1}\left(\kappa+1,\kappa+1;1;z^{2}\right)}},\label{eq:coefficientsPA}\\
C_{n}^{\left(-\kappa\right)} & = & \frac{z^{\left(n-\kappa\right)}\binom{n}{\kappa}}{\sqrt{_{2}F_{1}\left(\kappa+1,\kappa+1;1;z^{2}\right)}},\label{eq:coefficientsPS}
\end{eqnarray}
where $_{2}F_{1}\left(a,b;c;x\right)$ is the ordinary hypergeometric
function. The mean photon numbers of both the signal and idler modes
for these states are given by
\begin{eqnarray}
N_{S} & = & N_{I}=\sum_{n=n_{\pm}}^{\infty}\!\!\Big[C_{n}^{\left(\pm\kappa\right)}\Big]^{2}\!\!\left(n\pm\kappa\right).\label{eq:Ns}
\end{eqnarray}

\begin{figure*}
\begin{centering}
\includegraphics[scale=0.4]{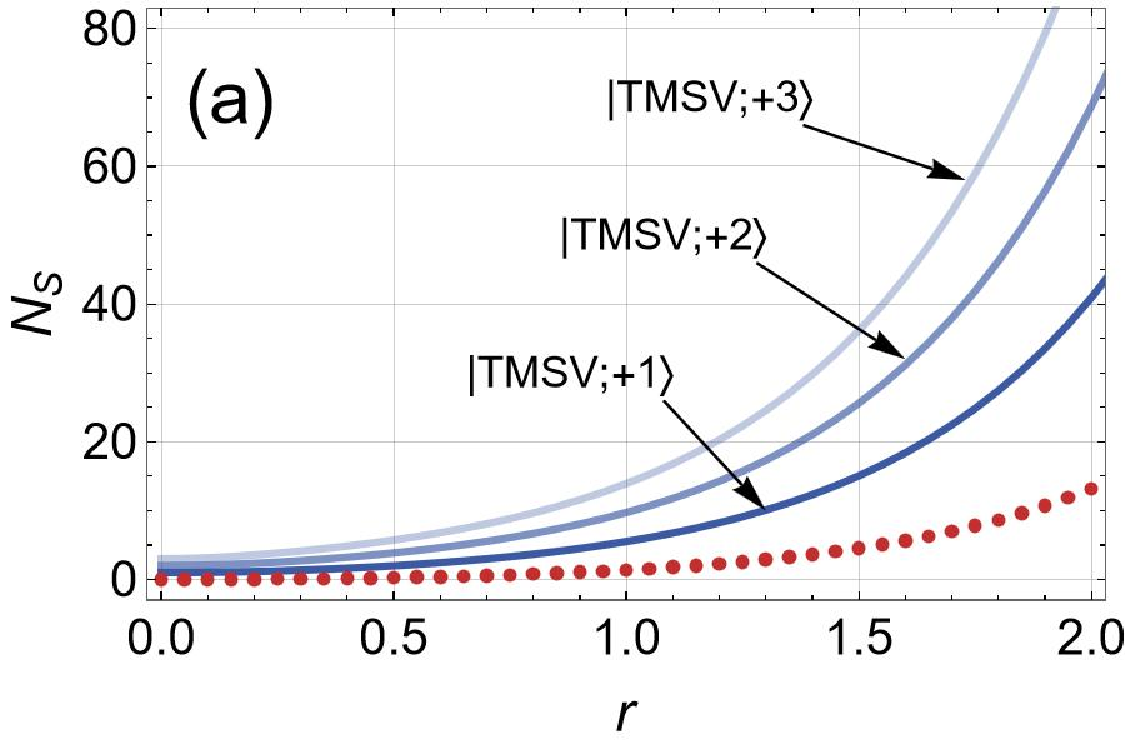} \includegraphics[scale=0.4]{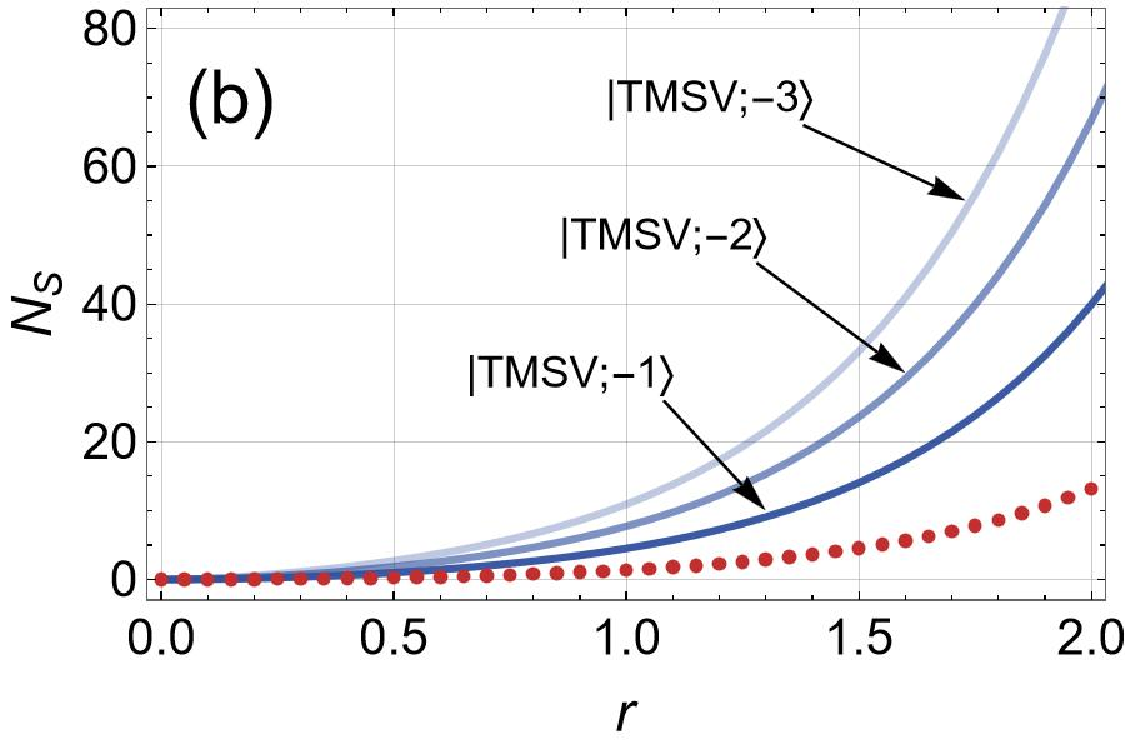}
\par\end{centering}
\begin{centering}
\includegraphics[scale=0.4]{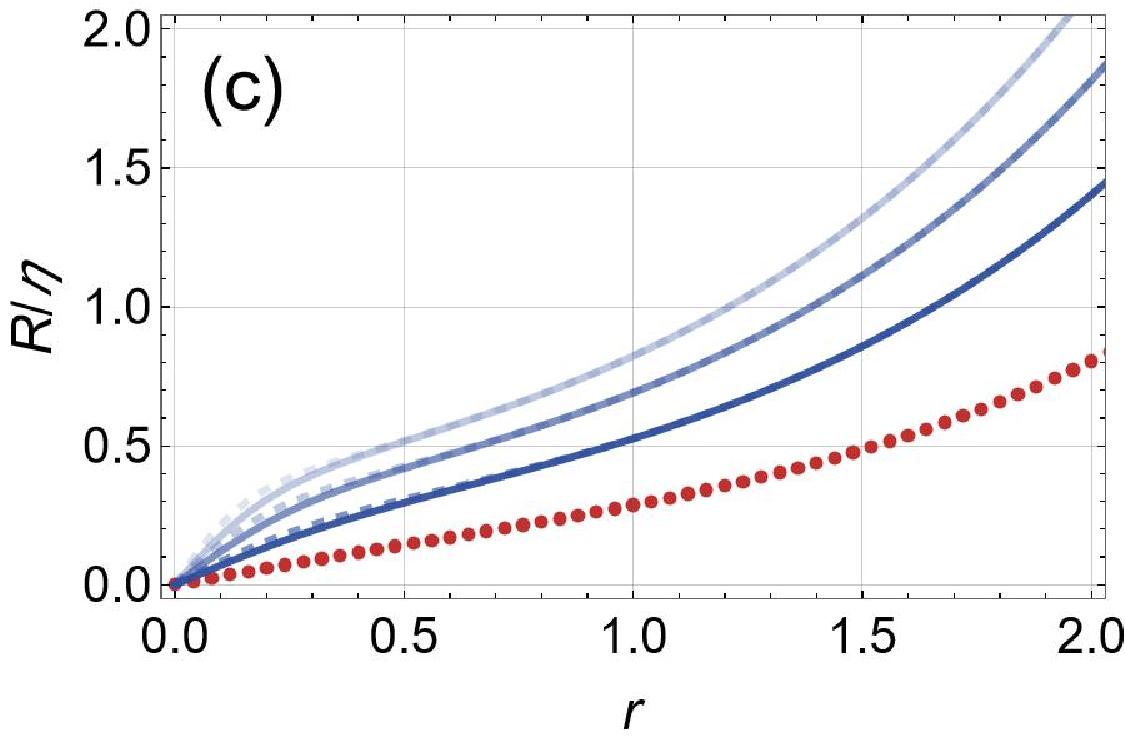} \includegraphics[scale=0.4]{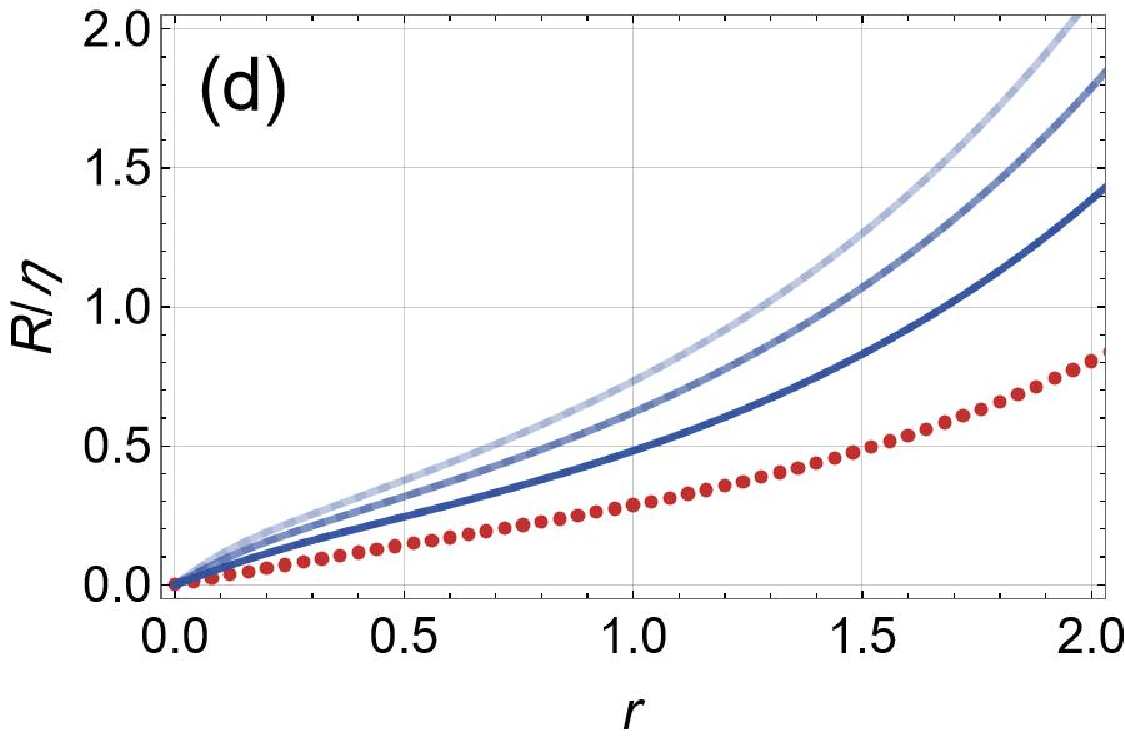}
\par\end{centering}
\centering{}\caption{Mean photon number of the signal mode vs the squeezing strength for
MPA-TMSV states (a) and MPS-TMSV states (b) with $N_{B}=10$. In both
(a) and (b), the solid lines ranging from darker blue at the bottom
to lighter blue at the top correspond to $\vert{\rm TMSV};\pm\kappa\rangle_{SI}$
with $\kappa=1,2,3$. The red dotted line represents the TMSV state
(i.e., $\kappa=0$). A comparison between $R/\eta$ (solid lines)
and $\sqrt{\mathcal{F}_{\pm\kappa}}/2$ (dashed lines) is depicted
in (c) and (d) corresponding to the same color schemes used in Figs.~\ref{fig:MeanPhotonNumberandSNR}(a)
and \ref{fig:MeanPhotonNumberandSNR}(b), respectively. \label{fig:MeanPhotonNumberandSNR}}
\end{figure*}

Figures.~\ref{fig:MeanPhotonNumberandSNR}(a) and \ref{fig:MeanPhotonNumberandSNR}(b)
show the signal strength as a function of $r$ for both cases, using
Eq.~\eqref{eq:Ns}. It is evident that both MPA-TMSV and MPS-TMSV
states share two common behaviors: (a) the signal strength increases
exponentially with $r$, and (b) the rate of increase for a given
$\kappa$ becomes more pronounced as $\kappa$ increases. The main
difference is that for MPA-TMSV states $N_{S}$ starts from a value
of $\kappa$, whereas for MPS-TMSV states it starts from zero. This
distinction arises because the TMSV state approaches a twin-vacuum
state as $r\!\rightarrow\!0$; under photon addition the vacuum state
becomes a Fock state, while under photon subtraction it remains unchanged.
This difference significantly affects the performance of MPA-TMSV
and MPS-TMSV states in QI tasks.

To evaluate the performance of these protocols, we compute the QFI
for the MPA-TMSV and MPS-TMSV states. With the formula given in \citep{Sanz2017PRL}
(also see Appendix B for detailed derivation), the analytical expressions
for these QFIs are derived as follows:
\begin{eqnarray}
\mathcal{F}_{\pm\kappa} & = & \frac{4}{1+N_{B}}\!\!\sum_{n=n_{\pm}+1}^{\infty}\!\!\frac{\Big[C_{n-1}^{\left(\pm\kappa\right)}C_{n}^{\left(\pm\kappa\right)}\Big]^{2}\!\!\left(n\pm\kappa\right)}{\Big[C_{n-1}^{\left(\pm\kappa\right)}\Big]^{2}+\Big[C_{n}^{\left(\pm\kappa\right)}\Big]^{2}\!\!\frac{N_{B}}{1+N_{B}}}.\quad\label{eq:PA=000026PS-QFI}
\end{eqnarray}
This result is valid for all states of the form given in Eq.~\eqref{eq:PA=000026PS-TMSV},
as the coefficients $C_{n}^{\left(\pm\kappa\right)}$ may be chosen
arbitrarily. Notably, for TMSV states ($\kappa=0$) where $C_{n}/C_{n-1}\!=\!N_{S}/\!\left(N_{S}+1\right)$
with $N_{S}\!=\!\sinh^{2}r$, the expression simplifies to 
\begin{equation}
\mathcal{F}_{0}=\frac{4N_{S}}{1+N_{B}}\frac{1}{1+\frac{N_{S}}{1+N_{S}}\frac{N_{B}}{1+N_{B}}},\label{eq:quantumQFI}
\end{equation}
which was first obtained in \citep{Sanz2017PRL}. It is evident that
the expression for $\mathcal{F}_{0}$ approximates $\mathcal{F}_{\alpha}$
in the limit of $N_{S}\!\rightarrow\!\infty$, behaving as a coherent
state.

Next, we discuss the attainability of the error probability bounds
predicted by Eq.~\eqref{eq:PA=000026PS-QFI} using the measurement
$\mathcal{O}\!=\!a_{R}a_{I}+a_{R}^{\dagger}a_{I}^{\dagger}$. This
measurement can be realized with linear optical elements and photon
counting \citep{Guha2009PRA,Sanz2017PRL}. Following the procedure
described in the previous subsection, one obtains
\begin{align}
\left\langle \mathcal{O}\right\rangle _{\!\rho_{\eta}} & =\eta\left\langle a_{S}a_{I}+a_{S}^{\dagger}a_{I}^{\dagger}\right\rangle _{\!S\!I},\\
\left\langle \mathcal{O}^{2}\right\rangle _{\!\rho_{\eta}} & =\left\langle a_{I}^{\dagger}a_{I}b^{\dagger}b+a_{I}a_{I}^{\dagger}bb^{\dagger}\right\rangle _{\!I\!B},
\end{align}
where $\left\langle \bullet\right\rangle _{\!S\!I}$ denotes the expectation
over $\vert{\rm TMSV};\pm\kappa\rangle_{SI}$ and $\left\langle \bullet\right\rangle _{\!I\!B}$
over $\rho_{I}\!\otimes\!\rho_{B}$. In the case of $\kappa\!=\!0$
(i.e., the TMSV state), it is straightforward to verify that this
measurement ensures $R/\eta\!=\!\sqrt{\mathcal{F}_{0}}/2$ with $\mu_{1}\!-\!\mu_{0}\!=\!2\eta\sqrt{N_{S}\left(1+N_{S}\right)}$
and $\sigma_{1}^{2}\!=\!\sigma_{0}^{2}\!=\!N_{S}N_{B}+\left(1+N_{S}\right)\left(1+N_{B}\right)$
with $N_{S}\!=\!\sinh^{2}r$, thus indicating that the measurement
$\mathcal{O}$ is optimal \citep{Sanz2017PRL}. For $\kappa\!\neq\!0$,
numerical calculations are employed due to analytical complexity.
We plot in Figs.~\ref{fig:MeanPhotonNumberandSNR}(c) and \ref{fig:MeanPhotonNumberandSNR}(d)
${\rm SNR/\eta}$ as a function of $r$, using:
\begin{eqnarray}
\mu_{1}-\mu_{0} & = & 2\eta\!\!\sum_{n=n_{\pm}+1}^{\infty}\!\!C_{n}^{\left(\pm\kappa\right)}C_{n-1}^{\left(\pm\kappa\right)}\!\left(n\pm\kappa\right),\\
\sigma_{1}^{2} & = & \sigma_{0}^{2}=N_{S}\!N_{B}+\left(1+N_{S}\right)\!\left(1+N_{B}\right),\quad
\end{eqnarray}
where $N_{S}$ is given by Eq.~\eqref{eq:Ns}. For comparison, the
quantity $\sqrt{\mathcal{F}_{\pm\kappa}}/2$ as determined by Eq.~\eqref{eq:PA=000026PS-QFI}
is also plotted. For MPA-TMSV states, $R/\eta$ and $\sqrt{\mathcal{F}_{\pm\kappa}}/2$
agree well for moderately large $r$, but show slight discrepancies
for small $r$. Conversely, for MPS-TMSV states, the two quantities
closely coincide over the entire range of $r$. These results indicate
that the measurement $\mathcal{O}$ is globally optimal for MPS-TMSV
states, but only partially optimal for MPA-TMSV states, depending
on $r$. Moreover, they have similar trends as those observed in the
signal strength (Figs.~\ref{fig:MeanPhotonNumberandSNR}(a) and \ref{fig:MeanPhotonNumberandSNR}(b)),
with the only notable difference that the signal strength for MPA-TMSV
states increases from a nonzero value. Finally, from Figs.~\ref{fig:MeanPhotonNumberandSNR}(c)
and \ref{fig:MeanPhotonNumberandSNR}(d), it is evident that both
the MPA-TMSV and MPS-TMSV states outperform the standard TMSV state
when the squeezing strength is constrained. This observation agrees
with the conclusion drawn in \citep{Zhang2014PRA-QI,Fan2018PRA}.
However, from another perspective, a contrary conclusion may also
be reached, highlighting the subtleties involved in assessing non-Gaussian
advantages in QI tasks.

\section{Further discussion on photon addition and photon subtraction \label{sec:Discussion}}

\subsection{Comparison under different constraints \label{subsec:constraints} }

To better understand the impact of photon addition and photon subtraction
on QI, we further discuss the scenarios considered above. In the previous
comparisons, the squeezing strength was constrained. However, such
comparisons are not entirely fair because the total energy consumed
in these protocols differs. The observed advantages of the non-Gaussian
protocols are primarily attributed to an increase in the mean photon
number as the number of added or subtracted photons increases, while
the squeezing strength remains fixed (see Figs.~\ref{fig:MeanPhotonNumberandSNR}(a)
and \ref{fig:MeanPhotonNumberandSNR}(b)).
\begin{figure*}
\begin{centering}
\includegraphics[scale=0.48]{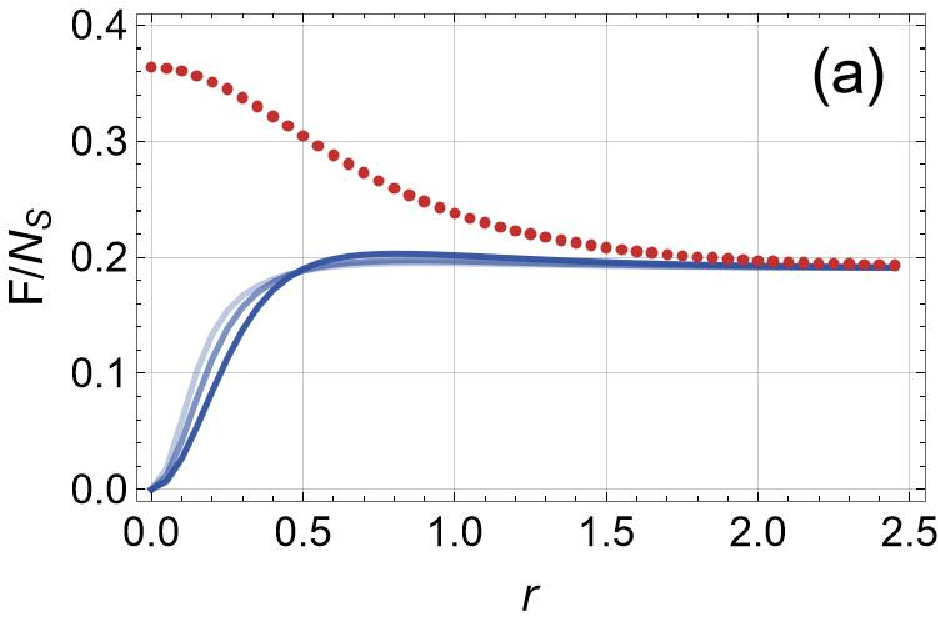} \includegraphics[scale=0.48]{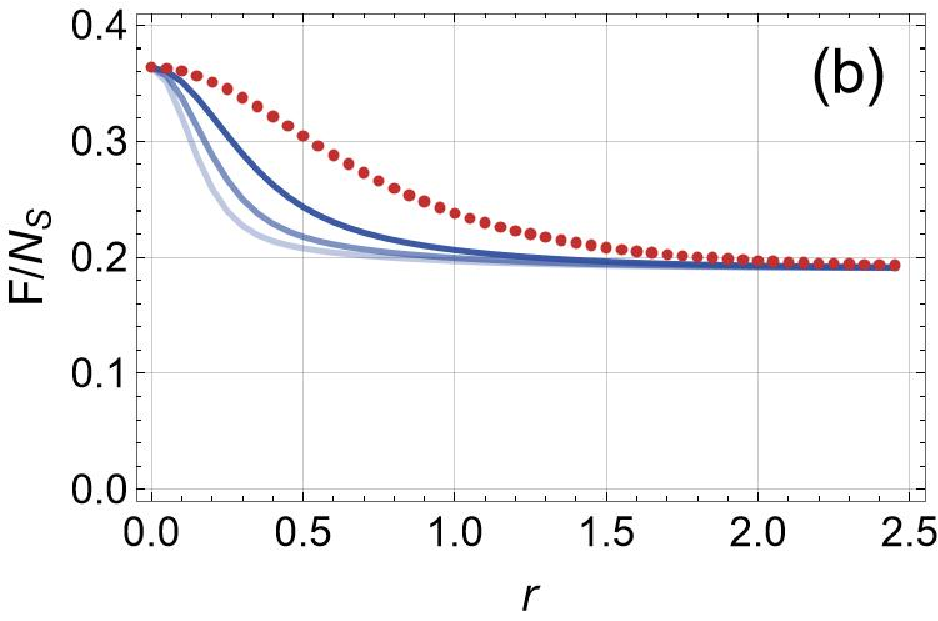}
\par\end{centering}
\begin{centering}
\includegraphics[scale=0.48]{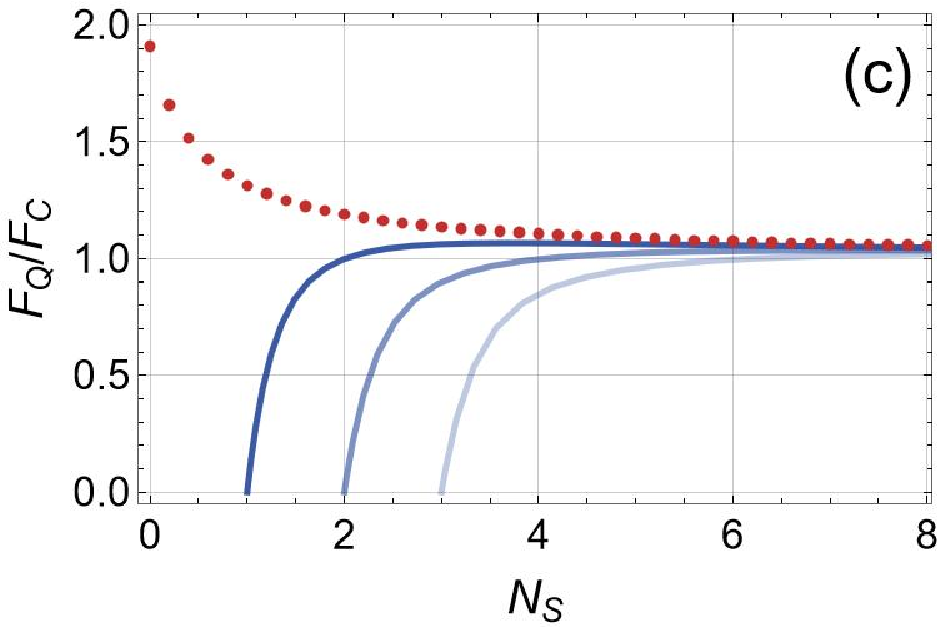} \includegraphics[scale=0.48]{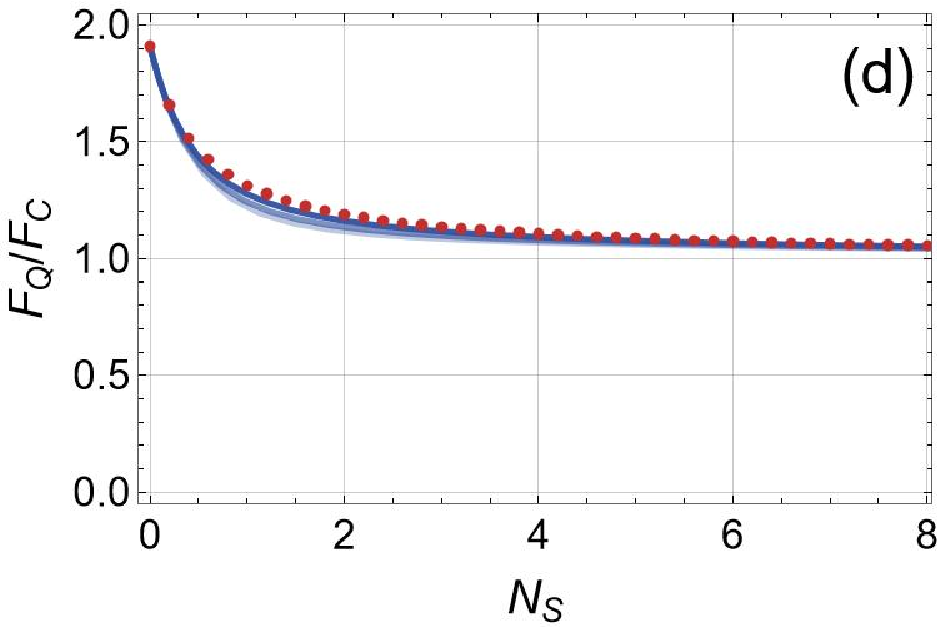}
\par\end{centering}
\centering{}\includegraphics[scale=0.48]{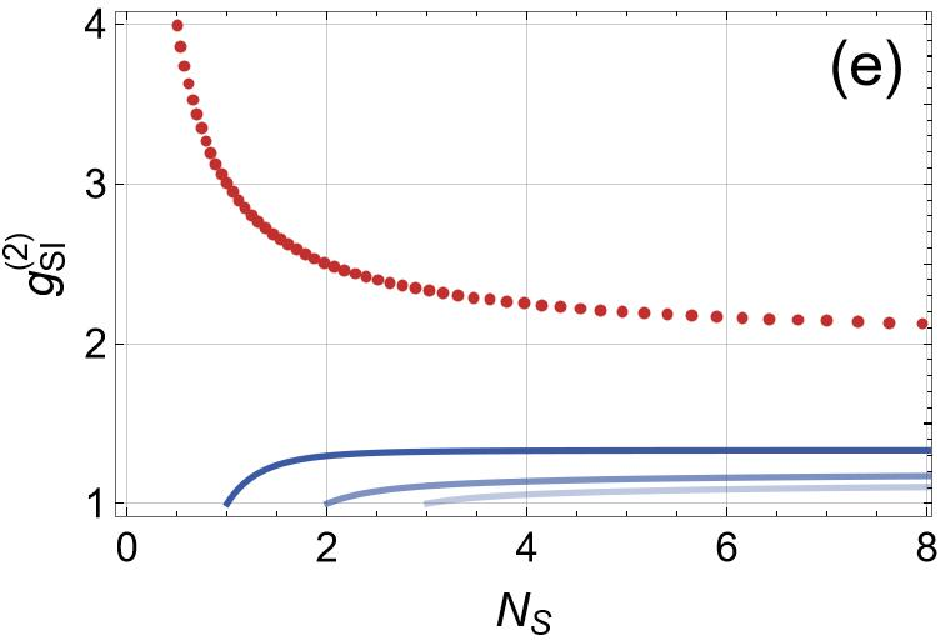} \includegraphics[scale=0.48]{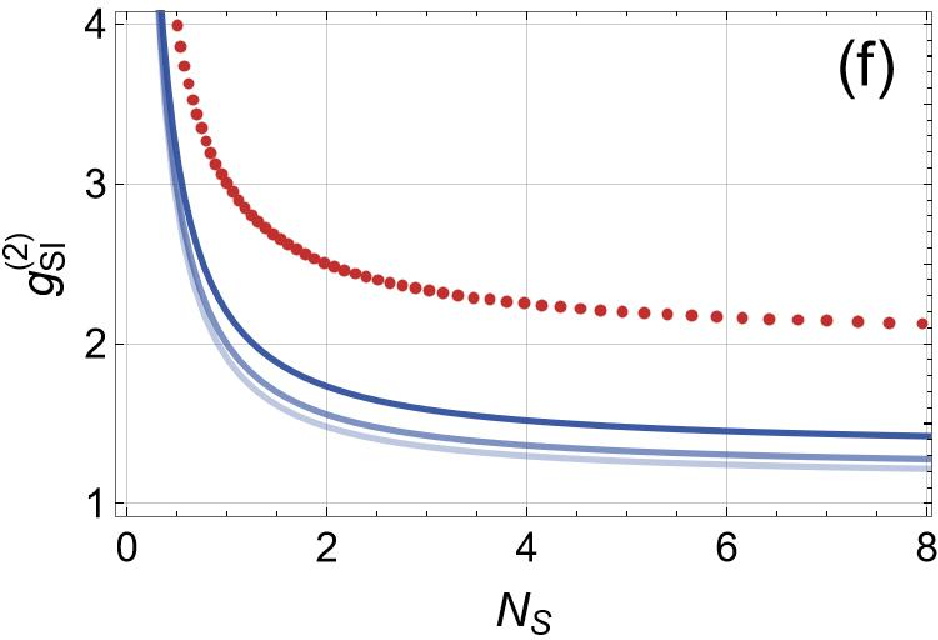}\caption{Averaged QFI versus squeezing strength (upper panel), quantum advantage
(parameterized by the ratio of QFI for QI to QFI for CI) (middle panel)
and normalized cross-correlation function (lower panel) versus signal
strength for MPA-TMSV states (left column) and MPS-TMSV states (right
column) with $N_{B}=10$. The color schemes employed are consistent
with those used in Figs.~\ref{fig:MeanPhotonNumberandSNR}(a) and
\ref{fig:MeanPhotonNumberandSNR}(b). \label{fig:Averaged-QFI}}
\end{figure*}

To enable a fair comparison, we plot the QFIs per photon $\mathcal{F}_{\pm\kappa}/N_{S}$
for both cases as a function of $r$ in Figs.~\ref{fig:Averaged-QFI}(a)
and \ref{fig:Averaged-QFI}(b). These plots reveal trends that differ
markedly from those  in Figs.~\ref{fig:MeanPhotonNumberandSNR}(c)
and \ref{fig:MeanPhotonNumberandSNR}(d), indicating the non-Gaussian
states do not outperform the TMSV state. Moreover, a clear difference
between the MPA-TMSV and MPS-TMSV states emerges. In Fig.~\ref{fig:Averaged-QFI}(a),
the averaged QFIs start at zero, then reach a plateau  around $0.2$,
and eventually merge with that of the TMSV state. In Fig.~\ref{fig:Averaged-QFI}(b),
the averaged QFIs decrease monotonically from $0.36$ (corresponding
to the value of $4/\!\left(1\!+\!N_{B}\right)$ with $N_{B}\!=\!10$)
to a plateau near $0.2$, merging with that of the TMSV state at larger
$r$. These results indicate that for MPA-TMSV states, there is no
advantage when $r$ is small, and they perform comparably to the TMSV
state as $r$ increases modestly. In contrast, the MPS-TMSV states
perform similarly to the TMSV state when $r$ approaches zero or is
modestly large, but they underperform in intermediate regimes. Notably,
adding or subtracting more photons to the TMSV state does not lead
to an increase in the averaged QFIs relative to the trends shown in
Fig.~\ref{fig:MeanPhotonNumberandSNR}(c) and \ref{fig:MeanPhotonNumberandSNR}(d).
The sharp difference between these two types of states arises from
their distinct structures at $r\!\rightarrow\!0$. In this limit,
the MPA-TMSV state lacks a vacuum component, whereas the MPS-TMSV
state includes it. 

To clarify this distinction, let us consider two simple entangled
states in the signal-idler mode by

\begin{eqnarray}
\vert\psi_{-}\rangle_{SI} & = & \sqrt{1-p}\vert00\rangle+\sqrt{p}\vert11\rangle,\\
\vert\psi_{+}\rangle_{SI} & = & \sqrt{1-p}\vert11\rangle+\sqrt{p}\vert22\rangle,
\end{eqnarray}
with $p\!\in\!\left[0,1\right]$. Here, the mean photon numbers are
$N_{S}^{-}\!=\!p$ for $\vert\psi_{-}\rangle_{SI}$ and $N_{S}^{+}\!=\!1+p$
for $\vert\psi_{+}\rangle_{SI}$. Setting $p\!=\!0$ yields states
corresponding to the MPS-TMSV and MPA-TMSV states in the limit of
$r\!\rightarrow\!0$, respectively. Their QFIs are given by
\begin{eqnarray}
\mathcal{F}_{\pm} & = & \frac{4x_{\pm}}{1+N_{B}}\frac{p\left(1-p\right)}{\left(1-p\right)+p\frac{N_{B}}{1+N_{B}}},
\end{eqnarray}
where $x_{-}\!=\!1$ and $x_{+}\!=\!2$. The two expressions differ
by a constant factor of $2$. For $\vert\psi_{+}\rangle_{SI}$, the
averaged QFI $\mathcal{F}_{+}/N_{S}^{+}$ equals zero at $p\!=\!0$
or $1$, and reaches its maximum $8\!\left[1\!+\!3N_{B}\!-\!\sqrt{8\left(N_{B}\!+\!N_{B}^{2}\right)}\right]\!/\!\left(1\!-\!N_{B}\right)^{2}$
at $p\!=\!\left[\sqrt{2\left(N_{B}\!+\!N_{B}^{2}\right)}\!-\!N_{B}\right]\!/\!\left(2\!+\!N_{B}\right)$.
In contrast, for $\vert\psi_{-}\rangle_{SI}$, the averaged QFI $\mathcal{F}_{-}/N_{S}^{-}$
attains its maximum $4/\!\left(1\!+\!N_{B}\right)$ at $p\!=\!0$.
These examples illustrate that while Holland-Burnett states are ineffective
for QI, a superposition of Holland-Burnett states of different photon
numbers can be useful. Assigning a higher weight to the vacuum component
in the superposition state offers greater advantages for target detection.

\subsection{Comparison of the correlation function \label{subsec:classical illumination-1} }

For further insight, we use the coherent state (given in Eq.~\eqref{eq:classicalQFI})
as a benchmark. Following \citep{Sanz2017PRL}, we define the ratio
of $\mathcal{F}_{\pm\kappa}/\mathcal{F}_{\alpha}$ to quantify the
quantum advantage achievable with the non-Gaussian states. Figures.~\ref{fig:Averaged-QFI}(c)
and \ref{fig:Averaged-QFI}(d) show this ratio for the MPA-TMSV and
MPS-TMSV states as a function of $N_{S}$, assuming the coherent state
has the same signal energy as the non-Gaussian states. The observed
trends are consistent with those in Figs.~\ref{fig:Averaged-QFI}(a)
and \ref{fig:Averaged-QFI}(b). Neither photon addition nor photon
subtraction provides an advantage over the protocol without these
modifications. In fact, the MPA-TMSV states appear to be ineffective
for QI, while the MPS-TMSV states show a similar advantage to the
TMSV state, albeit slightly less.

Finally, we provide insight into the these phenomena by examining
the normalized cross-correlation function, defined as \citep{Glauber1963PR}
\begin{eqnarray}
g_{SI}^{\left(2\right)} & \equiv & \frac{\big\langle a_{S}^{\dagger}a_{S}a_{I}^{\dagger}a_{I}\big\rangle}{\big\langle a_{S}^{\dagger}a_{S}\big\rangle\big\langle a_{I}^{\dagger}a_{I}\big\rangle}.\label{eq:correlation}
\end{eqnarray}
This function characterizes the intensity correlation between the
two distinct modes, and it can be measured using the Hanbury-Brown
and Twiss interferometer method \citep{BROWN1956nature}. In our cases,
where the two modes exhibit symmetric photon statistics ($\big\langle a_{S}^{\dagger}a_{S}\big\rangle\!=\!\big\langle a_{I}^{\dagger}a_{I}\big\rangle\!=\!\!N_{S}$),
the expression simplifies to $g_{SI}^{\left(2\right)}\!=\!\big\langle a_{S}^{\dagger}a_{S}a_{I}^{\dagger}a_{I}\big\rangle/N_{S}^{2}.$
For the TMSV state, we have $g_{SI}^{\left(2\right)}\!=\!2+1/N_{S}$.
The normalized cross-correlation functions for the MPA-TMSV and MPS-TMSV
states are obtained (see Appendix A for details) as

\begin{eqnarray}
g_{SI}^{\left(2\right)} & = & 1+\frac{\mathcal{N}_{\kappa+1,\kappa+1}^{+}\mathcal{N}_{\kappa,\kappa}^{+}-\left(\mathcal{N}_{\kappa+1,\kappa}^{+}\right)^{2}}{\left(\mathcal{N}_{\kappa+1,\kappa}^{+}-\mathcal{N}_{\kappa,\kappa}^{+}\right)^{2}},\label{eq:PhotonAdditionCorrelation}
\end{eqnarray}
and 
\begin{eqnarray}
g_{SI}^{\left(2\right)} & = & \frac{\mathcal{N}_{\kappa+1,\kappa+1}^{-}\mathcal{N}_{\kappa,\kappa}^{-}}{\left(\mathcal{N}_{\kappa+1,\kappa}^{-}\right)^{2}},\label{eq:PhotonSubtractionCorrelation}
\end{eqnarray}
respectively. Here, we introduce the notations
\begin{eqnarray}
\mathcal{N}_{\kappa,\iota}^{+} & = & \left(1-z^{2}\right)\!\sum_{n=0}^{\infty}\frac{z^{2n}\left(n+\kappa\right)!\left(n+\iota\right)!}{\left(n!\right)^{2}},
\end{eqnarray}
and 
\begin{eqnarray}
\mathcal{N}_{\kappa,\iota}^{-} & = & \left(1-z^{2}\right)\!\sum_{n=\kappa}^{\infty}\frac{z^{2n}\left(n!\right)^{2}}{\left(n-\kappa\right)!\left(n-\iota\right)!},\left(\kappa\geq\iota\right).
\end{eqnarray}
We plot these results in Figs.~\ref{fig:Averaged-QFI}(e) and \ref{fig:Averaged-QFI}(f)
as a function of $N_{S}.$ Interestingly, the trends are analogous
to those shown in Figs.~\ref{fig:Averaged-QFI}(c) and \ref{fig:Averaged-QFI}(d),
indicating that neither photon addition nor photon subtraction enhances
the correlations for a given signal strength. Experimental evidence
supporting this observation has been presented for the TMSV state
\citep{England2019PRA}. Hence, the normalized cross-correlation function
provides valuable insight into which resources contribute to the quantum
advantage in QI, a topic that has attracted significant attention
\citep{Weedbrook2016NJP,Bradshaw2017PRA,Yung2020npj,Kim2023QINP}. 

\section{Conclusion \label{sec:Conclution}}

The two figures of merit, SNR and QFI, which arise from state discrimination
theory and parameter estimation theory, respectively, establish upper
bounds on the minimum error probability in target detection tasks
when using local measurement strategies. These metrics have often
been considered separately to evaluate target detection protocols,
without revealing their intrinsic connection. In this paper, we dealt
with this problem by unveiling the relationship between SNR and QFI
under the assumption of low object reflectivity. This connection not
only enriches the theoretical link between state discrimination and
parameter estimation but also provides a clear operational criterion
for identifying optimal measurements in target detection tasks. 

Furthermore, we applied this formalism to two specific target detection
protocols: one within the framework of CI using generalized coherent
states, and the other within the QI framework using MPA-TMSV and MPS-TMSV
states. We derived the upper bounds on the minimum error probability
for these protocols and identified the optimal measurements needed
to reach these bounds. Additionally, we provided a deep insight into
the role of photon addition and photon subtraction in QI, demonstrating
that these non-Gaussian operations do not yield a performance improvement
over protocols without them. Finally, our paper suggests that the
normalized cross-correlation function may be responsible for the quantum
advantage in QI.

\section*{Acknowledgments}

We thank Xiao-Ming Lu and Yang Li for their helpful discussions and
the two anonymous referees for their enlightening comments and suggestions.
This work was supported by the National Natural Science Foundation
of China through Grant No. 12005106. In lovely memory of my father
by W.Z..

\section*{Appendix A: Derivation of QFI for CI\label{sec:AppendixAA}}

\makeatletter
\renewcommand{\theequation}{A\arabic{equation}}
\makeatother 
\setcounter{equation}{0}

We provide a detailed derivation of the QFI for CI using a generic
probe state $\vert\psi\rangle$. Using Eq.~\eqref{eq:SLD}, the QFI
in Eq.~\eqref{eq:QFI} can be rewritten as 
\begin{eqnarray}
\mathcal{F} & = & {\rm Tr}\!\left(\partial_{\eta}\rho_{\eta}\,L\right).\label{eq:QFI_pratialrho_sld}
\end{eqnarray}
In contrast to QI, the probe state in CI involves only the signal
mode $\rho_{S}\!=\!\vert\psi\rangle_{S}\langle\psi\vert$, omitting
the idler. In this case, $\rho_{\eta}$ (corresponding to Eq.~\eqref{eq:eta-density-matrix})
is given by
\begin{eqnarray}
\rho_{\eta} & = & {\rm Tr}_{S}\!\left[U\!\left(\eta\right)\vert\psi\rangle_{S}\langle\psi\vert\otimes\rho_{B}U^{\dagger}\!\left(\eta\right)\right].\label{eq:rho_CI}
\end{eqnarray}
In the limit $\eta\!\rightarrow\!0$, $\rho_{\eta}$ approximates
$\rho_{B}$. From Eq.~\eqref{eq:SLD} we have
\begin{eqnarray}
\langle\mu^{\prime}\vert\partial_{\eta}\rho_{\eta}\vert\mu\rangle & = & \frac{1}{2}\left(\varrho_{\mu^{\prime}}+\varrho_{\mu}\right)\langle\mu^{\prime}\vert L\vert\mu\rangle,
\end{eqnarray}
which leads to
\begin{eqnarray}
\langle\mu^{\prime}\vert L\vert\mu\rangle & = & \frac{2\langle\mu^{\prime}\vert\partial_{\eta}\rho_{\eta}\vert\mu\rangle}{\varrho_{\mu^{\prime}}+\varrho_{\mu}}.\label{eq:SLD_CI}
\end{eqnarray}
Expanding in the single-mode reflected basis $\left\{ \vert\mu\rangle\right\} _{\mu=0}^{\infty}$
and using Eq.~\eqref{eq:SLD_CI}, the QFI defined by Eq.~\eqref{eq:QFI_pratialrho_sld}
becomes
\begin{eqnarray}
\mathcal{F}_{{\rm CI}} & = & \sum_{\mu,\mu^{\prime}=0}^{\infty}\!\langle\mu\vert\partial_{\eta}\rho_{\eta}\vert\mu^{\prime}\rangle\langle\mu^{\prime}\vert L\vert\mu\rangle\nonumber \\
 & = & 2\sum_{\mu,\mu^{\prime}=0}^{\infty}\!\frac{\left|\langle\mu\vert\partial_{\eta}\rho_{\eta}\vert\mu^{\prime}\rangle\right|^{2}}{\varrho_{\mu}+\varrho_{\mu^{\prime}}}.\label{eq:QFI_CI}
\end{eqnarray}
To proceed, we identify $\left|\langle\mu\vert\partial_{\eta}\rho_{\eta}\vert\mu^{\prime}\rangle\right|^{2}$.
From Eq.~\eqref{eq:rho_CI}, in the limit $\eta\!\rightarrow\!0$,
we have 
\begin{eqnarray}
\partial_{\eta}\rho_{\eta} & = & {\rm Tr}_{S}\!\left[\left(a_{S}^{\dagger}b-a_{S}b^{\dagger}\right),\vert\psi\rangle_{S}\langle\psi\vert\otimes\rho_{B}\right]\nonumber \\
 & = & \langle\psi\vert a_{S}^{\dagger}\vert\psi\rangle\!\left(b\rho_{B}\!-\!\rho_{B}b\right)-\langle\psi\vert a_{S}\vert\psi\rangle\!\left(b^{\dagger}\rho_{B}\!-\!\rho_{B}b^{\dagger}\right).\quad\quad
\end{eqnarray}
This yields
\begin{align}
\langle\mu\vert\partial_{\eta} & \rho_{\eta}\vert\mu^{\prime}\rangle=\left(\varrho_{\mu^{\prime}}\!-\!\varrho_{\mu}\right)\times\nonumber \\
 & \left(\langle\psi\vert a_{S}^{\dagger}\vert\psi\rangle\sqrt{\mu^{\prime}}\delta_{\mu,\mu^{\prime}-1}-\langle\psi\vert a_{S}\vert\psi\rangle\sqrt{\mu^{\prime}+1}\delta_{\mu,\mu^{\prime}+1}\right).
\end{align}
Taking the modulus squared gives
\begin{align}
\big\vert\langle\mu\vert\partial_{\eta} & \rho_{\eta}\vert\mu^{\prime}\rangle\big\vert^{2}=\left|\langle\psi\vert a_{S}\vert\psi\rangle\right|^{2}\times\nonumber \\
 & \left(\varrho_{\mu^{\prime}}\!-\!\varrho_{\mu}\right)^{2}\left[\mu^{\prime}\delta_{\mu,\mu^{\prime}-1}+\left(\mu^{\prime}+1\right)\delta_{\mu,\mu^{\prime}+1}\right],
\end{align}
where the cross terms vanish due to $\delta_{\mu,\mu^{\prime}-1}\delta_{\mu,\mu^{\prime}+1}\!=\!0$.
Submitting into Eq.~\eqref{eq:QFI_CI} finally yields
\begin{align}
\mathcal{F}_{{\rm CI}}= & 2\left|\langle\psi\vert a_{S}\vert\psi\rangle\right|^{2}\times\nonumber \\
 & \left[\sum_{\mu=0}^{\infty}\frac{\left(\varrho_{\mu+1}-\varrho_{\mu}\right)^{2}}{\varrho_{\mu}+\varrho_{\mu+1}}\left(\mu+1\right)+\sum_{\mu=0}^{\infty}\frac{\left(\varrho_{\mu-1}-\varrho_{\mu}\right)^{2}}{\varrho_{\mu}+\varrho_{\mu-1}}\mu\right]\nonumber \\
= & 4\left|\langle\psi\vert a_{S}\vert\psi\rangle\right|^{2}\sum_{\mu=1}^{\infty}\frac{\left(\varrho_{\mu-1}-\varrho_{\mu}\right)^{2}}{\varrho_{\mu}+\varrho_{\mu-1}}\mu\nonumber \\
= & \frac{4\left|\langle\psi\vert a_{S}\vert\psi\rangle\right|^{2}}{2N_{B}+1},\label{eq:QFI_single}
\end{align}
where in the last equality we used the identities
\begin{equation}
\varrho_{\mu-1}=\varrho_{\mu}\!\left(\frac{N_{B}+1}{N_{B}}\right)\;\text{and}\;\sum_{\mu=1}^{\infty}\mu\varrho_{\mu}=N_{B}.\label{eq:rhoB_identities}
\end{equation}

\section*{Appendix B: Derivation of QFI for QI with  MPA-TMSV and MPS-TMSV
states \label{sec:AppendixBB}}

\makeatletter
\renewcommand{\theequation}{B\arabic{equation}}
\makeatother
\setcounter{equation}{0}

We now provide a detailed derivation of the QFI for the MPA-TMSV and
MPS-TMSV states. The derivation is analogous to that for CI, though
somewhat more involved. In the QI framework considered in the main
context, the detected state is given by Eq.~\eqref{eq:eta-density-matrix}
as
\begin{eqnarray}
\rho_{\eta} & = & \rho_{RI}\left(\eta\right)={\rm Tr}_{S}\!\left[U_{\eta}\rho_{SI}\otimes\rho_{B}U_{\eta}{}^{\dagger}\right],\label{eq:rho_reflectsdler}
\end{eqnarray}
where $\rho_{SI}\!=\!\vert{\rm TMSV};\pm\kappa\rangle_{SI}\langle{\rm TMSV};\pm\kappa\vert$
corresponds to the MPA-TMSV or MPS-TMSV states as defined in Eq.~\eqref{eq:PA=000026PS-TMSV}.
In the limit $\eta\!\rightarrow\!0$, $\rho_{\eta}$ approximates
$\rho_{\eta}\!=\!\rho_{I}\!\otimes\!\rho_{B}$, where $\rho_{I}$
is the reduced density matrix of the idler given by\begin{widetext}
\begin{equation}
\rho_{I}={\rm Tr}_{S}\!\left(\rho_{SI}\right)=\sum_{n=0}^{\infty}\!\left[C_{n}^{\left(\pm\kappa\right)}\right]^{2}\!\!h\left(n-n_{\pm}\right)\vert n\pm\kappa\rangle_{I}\langle n\pm\kappa\vert.\label{eq:rhoidler}
\end{equation}
We here shift the summation index from $n_{\pm}$ to zero, i.e., $\sum_{n=n_{\pm}}^{\infty}\!\!=\!\sum_{n=0}^{\infty}h\!\left(n-n_{\pm}\right)$,
by introducing the Heaviside function defined as $h\!\left(n\right)\!=\!\begin{cases}
0,\!\! & \!\!\!n\!<\!0\\
1,\!\! & \!\!\!n\!\geq\!0
\end{cases}$ for convenience in our calculations. From Eq.~\eqref{eq:SLD} and
using Eq.~\eqref{eq:rhoidler}, we have
\begin{eqnarray}
\langle n^{\prime},\mu^{\prime}\vert\partial_{\eta}\rho_{\eta}\vert n,\mu\rangle & = & \langle n^{\prime},\mu^{\prime}\vert\frac{1}{2}\!\left(\rho_{\eta}L+L\rho_{\eta}\right)\vert n,\mu\rangle\nonumber \\
 & = & \frac{1}{2}\!\left\{ \!\left[C_{n^{\prime}\mp\kappa}^{\left(\pm\kappa\right)}\right]^{2}\!\!\varrho_{\mu^{\prime}}h\left(n^{\prime}\mp\kappa-n_{\pm}\right)+\left[C_{n\mp\kappa}^{\left(\pm\kappa\right)}\right]^{2}\!\!\varrho_{\mu}h\left(n\mp\kappa-n_{\pm}\right)\right\} \!\langle n^{\prime},\mu^{\prime}\vert L\vert n,\mu\rangle.
\end{eqnarray}
Thus, we identify 
\begin{eqnarray}
\langle n^{\prime},\mu^{\prime}\vert L\vert n,\mu\rangle & = & \frac{2\langle n^{\prime},\mu^{\prime}\vert\partial_{\eta}\rho_{\eta}\vert n,\mu\rangle}{\left[C_{n^{\prime}\mp\kappa}^{\left(\pm\kappa\right)}\right]^{2}\!\!\varrho_{\mu^{\prime}}h\left(n^{\prime}\mp\kappa-n_{\pm}\right)+\left[C_{n\mp\kappa}^{\left(\pm\kappa\right)}\right]^{2}\!\!\varrho_{\mu}h\left(n\mp\kappa-n_{\pm}\right)},\label{eq:SLD_element}
\end{eqnarray}
Expanding the QFI as defined by Eq.~\eqref{eq:QFI_pratialrho_sld}
in the compound reflected-idler basis $\left\{ \vert n,\mu\rangle\right\} _{n,\mu=0}^{\infty}$
and using completeness relation $\sum_{n^{\prime},\mu^{\prime}=0}^{\infty}\vert n^{\prime},\mu^{\prime}\rangle\langle n^{\prime},\mu^{\prime}\vert\!=\!\openone$,
we obtain

\begin{eqnarray}
\mathcal{F_{\pm\kappa}} & = & \sum_{n,n^{\prime}=0}^{\infty}\sum_{\mu,\mu^{\prime}=0}^{\infty}\langle n,\mu\vert\partial_{\eta}\rho_{\eta}\vert n^{\prime},\mu^{\prime}\rangle\langle n^{\prime},\mu^{\prime}\vert L\vert n,\mu\rangle\nonumber \\
 & = & \sum_{n,n^{\prime}=n_{\pm}}^{\infty}\sum_{\mu,\mu^{\prime}=0}^{\infty}\frac{2\left|\langle n,\mu\vert\partial_{\eta}\rho_{\eta}\vert n^{\prime},\mu^{\prime}\rangle\right|^{2}}{\left[C_{n^{\prime}\mp\kappa}^{\left(\pm\kappa\right)}\right]^{2}\!\!\varrho_{\mu^{\prime}}h\left(n^{\prime}\mp\kappa\right)+\left[C_{n\mp\kappa}^{\left(\pm\kappa\right)}\right]^{2}\!\!\varrho_{\mu}h\left(n\mp\kappa\right)},\label{eq:QFI_partialrho}
\end{eqnarray}
where in the second equality we used Eq.~\eqref{eq:SLD_element}.
From Eq.~\eqref{eq:rho_reflectsdler}, in the limit $\eta\!\rightarrow\!0$,
we have
\begin{eqnarray}
\partial_{\eta}\rho_{\eta} & = & {\rm Tr}_{S}\left[\left(a_{S}^{\dagger}b-a_{S}b^{\dagger}\right),\rho_{SI}\otimes\rho_{B}\right]\nonumber \\
 & = & \sum_{m=n_{\pm}}^{\infty}\!\!C_{m}^{\left(\pm\kappa\right)}C_{m+1}^{\left(\pm\kappa\right)}\sqrt{m\pm\kappa+1}\vert m\pm\kappa\rangle_{I}\langle m\pm\kappa+1\vert\left(b\rho_{B}-\rho_{B}b\right)\nonumber \\
 &  & -\sum_{m=n_{\pm}+1}^{\infty}\!\!C_{m}^{\left(\pm\kappa\right)}C_{m-1}^{\left(\pm\kappa\right)}\sqrt{m\pm\kappa}\vert m\pm\kappa\rangle_{I}\langle m\pm\kappa-1\vert\left(b^{\dagger}\rho_{B}-\rho_{B}b^{\dagger}\right).
\end{eqnarray}
Multiplying $\langle n,\mu\vert$ and $\vert n^{\prime},\mu^{\prime}\rangle$
gives
\begin{eqnarray}
\langle n,\mu\vert\partial_{\eta}\rho_{\eta}\vert n^{\prime},\mu^{\prime}\rangle & = & \sum_{m=n_{\pm}}^{\infty}\!\!C_{m}^{\left(\pm\kappa\right)}C_{m+1}^{\left(\pm\kappa\right)}\left(\varrho_{\mu^{\prime}}-\varrho_{\mu}\right)\sqrt{m\pm\kappa+1}\sqrt{\mu^{\prime}}\;\delta_{n,m\pm\kappa}\delta_{n^{\prime},m\pm\kappa+1}\delta_{\mu,\mu^{\prime}-1}\nonumber \\
 &  & -\sum_{m=n_{\pm}+1}^{\infty}\!\!C_{m}^{\left(\pm\kappa\right)}C_{m-1}^{\left(\pm\kappa\right)}\left(\varrho_{\mu^{\prime}}-\varrho_{\mu}\right)\sqrt{m\pm\kappa}\sqrt{\mu^{\prime}+1}\;\delta_{n,m\pm\kappa}\delta_{n^{\prime},m\pm\kappa-1}\delta_{\mu,\mu^{\prime}+1},
\end{eqnarray}
where the Kronecker delta $\delta_{ij}$ is defined as $\delta_{ij}\!=\!\begin{cases}
1,\!\! & \!\!\!i\!=\!j\\
0,\!\! & \!\!\!i\!\neq\!j
\end{cases}$. Then, taking the modulus squared we arrive at 
\begin{eqnarray}
\left|\langle n,\mu\vert\partial_{\eta}\rho_{\eta}\vert n^{\prime},\mu^{\prime}\rangle\right|^{2} & = & \sum_{m=n_{\pm}}^{\infty}\!\!\left[C_{m}^{\left(\pm\kappa\right)}C_{m+1}^{\left(\pm\kappa\right)}\right]^{2}\!\!\left(\varrho_{\mu^{\prime}}-\varrho_{\mu}\right)^{2}\left(m\pm\kappa+1\right)\mu^{\prime}\;\delta_{n,m\pm\kappa}\delta_{n^{\prime},m\pm\kappa+1}\delta_{\mu,\mu^{\prime}-1}\nonumber \\
 &  & +\sum_{m=n_{\pm}+1}^{\infty}\!\!\left[C_{m}^{\left(\pm\kappa\right)}C_{m-1}^{\left(\pm\kappa\right)}\right]^{2}\!\!\left(\varrho_{\mu^{\prime}}-\varrho_{\mu}\right)^{2}\left(m\pm\kappa\right)\left(\mu^{\prime}+1\right)\;\delta_{n,m\pm\kappa}\delta_{n^{\prime},m\pm\kappa-1}\delta_{\mu,\mu^{\prime}+1}.
\end{eqnarray}
with cross terms vanishing due to $\delta_{\mu,\mu^{\prime}-1}\delta_{\mu,\mu^{\prime}+1}=0$.
Submitting this expression into Eq.~\eqref{eq:QFI_partialrho} finally
yields
\begin{eqnarray}
\mathcal{F_{\pm\kappa}} & = & 2\sum_{\mu=0}^{\infty}\sum_{n=n_{\pm}}^{\infty}\!\!\frac{\left[C_{n}^{\left(\pm\kappa\right)}C_{n+1}^{\left(\pm\kappa\right)}\right]^{2}\!\!\left(\varrho_{\mu+1}-\varrho_{\mu}\right)^{2}\!\!\left(n\pm\kappa+1\right)\left(\mu+1\right)}{\left[C_{n}^{\left(\pm\kappa\right)}\right]^{2}\!\!\varrho_{\mu}+\left[C_{n+1}^{\left(\pm\kappa\right)}\right]^{2}\!\!\varrho_{\mu+1}}+2\sum_{\mu=0}^{\infty}\sum_{n=n_{\pm}+1}^{\infty}\!\!\frac{\left[C_{n}^{\left(\pm\kappa\right)}C_{n-1}^{\left(\pm\kappa\right)}\right]^{2}\!\!\left(\varrho_{\mu-1}-\varrho_{\mu}\right)^{2}\left(n\pm\kappa\right)\mu}{\left[C_{n}^{\left(\pm\kappa\right)}\right]^{2}\!\!\varrho_{\mu}+\left[C_{n-1}^{\left(\pm\kappa\right)}\right]^{2}\!\!\varrho_{\mu-1}}\nonumber \\
 & = & 4\sum_{\mu=1}^{\infty}\sum_{n=n_{\pm}+1}^{\infty}\!\!\frac{\left[C_{n}^{\left(\pm\kappa\right)}C_{n-1}^{\left(\pm\kappa\right)}\right]^{2}\!\!\left(\varrho_{\mu-1}-\varrho_{\mu}\right)^{2}}{\left[C_{n}^{\left(\pm\kappa\right)}\right]^{2}\!\!\varrho_{\mu}+\left[C_{n-1}^{\left(\pm\kappa\right)}\right]^{2}\!\!\varrho_{\mu-1}}\left(n\pm\kappa\right)\mu,\label{eq:QFI_summ_um}
\end{eqnarray}
\end{widetext}where the last equality is obtained by the fact that
the two summation terms in the first expression are equivalent. This
equivalence can be easily verified by performing a suitable change
of the dummy index. Finally, by using the identities provided in Eq.~\eqref{eq:rhoB_identities},
we obtain the QFI for the MPA-TMSV and MPS-TMSV states as given in
Eq.~\eqref{eq:PA=000026PS-QFI} in the main text.

\section*{Appendix C: Derivation of the normalized cross-correlation function
for the MPA-TMSV and MPS-TMSV states\label{sec:AppendixC}}

\makeatletter
\renewcommand{\theequation}{C\arabic{equation}}
\makeatother 
\setcounter{equation}{0}

We now present a detailed derivation of the normalized cross-correlation
function for both the MPA-TMSV and MPS-TMSV states. To facilitate
this, we introduce a more general class of states that allow arbitrary
photon additions or subtractions in the signal and idler modes without
imposing the condition of equality. These states are given by \citep{Navarrete-Benlloch2012PRA}
\begin{eqnarray}
\vert{\rm TMSV};+\kappa,+\iota\rangle_{SI} & = & \frac{a_{S}^{\dagger\kappa}a_{I}^{\dagger l}}{\sqrt{\mathcal{N}_{\kappa,l}^{+}}}\vert{\rm TMSV}\rangle_{SI},\\
\vert{\rm TMSV};-\kappa,-\iota\rangle_{SI} & = & \frac{a_{S}^{\kappa}a_{I}^{l}}{\sqrt{\mathcal{N}_{\kappa,l}^{-}}}\vert{\rm TMSV}\rangle_{SI},
\end{eqnarray}
with the normalization factors $\mathcal{N}_{\kappa,\iota}^{+}$ and
$\mathcal{N}_{\kappa,\iota}^{-}$ given by \citep{Navarrete-Benlloch2012PRA}
\begin{eqnarray}
\mathcal{N}_{\kappa,\iota}^{+} & \equiv & \langle{\rm TMSV}\vert a_{S}^{\kappa}a_{I}^{\iota}a_{S}^{\dagger\kappa}a_{I}^{\dagger\iota}\vert{\rm TMSV}\rangle\nonumber \\
 & = & \left(1-z^{2}\right)\!\sum_{n=0}^{\infty}\frac{z^{2n}\left(n+\kappa\right)!\left(n+\iota\right)!}{\left(n!\right)^{2}},\label{eq:PhotonAdditioinNormalization}
\end{eqnarray}
and 
\begin{eqnarray}
\mathcal{N}_{\kappa,\iota}^{-} & \equiv & \langle{\rm TMSV}\vert a_{S}^{\dagger\kappa}a_{I}^{\dagger\iota}a_{S}^{\kappa}a_{I}^{\iota}\vert{\rm TMSV}\rangle\nonumber \\
 & = & \left(1-z^{2}\right)\!\sum_{n=\kappa}^{\infty}\frac{z^{2n}\left(n!\right)^{2}}{\left(n-\kappa\right)!\left(n-\iota\right)!},\left(\kappa\geq\iota\right).\label{eq:PhotonSubtractionNormalization}
\end{eqnarray}
When we set $\kappa\!=\!\iota$, these expressions reduce to those
used in the main text. The derivation of the normalized cross-correlation
function requires the computation of expectation values $\big\langle a_{S}^{\dagger}a_{S}\big\rangle$,
$\big\langle a_{I}^{\dagger}a_{I}\big\rangle$ and $\big\langle a_{S}^{\dagger}a_{S}a_{I}^{\dagger}a_{I}\big\rangle$.
In our cases, the first two are equal ($\big\langle a_{S}^{\dagger}a_{S}\big\rangle\!=\!\big\langle a_{I}^{\dagger}a_{I}\big\rangle\!=\!N_{S}$)
and have been obtained in Eq.~\eqref{eq:Ns}. Here, we provide alternative
expressions for these expectations in terms of the normalization factors.
For example, for $\vert{\rm TMSV};+\kappa\rangle_{\!S\!I}$, one can
directly derive
\begin{eqnarray}
\big\langle a_{S}^{\dagger}a_{S}\big\rangle & = & \frac{1}{\mathcal{N}_{\kappa,\kappa}^{+}}\langle{\rm TMSV}\vert a_{S}^{\kappa}a_{I}^{\iota}a_{S}^{\dagger}a_{S}a_{S}^{\dagger\kappa}a_{I}^{\dagger\iota}\vert{\rm TMSV}\rangle\nonumber \\
 & = & \frac{1}{\mathcal{N}_{\kappa,\kappa}^{+}}\langle{\rm TMSV}\vert a_{S}^{\kappa}\big(a_{S}a_{S}^{\dagger}-1\big)a_{S}^{\dagger\kappa}a_{I}^{\iota}a_{I}^{\dagger\iota}\vert{\rm TMSV}\rangle\nonumber \\
 & = & \frac{\mathcal{N}_{\kappa+1,\kappa}^{+}-\mathcal{N}_{\kappa,\kappa}^{+}}{\mathcal{N}_{\kappa,\kappa}^{+}}.
\end{eqnarray}
Following a similar procedure, one obtains 
\begin{eqnarray}
\big\langle a_{S}^{\dagger}a_{S}a_{I}^{\dagger}a_{I}\big\rangle & = & \frac{\mathcal{N}_{\kappa+1,\kappa+1}^{+}-2\mathcal{N}_{\kappa+1,\kappa}^{+}+\mathcal{N}_{\kappa,\kappa}^{+}}{\mathcal{N}_{\kappa,\kappa}^{+}}.
\end{eqnarray}
Submitting these results into Eq.~\eqref{eq:correlation} yields
the expression for the normalized cross-correlation function given
in Eq.~\eqref{eq:PhotonAdditionCorrelation}. Similarly, for $\vert{\rm TMSV};-\kappa\rangle_{SI}$,
we have 
\begin{eqnarray}
\big\langle a_{S}^{\dagger}a_{S}\big\rangle & = & \frac{\mathcal{N}_{\kappa+1,\kappa}^{-}}{\mathcal{N}_{\kappa,\kappa}^{-}},\\
\big\langle a_{S}^{\dagger}a_{S}a_{I}^{\dagger}a_{I}\big\rangle & = & \frac{\mathcal{N}_{\kappa+1,\kappa+1}^{-}}{\mathcal{N}_{\kappa,\kappa}^{-}},
\end{eqnarray}
which leads to Eq.~\eqref{eq:PhotonSubtractionCorrelation}.

\bibliographystyle{apsrev4-1}
\bibliography{C:/ZW/manuscripts/me/ZW}

\begin{thebibliography}{88}%
\makeatletter
\providecommand \@ifxundefined [1]{%
 \@ifx{#1\undefined}
}%
\providecommand \@ifnum [1]{%
 \ifnum #1\expandafter \@firstoftwo
 \else \expandafter \@secondoftwo
 \fi
}%
\providecommand \@ifx [1]{%
 \ifx #1\expandafter \@firstoftwo
 \else \expandafter \@secondoftwo
 \fi
}%
\providecommand \natexlab [1]{#1}%
\providecommand \enquote  [1]{``#1''}%
\providecommand \bibnamefont  [1]{#1}%
\providecommand \bibfnamefont [1]{#1}%
\providecommand \citenamefont [1]{#1}%
\providecommand \href@noop [0]{\@secondoftwo}%
\providecommand \href [0]{\begingroup \@sanitize@url \@href}%
\providecommand \@href[1]{\@@startlink{#1}\@@href}%
\providecommand \@@href[1]{\endgroup#1\@@endlink}%
\providecommand \@sanitize@url [0]{\catcode `\\12\catcode `\$12\catcode
  `\&12\catcode `\#12\catcode `\^12\catcode `\_12\catcode `\%12\relax}%
\providecommand \@@startlink[1]{}%
\providecommand \@@endlink[0]{}%
\providecommand \url  [0]{\begingroup\@sanitize@url \@url }%
\providecommand \@url [1]{\endgroup\@href {#1}{\urlprefix }}%
\providecommand \urlprefix  [0]{URL }%
\providecommand \Eprint [0]{\href }%
\providecommand \doibase [0]{http://dx.doi.org/}%
\providecommand \selectlanguage [0]{\@gobble}%
\providecommand \bibinfo  [0]{\@secondoftwo}%
\providecommand \bibfield  [0]{\@secondoftwo}%
\providecommand \translation [1]{[#1]}%
\providecommand \BibitemOpen [0]{}%
\providecommand \bibitemStop [0]{}%
\providecommand \bibitemNoStop [0]{.\EOS\space}%
\providecommand \EOS [0]{\spacefactor3000\relax}%
\providecommand \BibitemShut  [1]{\csname bibitem#1\endcsname}%
\let\auto@bib@innerbib\@empty
\bibitem [{\citenamefont {Lloyd}(2008)}]{Lloyd2008science}%
  \BibitemOpen
  \bibfield  {author} {\bibinfo {author} {\bibfnamefont {S.}~\bibnamefont
  {Lloyd}},\ }\bibfield  {title} {\emph {\bibinfo {title} {Enhanced sensitivity
  of photodetection via quantum illumination},\ }}\href {\doibase
  10.1126/science.1160627} {\bibfield  {journal} {\bibinfo  {journal}
  {Science}\ }\textbf {\bibinfo {volume} {321}},\ \bibinfo {pages} {1463}
  (\bibinfo {year} {2008})}\BibitemShut {NoStop}%
\bibitem [{\citenamefont {Pirandola}\ \emph {et~al.}(2018)\citenamefont
  {Pirandola}, \citenamefont {Bardhan}, \citenamefont {Gehring}, \citenamefont
  {Weedbrook},\ and\ \citenamefont {Lloyd}}]{Pirandola2018NPreview}%
  \BibitemOpen
  \bibfield  {author} {\bibinfo {author} {\bibfnamefont {S.}~\bibnamefont
  {Pirandola}}, \bibinfo {author} {\bibfnamefont {B.~R.}\ \bibnamefont
  {Bardhan}}, \bibinfo {author} {\bibfnamefont {T.}~\bibnamefont {Gehring}},
  \bibinfo {author} {\bibfnamefont {C.}~\bibnamefont {Weedbrook}}, \ and\
  \bibinfo {author} {\bibfnamefont {S.}~\bibnamefont {Lloyd}},\ }\bibfield
  {title} {\emph {\bibinfo {title} {Advances in photonic quantum sensing},\
  }}\href {https://doi.org/10.1038/s41566-018-0301-6} {\bibfield  {journal}
  {\bibinfo  {journal} {Nature Photonics}\ }\textbf {\bibinfo {volume} {12}},\
  \bibinfo {pages} {724} (\bibinfo {year} {2018})}\BibitemShut {NoStop}%
\bibitem [{\citenamefont {Shapiro}(2020)}]{Shapiro2020Review}%
  \BibitemOpen
  \bibfield  {author} {\bibinfo {author} {\bibfnamefont {J.~H.}\ \bibnamefont
  {Shapiro}},\ }\bibfield  {title} {\emph {\bibinfo {title} {The quantum
  illumination story},\ }}\href@noop {} {\bibfield  {journal} {\bibinfo
  {journal} {IEEE Aerospace and Electronic Systems Magazine}\ }\textbf
  {\bibinfo {volume} {35}},\ \bibinfo {pages} {8} (\bibinfo {year}
  {2020})}\BibitemShut {NoStop}%
\bibitem [{\citenamefont {Sorelli}\ \emph {et~al.}(2022)\citenamefont
  {Sorelli}, \citenamefont {Treps}, \citenamefont {Grosshans},\ and\
  \citenamefont {Boust}}]{Sorelli2022Review}%
  \BibitemOpen
  \bibfield  {author} {\bibinfo {author} {\bibfnamefont {G.}~\bibnamefont
  {Sorelli}}, \bibinfo {author} {\bibfnamefont {N.}~\bibnamefont {Treps}},
  \bibinfo {author} {\bibfnamefont {F.}~\bibnamefont {Grosshans}}, \ and\
  \bibinfo {author} {\bibfnamefont {F.}~\bibnamefont {Boust}},\ }\bibfield
  {title} {\emph {\bibinfo {title} {Detecting a target with quantum
  entanglement},\ }}\href@noop {} {\bibfield  {journal} {\bibinfo  {journal}
  {IEEE Aerospace and Electronic Systems Magazine}\ }\textbf {\bibinfo {volume}
  {37}},\ \bibinfo {pages} {68} (\bibinfo {year} {2022})}\BibitemShut {NoStop}%
\bibitem [{\citenamefont {Sacchi}(2005{\natexlab{a}})}]{Sacchi2005PRA}%
  \BibitemOpen
  \bibfield  {author} {\bibinfo {author} {\bibfnamefont {M.~F.}\ \bibnamefont
  {Sacchi}},\ }\bibfield  {title} {\emph {\bibinfo {title} {Optimal
  discrimination of quantum operations},\ }}\href {\doibase
  10.1103/PhysRevA.71.062340} {\bibfield  {journal} {\bibinfo  {journal} {Phys.
  Rev. A}\ }\textbf {\bibinfo {volume} {71}},\ \bibinfo {pages} {062340}
  (\bibinfo {year} {2005}{\natexlab{a}})}\BibitemShut {NoStop}%
\bibitem [{\citenamefont {Sacchi}(2005{\natexlab{b}})}]{Sacchi2005PRA-report}%
  \BibitemOpen
  \bibfield  {author} {\bibinfo {author} {\bibfnamefont {M.~F.}\ \bibnamefont
  {Sacchi}},\ }\bibfield  {title} {\emph {\bibinfo {title} {Entanglement can
  enhance the distinguishability of entanglement-breaking channels},\ }}\href
  {\doibase 10.1103/PhysRevA.72.014305} {\bibfield  {journal} {\bibinfo
  {journal} {Phys. Rev. A}\ }\textbf {\bibinfo {volume} {72}},\ \bibinfo
  {pages} {014305} (\bibinfo {year} {2005}{\natexlab{b}})}\BibitemShut
  {NoStop}%
\bibitem [{\citenamefont {Tan}\ \emph {et~al.}(2008)\citenamefont {Tan},
  \citenamefont {Erkmen}, \citenamefont {Giovannetti}, \citenamefont {Guha},
  \citenamefont {Lloyd}, \citenamefont {Maccone}, \citenamefont {Pirandola},\
  and\ \citenamefont {Shapiro}}]{Tan2008PRL}%
  \BibitemOpen
  \bibfield  {author} {\bibinfo {author} {\bibfnamefont {S.-H.}\ \bibnamefont
  {Tan}}, \bibinfo {author} {\bibfnamefont {B.~I.}\ \bibnamefont {Erkmen}},
  \bibinfo {author} {\bibfnamefont {V.}~\bibnamefont {Giovannetti}}, \bibinfo
  {author} {\bibfnamefont {S.}~\bibnamefont {Guha}}, \bibinfo {author}
  {\bibfnamefont {S.}~\bibnamefont {Lloyd}}, \bibinfo {author} {\bibfnamefont
  {L.}~\bibnamefont {Maccone}}, \bibinfo {author} {\bibfnamefont
  {S.}~\bibnamefont {Pirandola}}, \ and\ \bibinfo {author} {\bibfnamefont
  {J.~H.}\ \bibnamefont {Shapiro}},\ }\bibfield  {title} {\emph {\bibinfo
  {title} {Quantum illumination with gaussian states},\ }}\href {\doibase
  10.1103/PhysRevLett.101.253601} {\bibfield  {journal} {\bibinfo  {journal}
  {Phys. Rev. Lett.}\ }\textbf {\bibinfo {volume} {101}},\ \bibinfo {pages}
  {253601} (\bibinfo {year} {2008})}\BibitemShut {NoStop}%
\bibitem [{\citenamefont {Shapiro}\ and\ \citenamefont
  {Lloyd}(2009)}]{Shapiro2009NJP}%
  \BibitemOpen
  \bibfield  {author} {\bibinfo {author} {\bibfnamefont {J.~H.}\ \bibnamefont
  {Shapiro}}\ and\ \bibinfo {author} {\bibfnamefont {S.}~\bibnamefont
  {Lloyd}},\ }\bibfield  {title} {\emph {\bibinfo {title} {Quantum illumination
  versus coherent-state target detection},\ }}\href {\doibase
  10.1088/1367-2630/11/6/063045} {\bibfield  {journal} {\bibinfo  {journal}
  {New Journal of Physics}\ }\textbf {\bibinfo {volume} {11}},\ \bibinfo
  {pages} {063045} (\bibinfo {year} {2009})}\BibitemShut {NoStop}%
\bibitem [{\citenamefont {Guha}\ and\ \citenamefont
  {Erkmen}(2009)}]{Guha2009PRA}%
  \BibitemOpen
  \bibfield  {author} {\bibinfo {author} {\bibfnamefont {S.}~\bibnamefont
  {Guha}}\ and\ \bibinfo {author} {\bibfnamefont {B.~I.}\ \bibnamefont
  {Erkmen}},\ }\bibfield  {title} {\emph {\bibinfo {title} {Gaussian-state
  quantum-illumination receivers for target detection},\ }}\href {\doibase
  10.1103/PhysRevA.80.052310} {\bibfield  {journal} {\bibinfo  {journal} {Phys.
  Rev. A}\ }\textbf {\bibinfo {volume} {80}},\ \bibinfo {pages} {052310}
  (\bibinfo {year} {2009})}\BibitemShut {NoStop}%
\bibitem [{\citenamefont {Barzanjeh}\ \emph {et~al.}(2015)\citenamefont
  {Barzanjeh}, \citenamefont {Guha}, \citenamefont {Weedbrook}, \citenamefont
  {Vitali}, \citenamefont {Shapiro},\ and\ \citenamefont
  {Pirandola}}]{Barzanjeh2015PRL}%
  \BibitemOpen
  \bibfield  {author} {\bibinfo {author} {\bibfnamefont {S.}~\bibnamefont
  {Barzanjeh}}, \bibinfo {author} {\bibfnamefont {S.}~\bibnamefont {Guha}},
  \bibinfo {author} {\bibfnamefont {C.}~\bibnamefont {Weedbrook}}, \bibinfo
  {author} {\bibfnamefont {D.}~\bibnamefont {Vitali}}, \bibinfo {author}
  {\bibfnamefont {J.~H.}\ \bibnamefont {Shapiro}}, \ and\ \bibinfo {author}
  {\bibfnamefont {S.}~\bibnamefont {Pirandola}},\ }\bibfield  {title} {\emph
  {\bibinfo {title} {Microwave quantum illumination},\ }}\href {\doibase
  10.1103/PhysRevLett.114.080503} {\bibfield  {journal} {\bibinfo  {journal}
  {Phys. Rev. Lett.}\ }\textbf {\bibinfo {volume} {114}},\ \bibinfo {pages}
  {080503} (\bibinfo {year} {2015})}\BibitemShut {NoStop}%
\bibitem [{\citenamefont {Zhuang}\ \emph
  {et~al.}(2017{\natexlab{a}})\citenamefont {Zhuang}, \citenamefont {Zhang},\
  and\ \citenamefont {Shapiro}}]{Zhuang2017PRL}%
  \BibitemOpen
  \bibfield  {author} {\bibinfo {author} {\bibfnamefont {Q.}~\bibnamefont
  {Zhuang}}, \bibinfo {author} {\bibfnamefont {Z.}~\bibnamefont {Zhang}}, \
  and\ \bibinfo {author} {\bibfnamefont {J.~H.}\ \bibnamefont {Shapiro}},\
  }\bibfield  {title} {\emph {\bibinfo {title} {Optimum mixed-state
  discrimination for noisy entanglement-enhanced sensing},\ }}\href {\doibase
  10.1103/PhysRevLett.118.040801} {\bibfield  {journal} {\bibinfo  {journal}
  {Phys. Rev. Lett.}\ }\textbf {\bibinfo {volume} {118}},\ \bibinfo {pages}
  {040801} (\bibinfo {year} {2017}{\natexlab{a}})}\BibitemShut {NoStop}%
\bibitem [{\citenamefont {Zhuang}\ \emph
  {et~al.}(2017{\natexlab{b}})\citenamefont {Zhuang}, \citenamefont {Zhang},\
  and\ \citenamefont {Shapiro}}]{Zhuang2017JOSAB}%
  \BibitemOpen
  \bibfield  {author} {\bibinfo {author} {\bibfnamefont {Q.}~\bibnamefont
  {Zhuang}}, \bibinfo {author} {\bibfnamefont {Z.}~\bibnamefont {Zhang}}, \
  and\ \bibinfo {author} {\bibfnamefont {J.~H.}\ \bibnamefont {Shapiro}},\
  }\bibfield  {title} {\emph {\bibinfo {title} {Entanglement-enhanced
  neyman-pearson target detection using quantum illumination},\ }}\href
  {\doibase 10.1364/JOSAB.34.001567} {\bibfield  {journal} {\bibinfo  {journal}
  {J. Opt. Soc. Am. B}\ }\textbf {\bibinfo {volume} {34}},\ \bibinfo {pages}
  {1567} (\bibinfo {year} {2017}{\natexlab{b}})}\BibitemShut {NoStop}%
\bibitem [{\citenamefont {Yung}\ \emph {et~al.}(2020)\citenamefont {Yung},
  \citenamefont {Meng}, \citenamefont {Zhang},\ and\ \citenamefont
  {Zhao}}]{Yung2020npj}%
  \BibitemOpen
  \bibfield  {author} {\bibinfo {author} {\bibfnamefont {M.-H.}\ \bibnamefont
  {Yung}}, \bibinfo {author} {\bibfnamefont {F.}~\bibnamefont {Meng}}, \bibinfo
  {author} {\bibfnamefont {X.-M.}\ \bibnamefont {Zhang}}, \ and\ \bibinfo
  {author} {\bibfnamefont {M.-J.}\ \bibnamefont {Zhao}},\ }\bibfield  {title}
  {\emph {\bibinfo {title} {One-shot detection limits of quantum illumination
  with discrete signals},\ }}\href {https://doi.org/10.1038/s41534-020-00303-z}
  {\bibfield  {journal} {\bibinfo  {journal} {npj Quantum Information}\
  }\textbf {\bibinfo {volume} {6}},\ \bibinfo {pages} {75} (\bibinfo {year}
  {2020})}\BibitemShut {NoStop}%
\bibitem [{\citenamefont {Lee}\ \emph {et~al.}(2021)\citenamefont {Lee},
  \citenamefont {Ihn},\ and\ \citenamefont {Kim}}]{Lee2021PRA}%
  \BibitemOpen
  \bibfield  {author} {\bibinfo {author} {\bibfnamefont {S.-Y.}\ \bibnamefont
  {Lee}}, \bibinfo {author} {\bibfnamefont {Y.~S.}\ \bibnamefont {Ihn}}, \ and\
  \bibinfo {author} {\bibfnamefont {Z.}~\bibnamefont {Kim}},\ }\bibfield
  {title} {\emph {\bibinfo {title} {Quantum illumination via quantum-enhanced
  sensing},\ }}\href {\doibase 10.1103/PhysRevA.103.012411} {\bibfield
  {journal} {\bibinfo  {journal} {Phys. Rev. A}\ }\textbf {\bibinfo {volume}
  {103}},\ \bibinfo {pages} {012411} (\bibinfo {year} {2021})}\BibitemShut
  {NoStop}%
\bibitem [{\citenamefont {Jo}\ \emph {et~al.}(2021)\citenamefont {Jo},
  \citenamefont {Lee}, \citenamefont {Ihn}, \citenamefont {Kim},\ and\
  \citenamefont {Lee}}]{Jo2021PRR}%
  \BibitemOpen
  \bibfield  {author} {\bibinfo {author} {\bibfnamefont {Y.}~\bibnamefont
  {Jo}}, \bibinfo {author} {\bibfnamefont {S.}~\bibnamefont {Lee}}, \bibinfo
  {author} {\bibfnamefont {Y.~S.}\ \bibnamefont {Ihn}}, \bibinfo {author}
  {\bibfnamefont {Z.}~\bibnamefont {Kim}}, \ and\ \bibinfo {author}
  {\bibfnamefont {S.-Y.}\ \bibnamefont {Lee}},\ }\bibfield  {title} {\emph
  {\bibinfo {title} {Quantum illumination receiver using double homodyne
  detection},\ }}\href {\doibase 10.1103/PhysRevResearch.3.013006} {\bibfield
  {journal} {\bibinfo  {journal} {Phys. Rev. Research}\ }\textbf {\bibinfo
  {volume} {3}},\ \bibinfo {pages} {013006} (\bibinfo {year}
  {2021})}\BibitemShut {NoStop}%
\bibitem [{\citenamefont {{Gallego Torromé}}\ and\ \citenamefont
  {Barzanjeh}(2024)}]{GallegoTorrome2024review}%
  \BibitemOpen
  \bibfield  {author} {\bibinfo {author} {\bibfnamefont {R.}~\bibnamefont
  {{Gallego Torromé}}}\ and\ \bibinfo {author} {\bibfnamefont
  {S.}~\bibnamefont {Barzanjeh}},\ }\bibfield  {title} {\emph {\bibinfo {title}
  {Advances in quantum radar and quantum lidar},\ }}\href {\doibase
  https://doi.org/10.1016/j.pquantelec.2023.100497} {\bibfield  {journal}
  {\bibinfo  {journal} {Progress in Quantum Electronics}\ }\textbf {\bibinfo
  {volume} {93}},\ \bibinfo {pages} {100497} (\bibinfo {year}
  {2024})}\BibitemShut {NoStop}%
\bibitem [{\citenamefont {Kronowetter}\ \emph {et~al.}(2024)\citenamefont
  {Kronowetter}, \citenamefont {W\"urth}, \citenamefont {Utschick},
  \citenamefont {Gross},\ and\ \citenamefont {Fedorov}}]{Kronowetter2024PRApp}%
  \BibitemOpen
  \bibfield  {author} {\bibinfo {author} {\bibfnamefont {F.}~\bibnamefont
  {Kronowetter}}, \bibinfo {author} {\bibfnamefont {M.}~\bibnamefont
  {W\"urth}}, \bibinfo {author} {\bibfnamefont {W.}~\bibnamefont {Utschick}},
  \bibinfo {author} {\bibfnamefont {R.}~\bibnamefont {Gross}}, \ and\ \bibinfo
  {author} {\bibfnamefont {K.}~\bibnamefont {Fedorov}},\ }\bibfield  {title}
  {\emph {\bibinfo {title} {Imperfect photon detection in quantum
  illumination},\ }}\href {\doibase 10.1103/PhysRevApplied.21.014007}
  {\bibfield  {journal} {\bibinfo  {journal} {Phys. Rev. Appl.}\ }\textbf
  {\bibinfo {volume} {21}},\ \bibinfo {pages} {014007} (\bibinfo {year}
  {2024})}\BibitemShut {NoStop}%
\bibitem [{\citenamefont {Wei}\ \emph {et~al.}(2024)\citenamefont {Wei},
  \citenamefont {Albarelli}, \citenamefont {Li},\ and\ \citenamefont
  {Giovannetti}}]{wei2024quantumilluminationadvantagequantum}%
  \BibitemOpen
  \bibfield  {author} {\bibinfo {author} {\bibfnamefont {R.}~\bibnamefont
  {Wei}}, \bibinfo {author} {\bibfnamefont {F.}~\bibnamefont {Albarelli}},
  \bibinfo {author} {\bibfnamefont {J.}~\bibnamefont {Li}}, \ and\ \bibinfo
  {author} {\bibfnamefont {V.}~\bibnamefont {Giovannetti}},\ }\href
  {https://arxiv.org/abs/2411.14414} {\bibinfo {title} {Quantum illumination
  advantage in quantum doppler radar},\ } (\bibinfo {year} {2024}),\ \Eprint
  {http://arxiv.org/abs/2411.14414} {arXiv:2411.14414 [quant-ph]} \BibitemShut
  {NoStop}%
\bibitem [{\citenamefont {Lopaeva}\ \emph {et~al.}(2013)\citenamefont
  {Lopaeva}, \citenamefont {Ruo~Berchera}, \citenamefont {Degiovanni},
  \citenamefont {Olivares}, \citenamefont {Brida},\ and\ \citenamefont
  {Genovese}}]{Lopaeva2013PRL}%
  \BibitemOpen
  \bibfield  {author} {\bibinfo {author} {\bibfnamefont {E.~D.}\ \bibnamefont
  {Lopaeva}}, \bibinfo {author} {\bibfnamefont {I.}~\bibnamefont
  {Ruo~Berchera}}, \bibinfo {author} {\bibfnamefont {I.~P.}\ \bibnamefont
  {Degiovanni}}, \bibinfo {author} {\bibfnamefont {S.}~\bibnamefont
  {Olivares}}, \bibinfo {author} {\bibfnamefont {G.}~\bibnamefont {Brida}}, \
  and\ \bibinfo {author} {\bibfnamefont {M.}~\bibnamefont {Genovese}},\
  }\bibfield  {title} {\emph {\bibinfo {title} {Experimental realization of
  quantum illumination},\ }}\href {\doibase 10.1103/PhysRevLett.110.153603}
  {\bibfield  {journal} {\bibinfo  {journal} {Phys. Rev. Lett.}\ }\textbf
  {\bibinfo {volume} {110}},\ \bibinfo {pages} {153603} (\bibinfo {year}
  {2013})}\BibitemShut {NoStop}%
\bibitem [{\citenamefont {Zhang}\ \emph {et~al.}(2015)\citenamefont {Zhang},
  \citenamefont {Mouradian}, \citenamefont {Wong},\ and\ \citenamefont
  {Shapiro}}]{Zhang2015PRL-QI}%
  \BibitemOpen
  \bibfield  {author} {\bibinfo {author} {\bibfnamefont {Z.}~\bibnamefont
  {Zhang}}, \bibinfo {author} {\bibfnamefont {S.}~\bibnamefont {Mouradian}},
  \bibinfo {author} {\bibfnamefont {F.~N.~C.}\ \bibnamefont {Wong}}, \ and\
  \bibinfo {author} {\bibfnamefont {J.~H.}\ \bibnamefont {Shapiro}},\
  }\bibfield  {title} {\emph {\bibinfo {title} {Entanglement-enhanced sensing
  in a lossy and noisy environment},\ }}\href {\doibase
  10.1103/PhysRevLett.114.110506} {\bibfield  {journal} {\bibinfo  {journal}
  {Phys. Rev. Lett.}\ }\textbf {\bibinfo {volume} {114}},\ \bibinfo {pages}
  {110506} (\bibinfo {year} {2015})}\BibitemShut {NoStop}%
\bibitem [{\citenamefont {England}\ \emph {et~al.}(2019)\citenamefont
  {England}, \citenamefont {Balaji},\ and\ \citenamefont
  {Sussman}}]{England2019PRA}%
  \BibitemOpen
  \bibfield  {author} {\bibinfo {author} {\bibfnamefont {D.~G.}\ \bibnamefont
  {England}}, \bibinfo {author} {\bibfnamefont {B.}~\bibnamefont {Balaji}}, \
  and\ \bibinfo {author} {\bibfnamefont {B.~J.}\ \bibnamefont {Sussman}},\
  }\bibfield  {title} {\emph {\bibinfo {title} {Quantum-enhanced standoff
  detection using correlated photon pairs},\ }}\href {\doibase
  10.1103/PhysRevA.99.023828} {\bibfield  {journal} {\bibinfo  {journal} {Phys.
  Rev. A}\ }\textbf {\bibinfo {volume} {99}},\ \bibinfo {pages} {023828}
  (\bibinfo {year} {2019})}\BibitemShut {NoStop}%
\bibitem [{\citenamefont {Xu}\ \emph {et~al.}(2021)\citenamefont {Xu},
  \citenamefont {Zhang}, \citenamefont {Xu}, \citenamefont {Jiang},
  \citenamefont {Yung},\ and\ \citenamefont {Zhang}}]{Xu2021PRL}%
  \BibitemOpen
  \bibfield  {author} {\bibinfo {author} {\bibfnamefont {F.}~\bibnamefont
  {Xu}}, \bibinfo {author} {\bibfnamefont {X.-M.}\ \bibnamefont {Zhang}},
  \bibinfo {author} {\bibfnamefont {L.}~\bibnamefont {Xu}}, \bibinfo {author}
  {\bibfnamefont {T.}~\bibnamefont {Jiang}}, \bibinfo {author} {\bibfnamefont
  {M.-H.}\ \bibnamefont {Yung}}, \ and\ \bibinfo {author} {\bibfnamefont
  {L.}~\bibnamefont {Zhang}},\ }\bibfield  {title} {\emph {\bibinfo {title}
  {Experimental quantum target detection approaching the fundamental helstrom
  limit},\ }}\href {\doibase 10.1103/PhysRevLett.127.040504} {\bibfield
  {journal} {\bibinfo  {journal} {Phys. Rev. Lett.}\ }\textbf {\bibinfo
  {volume} {127}},\ \bibinfo {pages} {040504} (\bibinfo {year}
  {2021})}\BibitemShut {NoStop}%
\bibitem [{\citenamefont {Bradshaw}\ \emph {et~al.}(2021)\citenamefont
  {Bradshaw}, \citenamefont {Conlon}, \citenamefont {Tserkis}, \citenamefont
  {Gu}, \citenamefont {Lam},\ and\ \citenamefont {Assad}}]{Bradshaw2021PRA}%
  \BibitemOpen
  \bibfield  {author} {\bibinfo {author} {\bibfnamefont {M.}~\bibnamefont
  {Bradshaw}}, \bibinfo {author} {\bibfnamefont {L.~O.}\ \bibnamefont
  {Conlon}}, \bibinfo {author} {\bibfnamefont {S.}~\bibnamefont {Tserkis}},
  \bibinfo {author} {\bibfnamefont {M.}~\bibnamefont {Gu}}, \bibinfo {author}
  {\bibfnamefont {P.~K.}\ \bibnamefont {Lam}}, \ and\ \bibinfo {author}
  {\bibfnamefont {S.~M.}\ \bibnamefont {Assad}},\ }\bibfield  {title} {\emph
  {\bibinfo {title} {Optimal probes for continuous-variable quantum
  illumination},\ }}\href {\doibase 10.1103/PhysRevA.103.062413} {\bibfield
  {journal} {\bibinfo  {journal} {Phys. Rev. A}\ }\textbf {\bibinfo {volume}
  {103}},\ \bibinfo {pages} {062413} (\bibinfo {year} {2021})}\BibitemShut
  {NoStop}%
\bibitem [{\citenamefont {Kim}\ \emph {et~al.}(2023{\natexlab{a}})\citenamefont
  {Kim}, \citenamefont {Jo}, \citenamefont {Kim}, \citenamefont {Jeong},
  \citenamefont {Kim}, \citenamefont {Park}, \citenamefont {Kim},\ and\
  \citenamefont {Lee}}]{Kim2023PRR}%
  \BibitemOpen
  \bibfield  {author} {\bibinfo {author} {\bibfnamefont {D.~H.}\ \bibnamefont
  {Kim}}, \bibinfo {author} {\bibfnamefont {Y.}~\bibnamefont {Jo}}, \bibinfo
  {author} {\bibfnamefont {D.~Y.}\ \bibnamefont {Kim}}, \bibinfo {author}
  {\bibfnamefont {T.}~\bibnamefont {Jeong}}, \bibinfo {author} {\bibfnamefont
  {J.}~\bibnamefont {Kim}}, \bibinfo {author} {\bibfnamefont {N.~H.}\
  \bibnamefont {Park}}, \bibinfo {author} {\bibfnamefont {Z.}~\bibnamefont
  {Kim}}, \ and\ \bibinfo {author} {\bibfnamefont {S.-Y.}\ \bibnamefont
  {Lee}},\ }\bibfield  {title} {\emph {\bibinfo {title} {Gaussian quantum
  illumination via monotone metrics},\ }}\href {\doibase
  10.1103/PhysRevResearch.5.033010} {\bibfield  {journal} {\bibinfo  {journal}
  {Phys. Rev. Res.}\ }\textbf {\bibinfo {volume} {5}},\ \bibinfo {pages}
  {033010} (\bibinfo {year} {2023}{\natexlab{a}})}\BibitemShut {NoStop}%
\bibitem [{\citenamefont {Calsamiglia}\ \emph {et~al.}(2008)\citenamefont
  {Calsamiglia}, \citenamefont {Mu\~noz Tapia}, \citenamefont {Masanes},
  \citenamefont {Acin},\ and\ \citenamefont {Bagan}}]{Calsamiglia2008PRA}%
  \BibitemOpen
  \bibfield  {author} {\bibinfo {author} {\bibfnamefont {J.}~\bibnamefont
  {Calsamiglia}}, \bibinfo {author} {\bibfnamefont {R.}~\bibnamefont {Mu\~noz
  Tapia}}, \bibinfo {author} {\bibfnamefont {L.}~\bibnamefont {Masanes}},
  \bibinfo {author} {\bibfnamefont {A.}~\bibnamefont {Acin}}, \ and\ \bibinfo
  {author} {\bibfnamefont {E.}~\bibnamefont {Bagan}},\ }\bibfield  {title}
  {\emph {\bibinfo {title} {Quantum chernoff bound as a measure of
  distinguishability between density matrices: Application to qubit and
  gaussian states},\ }}\href {\doibase 10.1103/PhysRevA.77.032311} {\bibfield
  {journal} {\bibinfo  {journal} {Phys. Rev. A}\ }\textbf {\bibinfo {volume}
  {77}},\ \bibinfo {pages} {032311} (\bibinfo {year} {2008})}\BibitemShut
  {NoStop}%
\bibitem [{\citenamefont {Sanz}\ \emph {et~al.}(2017)\citenamefont {Sanz},
  \citenamefont {Las~Heras}, \citenamefont {Garc\'{\i}a-Ripoll}, \citenamefont
  {Solano},\ and\ \citenamefont {Di~Candia}}]{Sanz2017PRL}%
  \BibitemOpen
  \bibfield  {author} {\bibinfo {author} {\bibfnamefont {M.}~\bibnamefont
  {Sanz}}, \bibinfo {author} {\bibfnamefont {U.}~\bibnamefont {Las~Heras}},
  \bibinfo {author} {\bibfnamefont {J.~J.}\ \bibnamefont {Garc\'{\i}a-Ripoll}},
  \bibinfo {author} {\bibfnamefont {E.}~\bibnamefont {Solano}}, \ and\ \bibinfo
  {author} {\bibfnamefont {R.}~\bibnamefont {Di~Candia}},\ }\bibfield  {title}
  {\emph {\bibinfo {title} {Quantum estimation methods for quantum
  illumination},\ }}\href {\doibase 10.1103/PhysRevLett.118.070803} {\bibfield
  {journal} {\bibinfo  {journal} {Phys. Rev. Lett.}\ }\textbf {\bibinfo
  {volume} {118}},\ \bibinfo {pages} {070803} (\bibinfo {year}
  {2017})}\BibitemShut {NoStop}%
\bibitem [{\citenamefont {Barzanjeh}\ \emph {et~al.}(2020)\citenamefont
  {Barzanjeh}, \citenamefont {Pirandola}, \citenamefont {Vitali},\ and\
  \citenamefont {Fink}}]{Barzanjeh2020SciAdv}%
  \BibitemOpen
  \bibfield  {author} {\bibinfo {author} {\bibfnamefont {S.}~\bibnamefont
  {Barzanjeh}}, \bibinfo {author} {\bibfnamefont {S.}~\bibnamefont
  {Pirandola}}, \bibinfo {author} {\bibfnamefont {D.}~\bibnamefont {Vitali}}, \
  and\ \bibinfo {author} {\bibfnamefont {J.~M.}\ \bibnamefont {Fink}},\
  }\bibfield  {title} {\emph {\bibinfo {title} {Microwave quantum illumination
  using a digital receiver},\ }}\href {\doibase 10.1126/sciadv.abb0451}
  {\bibfield  {journal} {\bibinfo  {journal} {Science Advances}\ }\textbf
  {\bibinfo {volume} {6}},\ \bibinfo {pages} {eabb0451} (\bibinfo {year}
  {2020})},\ \Eprint
  {http://arxiv.org/abs/https://www.science.org/doi/pdf/10.1126/sciadv.abb0451}
  {https://www.science.org/doi/pdf/10.1126/sciadv.abb0451} \BibitemShut
  {NoStop}%
\bibitem [{\citenamefont {Shi}\ \emph {et~al.}(2023)\citenamefont {Shi},
  \citenamefont {Zhang},\ and\ \citenamefont
  {Zhuang}}]{shi2023fulfillingentanglementsoptimaladvantage}%
  \BibitemOpen
  \bibfield  {author} {\bibinfo {author} {\bibfnamefont {H.}~\bibnamefont
  {Shi}}, \bibinfo {author} {\bibfnamefont {B.}~\bibnamefont {Zhang}}, \ and\
  \bibinfo {author} {\bibfnamefont {Q.}~\bibnamefont {Zhuang}},\ }\href
  {https://arxiv.org/abs/2207.06609} {\bibinfo {title} {Fulfilling
  entanglement's optimal advantage via converting correlation to coherence},\ }
  (\bibinfo {year} {2023}),\ \Eprint {http://arxiv.org/abs/2207.06609}
  {arXiv:2207.06609 [quant-ph]} \BibitemShut {NoStop}%
\bibitem [{\citenamefont {Titulaer}\ and\ \citenamefont
  {Glauber}(1966)}]{Titulaer1966PR}%
  \BibitemOpen
  \bibfield  {author} {\bibinfo {author} {\bibfnamefont {U.~M.}\ \bibnamefont
  {Titulaer}}\ and\ \bibinfo {author} {\bibfnamefont {R.~J.}\ \bibnamefont
  {Glauber}},\ }\bibfield  {title} {\emph {\bibinfo {title} {Density operators
  for coherent fields},\ }}\href {\doibase 10.1103/PhysRev.145.1041} {\bibfield
   {journal} {\bibinfo  {journal} {Phys. Rev.}\ }\textbf {\bibinfo {volume}
  {145}},\ \bibinfo {pages} {1041} (\bibinfo {year} {1966})}\BibitemShut
  {NoStop}%
\bibitem [{\citenamefont {Bialynicka-Birula}(1968)}]{BialynickaBirula1968PR}%
  \BibitemOpen
  \bibfield  {author} {\bibinfo {author} {\bibfnamefont {Z.}~\bibnamefont
  {Bialynicka-Birula}},\ }\bibfield  {title} {\emph {\bibinfo {title}
  {Properties of the generalized coherent state},\ }}\href {\doibase
  10.1103/PhysRev.173.1207} {\bibfield  {journal} {\bibinfo  {journal} {Phys.
  Rev.}\ }\textbf {\bibinfo {volume} {173}},\ \bibinfo {pages} {1207} (\bibinfo
  {year} {1968})}\BibitemShut {NoStop}%
\bibitem [{\citenamefont {Stoler}(1971)}]{Stoler1971PRD}%
  \BibitemOpen
  \bibfield  {author} {\bibinfo {author} {\bibfnamefont {D.}~\bibnamefont
  {Stoler}},\ }\bibfield  {title} {\emph {\bibinfo {title} {Generalized
  coherent states},\ }}\href {\doibase 10.1103/PhysRevD.4.2309} {\bibfield
  {journal} {\bibinfo  {journal} {Phys. Rev. D}\ }\textbf {\bibinfo {volume}
  {4}},\ \bibinfo {pages} {2309} (\bibinfo {year} {1971})}\BibitemShut
  {NoStop}%
\bibitem [{\citenamefont {Yurke}\ and\ \citenamefont
  {Stoler}(1986)}]{Yurke1986PRL}%
  \BibitemOpen
  \bibfield  {author} {\bibinfo {author} {\bibfnamefont {B.}~\bibnamefont
  {Yurke}}\ and\ \bibinfo {author} {\bibfnamefont {D.}~\bibnamefont {Stoler}},\
  }\bibfield  {title} {\emph {\bibinfo {title} {Generating quantum mechanical
  superpositions of macroscopically distinguishable states via amplitude
  dispersion},\ }}\href {\doibase 10.1103/PhysRevLett.57.13} {\bibfield
  {journal} {\bibinfo  {journal} {Phys. Rev. Lett.}\ }\textbf {\bibinfo
  {volume} {57}},\ \bibinfo {pages} {13} (\bibinfo {year} {1986})}\BibitemShut
  {NoStop}%
\bibitem [{\citenamefont {Uria}\ \emph {et~al.}(2023)\citenamefont {Uria},
  \citenamefont {Maldonado-Trapp}, \citenamefont {Hermann-Avigliano},\ and\
  \citenamefont {Solano}}]{Uria2023PRR}%
  \BibitemOpen
  \bibfield  {author} {\bibinfo {author} {\bibfnamefont {M.}~\bibnamefont
  {Uria}}, \bibinfo {author} {\bibfnamefont {A.}~\bibnamefont
  {Maldonado-Trapp}}, \bibinfo {author} {\bibfnamefont {C.}~\bibnamefont
  {Hermann-Avigliano}}, \ and\ \bibinfo {author} {\bibfnamefont
  {P.}~\bibnamefont {Solano}},\ }\bibfield  {title} {\emph {\bibinfo {title}
  {Emergence of non-gaussian coherent states through nonlinear interactions},\
  }}\href {\doibase 10.1103/PhysRevResearch.5.013165} {\bibfield  {journal}
  {\bibinfo  {journal} {Phys. Rev. Res.}\ }\textbf {\bibinfo {volume} {5}},\
  \bibinfo {pages} {013165} (\bibinfo {year} {2023})}\BibitemShut {NoStop}%
\bibitem [{\citenamefont {Walschaers}(2021)}]{Walschaers2021PRXQuantum}%
  \BibitemOpen
  \bibfield  {author} {\bibinfo {author} {\bibfnamefont {M.}~\bibnamefont
  {Walschaers}},\ }\bibfield  {title} {\emph {\bibinfo {title} {Non-gaussian
  quantum states and where to find them},\ }}\href {\doibase
  10.1103/PRXQuantum.2.030204} {\bibfield  {journal} {\bibinfo  {journal} {PRX
  Quantum}\ }\textbf {\bibinfo {volume} {2}},\ \bibinfo {pages} {030204}
  (\bibinfo {year} {2021})}\BibitemShut {NoStop}%
\bibitem [{\citenamefont {Lewis-Swan}\ \emph {et~al.}(2020)\citenamefont
  {Lewis-Swan}, \citenamefont {Barberena}, \citenamefont {Muniz}, \citenamefont
  {Cline}, \citenamefont {Young}, \citenamefont {Thompson},\ and\ \citenamefont
  {Rey}}]{LewisSwan2020PRL}%
  \BibitemOpen
  \bibfield  {author} {\bibinfo {author} {\bibfnamefont {R.~J.}\ \bibnamefont
  {Lewis-Swan}}, \bibinfo {author} {\bibfnamefont {D.}~\bibnamefont
  {Barberena}}, \bibinfo {author} {\bibfnamefont {J.~A.}\ \bibnamefont
  {Muniz}}, \bibinfo {author} {\bibfnamefont {J.~R.~K.}\ \bibnamefont {Cline}},
  \bibinfo {author} {\bibfnamefont {D.}~\bibnamefont {Young}}, \bibinfo
  {author} {\bibfnamefont {J.~K.}\ \bibnamefont {Thompson}}, \ and\ \bibinfo
  {author} {\bibfnamefont {A.~M.}\ \bibnamefont {Rey}},\ }\bibfield  {title}
  {\emph {\bibinfo {title} {Protocol for precise field sensing in the optical
  domain with cold atoms in a cavity},\ }}\href {\doibase
  10.1103/PhysRevLett.124.193602} {\bibfield  {journal} {\bibinfo  {journal}
  {Phys. Rev. Lett.}\ }\textbf {\bibinfo {volume} {124}},\ \bibinfo {pages}
  {193602} (\bibinfo {year} {2020})}\BibitemShut {NoStop}%
\bibitem [{\citenamefont {Zhang}\ \emph {et~al.}(2014)\citenamefont {Zhang},
  \citenamefont {Guo}, \citenamefont {Bao}, \citenamefont {Shi}, \citenamefont
  {Jin}, \citenamefont {Zou},\ and\ \citenamefont {Guo}}]{Zhang2014PRA-QI}%
  \BibitemOpen
  \bibfield  {author} {\bibinfo {author} {\bibfnamefont {S.}~\bibnamefont
  {Zhang}}, \bibinfo {author} {\bibfnamefont {J.}~\bibnamefont {Guo}}, \bibinfo
  {author} {\bibfnamefont {W.}~\bibnamefont {Bao}}, \bibinfo {author}
  {\bibfnamefont {J.}~\bibnamefont {Shi}}, \bibinfo {author} {\bibfnamefont
  {C.}~\bibnamefont {Jin}}, \bibinfo {author} {\bibfnamefont {X.}~\bibnamefont
  {Zou}}, \ and\ \bibinfo {author} {\bibfnamefont {G.}~\bibnamefont {Guo}},\
  }\bibfield  {title} {\emph {\bibinfo {title} {Quantum illumination with
  photon-subtracted continuous-variable entanglement},\ }}\href {\doibase
  10.1103/PhysRevA.89.062309} {\bibfield  {journal} {\bibinfo  {journal} {Phys.
  Rev. A}\ }\textbf {\bibinfo {volume} {89}},\ \bibinfo {pages} {062309}
  (\bibinfo {year} {2014})}\BibitemShut {NoStop}%
\bibitem [{\citenamefont {Fan}\ and\ \citenamefont
  {Zubairy}(2018)}]{Fan2018PRA}%
  \BibitemOpen
  \bibfield  {author} {\bibinfo {author} {\bibfnamefont {L.}~\bibnamefont
  {Fan}}\ and\ \bibinfo {author} {\bibfnamefont {M.~S.}\ \bibnamefont
  {Zubairy}},\ }\bibfield  {title} {\emph {\bibinfo {title} {Quantum
  illumination using non-gaussian states generated by photon subtraction and
  photon addition},\ }}\href {\doibase 10.1103/PhysRevA.98.012319} {\bibfield
  {journal} {\bibinfo  {journal} {Phys. Rev. A}\ }\textbf {\bibinfo {volume}
  {98}},\ \bibinfo {pages} {012319} (\bibinfo {year} {2018})}\BibitemShut
  {NoStop}%
\bibitem [{\citenamefont {Yurke}\ \emph {et~al.}(1986)\citenamefont {Yurke},
  \citenamefont {McCall},\ and\ \citenamefont {Klauder}}]{Yurke1986PRA}%
  \BibitemOpen
  \bibfield  {author} {\bibinfo {author} {\bibfnamefont {B.}~\bibnamefont
  {Yurke}}, \bibinfo {author} {\bibfnamefont {S.~L.}\ \bibnamefont {McCall}}, \
  and\ \bibinfo {author} {\bibfnamefont {J.~R.}\ \bibnamefont {Klauder}},\
  }\bibfield  {title} {\emph {\bibinfo {title} {{SU}(2) and {SU}(1,1)
  interferometers},\ }}\href {\doibase 10.1103/physreva.33.4033} {\bibfield
  {journal} {\bibinfo  {journal} {Phys. Rev. A}\ }\textbf {\bibinfo {volume}
  {33}},\ \bibinfo {pages} {4033} (\bibinfo {year} {1986})}\BibitemShut
  {NoStop}%
\bibitem [{\citenamefont {Audenaert}\ \emph {et~al.}(2007)\citenamefont
  {Audenaert}, \citenamefont {Calsamiglia}, \citenamefont {Mu\~noz Tapia},
  \citenamefont {Bagan}, \citenamefont {Masanes}, \citenamefont {Acin},\ and\
  \citenamefont {Verstraete}}]{Audenaert2007PRL}%
  \BibitemOpen
  \bibfield  {author} {\bibinfo {author} {\bibfnamefont {K.~M.~R.}\
  \bibnamefont {Audenaert}}, \bibinfo {author} {\bibfnamefont {J.}~\bibnamefont
  {Calsamiglia}}, \bibinfo {author} {\bibfnamefont {R.}~\bibnamefont {Mu\~noz
  Tapia}}, \bibinfo {author} {\bibfnamefont {E.}~\bibnamefont {Bagan}},
  \bibinfo {author} {\bibfnamefont {L.}~\bibnamefont {Masanes}}, \bibinfo
  {author} {\bibfnamefont {A.}~\bibnamefont {Acin}}, \ and\ \bibinfo {author}
  {\bibfnamefont {F.}~\bibnamefont {Verstraete}},\ }\bibfield  {title} {\emph
  {\bibinfo {title} {Discriminating states: The quantum chernoff bound},\
  }}\href {\doibase 10.1103/PhysRevLett.98.160501} {\bibfield  {journal}
  {\bibinfo  {journal} {Phys. Rev. Lett.}\ }\textbf {\bibinfo {volume} {98}},\
  \bibinfo {pages} {160501} (\bibinfo {year} {2007})}\BibitemShut {NoStop}%
\bibitem [{\citenamefont {Braunstein}\ and\ \citenamefont
  {Caves}(1994)}]{Braunstein1994PRL}%
  \BibitemOpen
  \bibfield  {author} {\bibinfo {author} {\bibfnamefont {S.~L.}\ \bibnamefont
  {Braunstein}}\ and\ \bibinfo {author} {\bibfnamefont {C.~M.}\ \bibnamefont
  {Caves}},\ }\bibfield  {title} {\emph {\bibinfo {title} {Statistical distance
  and the geometry of quantum states},\ }}\href {\doibase
  10.1103/PhysRevLett.72.3439} {\bibfield  {journal} {\bibinfo  {journal}
  {Phys. Rev. Lett.}\ }\textbf {\bibinfo {volume} {72}},\ \bibinfo {pages}
  {3439} (\bibinfo {year} {1994})}\BibitemShut {NoStop}%
\bibitem [{\citenamefont {Woodworth}\ \emph {et~al.}(2020)\citenamefont
  {Woodworth}, \citenamefont {Chan}, \citenamefont {Hermann-Avigliano},\ and\
  \citenamefont {Marino}}]{Woodworth2020PRA}%
  \BibitemOpen
  \bibfield  {author} {\bibinfo {author} {\bibfnamefont {T.~S.}\ \bibnamefont
  {Woodworth}}, \bibinfo {author} {\bibfnamefont {K.~W.~C.}\ \bibnamefont
  {Chan}}, \bibinfo {author} {\bibfnamefont {C.}~\bibnamefont
  {Hermann-Avigliano}}, \ and\ \bibinfo {author} {\bibfnamefont {A.~M.}\
  \bibnamefont {Marino}},\ }\bibfield  {title} {\emph {\bibinfo {title}
  {Transmission estimation at the cram\'er-rao bound for squeezed states of
  light in the presence of loss and imperfect detection},\ }}\href {\doibase
  10.1103/PhysRevA.102.052603} {\bibfield  {journal} {\bibinfo  {journal}
  {Phys. Rev. A}\ }\textbf {\bibinfo {volume} {102}},\ \bibinfo {pages}
  {052603} (\bibinfo {year} {2020})}\BibitemShut {NoStop}%
\bibitem [{\citenamefont {Woodworth}\ \emph {et~al.}(2022)\citenamefont
  {Woodworth}, \citenamefont {Hermann-Avigliano}, \citenamefont {Chan},\ and\
  \citenamefont {Marino}}]{Woodworth2022EPJ}%
  \BibitemOpen
  \bibfield  {author} {\bibinfo {author} {\bibfnamefont {T.~S.}\ \bibnamefont
  {Woodworth}}, \bibinfo {author} {\bibfnamefont {C.}~\bibnamefont
  {Hermann-Avigliano}}, \bibinfo {author} {\bibfnamefont {K.~W.~C.}\
  \bibnamefont {Chan}}, \ and\ \bibinfo {author} {\bibfnamefont {A.~M.}\
  \bibnamefont {Marino}},\ }\bibfield  {title} {\emph {\bibinfo {title}
  {Transmission estimation at the quantum cramér-rao bound with macroscopic
  quantum light},\ }}\href {\doibase 10.1140/epjqt/s40507-022-00154-x}
  {\bibfield  {journal} {\bibinfo  {journal} {EPJ Quantum Technology}\ }\textbf
  {\bibinfo {volume} {9}},\ \bibinfo {pages} {38} (\bibinfo {year}
  {2022})}\BibitemShut {NoStop}%
\bibitem [{\citenamefont {Dowran}\ \emph {et~al.}(2021)\citenamefont {Dowran},
  \citenamefont {Woodworth}, \citenamefont {Kumar},\ and\ \citenamefont
  {Marino}}]{Dowran2021QST}%
  \BibitemOpen
  \bibfield  {author} {\bibinfo {author} {\bibfnamefont {M.}~\bibnamefont
  {Dowran}}, \bibinfo {author} {\bibfnamefont {T.~S.}\ \bibnamefont
  {Woodworth}}, \bibinfo {author} {\bibfnamefont {A.}~\bibnamefont {Kumar}}, \
  and\ \bibinfo {author} {\bibfnamefont {A.~M.}\ \bibnamefont {Marino}},\
  }\bibfield  {title} {\emph {\bibinfo {title} {Fundamental sensitivity bounds
  for quantum enhanced optical resonance sensors based on transmission and
  phase estimation},\ }}\href {\doibase 10.1088/2058-9565/ac3550} {\bibfield
  {journal} {\bibinfo  {journal} {Quantum Science and Technology}\ }\textbf
  {\bibinfo {volume} {7}},\ \bibinfo {pages} {015011} (\bibinfo {year}
  {2021})}\BibitemShut {NoStop}%
\bibitem [{\citenamefont {Monras}\ and\ \citenamefont
  {Paris}(2007)}]{Monras2007PRL}%
  \BibitemOpen
  \bibfield  {author} {\bibinfo {author} {\bibfnamefont {A.}~\bibnamefont
  {Monras}}\ and\ \bibinfo {author} {\bibfnamefont {M.~G.~A.}\ \bibnamefont
  {Paris}},\ }\bibfield  {title} {\emph {\bibinfo {title} {Optimal quantum
  estimation of loss in bosonic channels},\ }}\href {\doibase
  10.1103/PhysRevLett.98.160401} {\bibfield  {journal} {\bibinfo  {journal}
  {Phys. Rev. Lett.}\ }\textbf {\bibinfo {volume} {98}},\ \bibinfo {pages}
  {160401} (\bibinfo {year} {2007})}\BibitemShut {NoStop}%
\bibitem [{\citenamefont {Adesso}\ \emph {et~al.}(2009)\citenamefont {Adesso},
  \citenamefont {Dell'Anno}, \citenamefont {De~Siena}, \citenamefont
  {Illuminati},\ and\ \citenamefont {Souza}}]{Adesso2009PRA}%
  \BibitemOpen
  \bibfield  {author} {\bibinfo {author} {\bibfnamefont {G.}~\bibnamefont
  {Adesso}}, \bibinfo {author} {\bibfnamefont {F.}~\bibnamefont {Dell'Anno}},
  \bibinfo {author} {\bibfnamefont {S.}~\bibnamefont {De~Siena}}, \bibinfo
  {author} {\bibfnamefont {F.}~\bibnamefont {Illuminati}}, \ and\ \bibinfo
  {author} {\bibfnamefont {L.~A.~M.}\ \bibnamefont {Souza}},\ }\bibfield
  {title} {\emph {\bibinfo {title} {Optimal estimation of losses at the
  ultimate quantum limit with non-gaussian states},\ }}\href {\doibase
  10.1103/PhysRevA.79.040305} {\bibfield  {journal} {\bibinfo  {journal} {Phys.
  Rev. A}\ }\textbf {\bibinfo {volume} {79}},\ \bibinfo {pages} {040305}
  (\bibinfo {year} {2009})}\BibitemShut {NoStop}%
\bibitem [{\citenamefont {Invernizzi}\ \emph {et~al.}(2011)\citenamefont
  {Invernizzi}, \citenamefont {Paris},\ and\ \citenamefont
  {Pirandola}}]{Invernizzi2011PRA}%
  \BibitemOpen
  \bibfield  {author} {\bibinfo {author} {\bibfnamefont {C.}~\bibnamefont
  {Invernizzi}}, \bibinfo {author} {\bibfnamefont {M.~G.~A.}\ \bibnamefont
  {Paris}}, \ and\ \bibinfo {author} {\bibfnamefont {S.}~\bibnamefont
  {Pirandola}},\ }\bibfield  {title} {\emph {\bibinfo {title} {Optimal
  detection of losses by thermal probes},\ }}\href {\doibase
  10.1103/PhysRevA.84.022334} {\bibfield  {journal} {\bibinfo  {journal} {Phys.
  Rev. A}\ }\textbf {\bibinfo {volume} {84}},\ \bibinfo {pages} {022334}
  (\bibinfo {year} {2011})}\BibitemShut {NoStop}%
\bibitem [{\citenamefont {Fuchs}\ and\ \citenamefont {van~de
  Graaf}(1999)}]{Fuchs1999IEEE}%
  \BibitemOpen
  \bibfield  {author} {\bibinfo {author} {\bibfnamefont {C.}~\bibnamefont
  {Fuchs}}\ and\ \bibinfo {author} {\bibfnamefont {J.}~\bibnamefont {van~de
  Graaf}},\ }\bibfield  {title} {\emph {\bibinfo {title} {Cryptographic
  distinguishability measures for quantum-mechanical states},\ }}\href
  {\doibase 10.1109/18.761271} {\bibfield  {journal} {\bibinfo  {journal} {IEEE
  Transactions on Information Theory}\ }\textbf {\bibinfo {volume} {45}},\
  \bibinfo {pages} {1216} (\bibinfo {year} {1999})}\BibitemShut {NoStop}%
\bibitem [{\citenamefont {Trees}\ \emph {et~al.}(2013)\citenamefont {Trees},
  \citenamefont {Bell},\ and\ \citenamefont {Tian}}]{Trees2013Book}%
  \BibitemOpen
  \bibfield  {author} {\bibinfo {author} {\bibfnamefont {H.~L.~V.}\
  \bibnamefont {Trees}}, \bibinfo {author} {\bibfnamefont {K.~L.}\ \bibnamefont
  {Bell}}, \ and\ \bibinfo {author} {\bibfnamefont {Z.}~\bibnamefont {Tian}},\
  }\href@noop {} {\emph {\bibinfo {title} {Detection, estimation and modulation
  theory Part I: Detection, estimation, and filtering theory, Second
  Edition}}}\ (\bibinfo  {publisher} {Wiley},\ \bibinfo {year}
  {2013})\BibitemShut {NoStop}%
\bibitem [{\citenamefont {Helstrom}(1976)}]{Helstrom1976Book}%
  \BibitemOpen
  \bibfield  {author} {\bibinfo {author} {\bibfnamefont {C.~W.}\ \bibnamefont
  {Helstrom}},\ }\href@noop {} {\emph {\bibinfo {title} {Quantum Detection and
  Estimation Theory}}}\ (\bibinfo  {publisher} {Academic, New York},\ \bibinfo
  {year} {1976})\BibitemShut {NoStop}%
\bibitem [{\citenamefont {Holevo}(1982)}]{Holevo1982Book}%
  \BibitemOpen
  \bibfield  {author} {\bibinfo {author} {\bibfnamefont {A.~S.}\ \bibnamefont
  {Holevo}},\ }\href@noop {} {\emph {\bibinfo {title} {Probabilistic and
  Statistical Aspects of Quantum Theory}}}\ (\bibinfo  {publisher}
  {North-Holland, Amsterdam},\ \bibinfo {year} {1982})\BibitemShut {NoStop}%
\bibitem [{\citenamefont {Zhong}\ \emph {et~al.}(2014)\citenamefont {Zhong},
  \citenamefont {Lu}, \citenamefont {Jing},\ and\ \citenamefont
  {Wang}}]{Zhong2014JPA}%
  \BibitemOpen
  \bibfield  {author} {\bibinfo {author} {\bibfnamefont {W.}~\bibnamefont
  {Zhong}}, \bibinfo {author} {\bibfnamefont {X.~M.}\ \bibnamefont {Lu}},
  \bibinfo {author} {\bibfnamefont {X.~X.}\ \bibnamefont {Jing}}, \ and\
  \bibinfo {author} {\bibfnamefont {X.~G.}\ \bibnamefont {Wang}},\ }\bibfield
  {title} {\emph {\bibinfo {title} {Optimal condition for measurement
  observable via error-propagation},\ }}\href
  {http://stacks.iop.org/1751-8121/47/i=38/a=385304} {\bibfield  {journal}
  {\bibinfo  {journal} {J. Phys. A: Math. Theor.}\ }\textbf {\bibinfo {volume}
  {47}},\ \bibinfo {pages} {385304} (\bibinfo {year} {2014})}\BibitemShut
  {NoStop}%
\bibitem [{\citenamefont {Noh}\ \emph {et~al.}(2022)\citenamefont {Noh},
  \citenamefont {Lee},\ and\ \citenamefont {Lee}}]{Noh2022JOSAB}%
  \BibitemOpen
  \bibfield  {author} {\bibinfo {author} {\bibfnamefont {C.}~\bibnamefont
  {Noh}}, \bibinfo {author} {\bibfnamefont {C.}~\bibnamefont {Lee}}, \ and\
  \bibinfo {author} {\bibfnamefont {S.-Y.}\ \bibnamefont {Lee}},\ }\bibfield
  {title} {\emph {\bibinfo {title} {Quantum illumination with definite
  photon-number entangled states},\ }}\href {\doibase 10.1364/JOSAB.455994}
  {\bibfield  {journal} {\bibinfo  {journal} {J. Opt. Soc. Am. B}\ }\textbf
  {\bibinfo {volume} {39}},\ \bibinfo {pages} {1316} (\bibinfo {year}
  {2022})}\BibitemShut {NoStop}%
\bibitem [{\citenamefont {Pirandola}\ and\ \citenamefont
  {Lloyd}(2008)}]{Pirandola2008PRA}%
  \BibitemOpen
  \bibfield  {author} {\bibinfo {author} {\bibfnamefont {S.}~\bibnamefont
  {Pirandola}}\ and\ \bibinfo {author} {\bibfnamefont {S.}~\bibnamefont
  {Lloyd}},\ }\bibfield  {title} {\emph {\bibinfo {title} {Computable bounds
  for the discrimination of gaussian states},\ }}\href {\doibase
  10.1103/PhysRevA.78.012331} {\bibfield  {journal} {\bibinfo  {journal} {Phys.
  Rev. A}\ }\textbf {\bibinfo {volume} {78}},\ \bibinfo {pages} {012331}
  (\bibinfo {year} {2008})}\BibitemShut {NoStop}%
\bibitem [{\citenamefont {Kholevo}(1974)}]{Kholevo1974TPA}%
  \BibitemOpen
  \bibfield  {author} {\bibinfo {author} {\bibfnamefont {A.~S.}\ \bibnamefont
  {Kholevo}},\ }\bibfield  {title} {\emph {\bibinfo {title} {A generalization
  of the rao-cramer inequality},\ }}\href {\doibase 10.1137/1118039} {\bibfield
   {journal} {\bibinfo  {journal} {Theory of Probability \& Its Applications}\
  }\textbf {\bibinfo {volume} {18}},\ \bibinfo {pages} {359} (\bibinfo {year}
  {1974})},\ \Eprint {http://arxiv.org/abs/https://doi.org/10.1137/1118039}
  {https://doi.org/10.1137/1118039} \BibitemShut {NoStop}%
\bibitem [{\citenamefont {Uys}\ and\ \citenamefont
  {Meystre}(2007)}]{Uys2007PRA}%
  \BibitemOpen
  \bibfield  {author} {\bibinfo {author} {\bibfnamefont {H.}~\bibnamefont
  {Uys}}\ and\ \bibinfo {author} {\bibfnamefont {P.}~\bibnamefont {Meystre}},\
  }\bibfield  {title} {\emph {\bibinfo {title} {Quantum states for
  heisenberg-limited interferometry},\ }}\href {\doibase
  10.1103/PhysRevA.76.013804} {\bibfield  {journal} {\bibinfo  {journal} {Phys.
  Rev. A}\ }\textbf {\bibinfo {volume} {76}},\ \bibinfo {pages} {013804}
  (\bibinfo {year} {2007})}\BibitemShut {NoStop}%
\bibitem [{\citenamefont {Agarwal}\ and\ \citenamefont
  {Tara}(1992)}]{Agarwal1992PRA}%
  \BibitemOpen
  \bibfield  {author} {\bibinfo {author} {\bibfnamefont {G.~S.}\ \bibnamefont
  {Agarwal}}\ and\ \bibinfo {author} {\bibfnamefont {K.}~\bibnamefont {Tara}},\
  }\bibfield  {title} {\emph {\bibinfo {title} {Nonclassical character of
  states exhibiting no squeezing or sub-poissonian statistics},\ }}\href
  {\doibase 10.1103/PhysRevA.46.485} {\bibfield  {journal} {\bibinfo  {journal}
  {Phys. Rev. A}\ }\textbf {\bibinfo {volume} {46}},\ \bibinfo {pages} {485}
  (\bibinfo {year} {1992})}\BibitemShut {NoStop}%
\bibitem [{\citenamefont {Andersen}\ \emph {et~al.}(2016)\citenamefont
  {Andersen}, \citenamefont {Gehring}, \citenamefont {Marquardt},\ and\
  \citenamefont {Leuchs}}]{Andersen2016PS}%
  \BibitemOpen
  \bibfield  {author} {\bibinfo {author} {\bibfnamefont {U.~L.}\ \bibnamefont
  {Andersen}}, \bibinfo {author} {\bibfnamefont {T.}~\bibnamefont {Gehring}},
  \bibinfo {author} {\bibfnamefont {C.}~\bibnamefont {Marquardt}}, \ and\
  \bibinfo {author} {\bibfnamefont {G.}~\bibnamefont {Leuchs}},\ }\bibfield
  {title} {\emph {\bibinfo {title} {30 years of squeezed light generation},\
  }}\href {\doibase 10.1088/0031-8949/91/5/053001} {\bibfield  {journal}
  {\bibinfo  {journal} {Physica Scripta}\ }\textbf {\bibinfo {volume} {91}},\
  \bibinfo {pages} {053001} (\bibinfo {year} {2016})}\BibitemShut {NoStop}%
\bibitem [{\citenamefont {Uria}\ \emph {et~al.}(2020)\citenamefont {Uria},
  \citenamefont {Solano},\ and\ \citenamefont
  {Hermann-Avigliano}}]{Uria2020PRL}%
  \BibitemOpen
  \bibfield  {author} {\bibinfo {author} {\bibfnamefont {M.}~\bibnamefont
  {Uria}}, \bibinfo {author} {\bibfnamefont {P.}~\bibnamefont {Solano}}, \ and\
  \bibinfo {author} {\bibfnamefont {C.}~\bibnamefont {Hermann-Avigliano}},\
  }\bibfield  {title} {\emph {\bibinfo {title} {Deterministic generation of
  large fock states},\ }}\href {\doibase 10.1103/PhysRevLett.125.093603}
  {\bibfield  {journal} {\bibinfo  {journal} {Phys. Rev. Lett.}\ }\textbf
  {\bibinfo {volume} {125}},\ \bibinfo {pages} {093603} (\bibinfo {year}
  {2020})}\BibitemShut {NoStop}%
\bibitem [{\citenamefont {Klimov}\ and\ \citenamefont
  {Chumakov}(2009)}]{Klimov2009Book}%
  \BibitemOpen
  \bibfield  {author} {\bibinfo {author} {\bibfnamefont {A.~B.}\ \bibnamefont
  {Klimov}}\ and\ \bibinfo {author} {\bibfnamefont {S.~M.}\ \bibnamefont
  {Chumakov}},\ }\href@noop {} {\emph {\bibinfo {title} {A group-theoretical
  approach to quantum optics: models of atom-field interactions}}}\ (\bibinfo
  {publisher} {John Wiley \& Sons},\ \bibinfo {year} {2009})\BibitemShut
  {NoStop}%
\bibitem [{\citenamefont {Navarrete-Benlloch}\ \emph
  {et~al.}(2012)\citenamefont {Navarrete-Benlloch}, \citenamefont
  {Garc\'{\i}a-Patr\'on}, \citenamefont {Shapiro},\ and\ \citenamefont
  {Cerf}}]{Navarrete-Benlloch2012PRA}%
  \BibitemOpen
  \bibfield  {author} {\bibinfo {author} {\bibfnamefont {C.}~\bibnamefont
  {Navarrete-Benlloch}}, \bibinfo {author} {\bibfnamefont {R.}~\bibnamefont
  {Garc\'{\i}a-Patr\'on}}, \bibinfo {author} {\bibfnamefont {J.~H.}\
  \bibnamefont {Shapiro}}, \ and\ \bibinfo {author} {\bibfnamefont {N.~J.}\
  \bibnamefont {Cerf}},\ }\bibfield  {title} {\emph {\bibinfo {title}
  {Enhancing quantum entanglement by photon addition and subtraction},\ }}\href
  {\doibase 10.1103/PhysRevA.86.012328} {\bibfield  {journal} {\bibinfo
  {journal} {Phys. Rev. A}\ }\textbf {\bibinfo {volume} {86}},\ \bibinfo
  {pages} {012328} (\bibinfo {year} {2012})}\BibitemShut {NoStop}%
\bibitem [{\citenamefont {Barnett}\ \emph {et~al.}(2018)\citenamefont
  {Barnett}, \citenamefont {Ferenczi}, \citenamefont {Gilson},\ and\
  \citenamefont {Speirits}}]{Barnett2018PRA}%
  \BibitemOpen
  \bibfield  {author} {\bibinfo {author} {\bibfnamefont {S.~M.}\ \bibnamefont
  {Barnett}}, \bibinfo {author} {\bibfnamefont {G.}~\bibnamefont {Ferenczi}},
  \bibinfo {author} {\bibfnamefont {C.~R.}\ \bibnamefont {Gilson}}, \ and\
  \bibinfo {author} {\bibfnamefont {F.~C.}\ \bibnamefont {Speirits}},\
  }\bibfield  {title} {\emph {\bibinfo {title} {Statistics of photon-subtracted
  and photon-added states},\ }}\href {\doibase 10.1103/PhysRevA.98.013809}
  {\bibfield  {journal} {\bibinfo  {journal} {Phys. Rev. A}\ }\textbf {\bibinfo
  {volume} {98}},\ \bibinfo {pages} {013809} (\bibinfo {year}
  {2018})}\BibitemShut {NoStop}%
\bibitem [{\citenamefont {Dakna}\ \emph {et~al.}(1997)\citenamefont {Dakna},
  \citenamefont {Anhut}, \citenamefont {Opatrn\'y}, \citenamefont {Kn\"oll},\
  and\ \citenamefont {Welsch}}]{Dakna1997PRA}%
  \BibitemOpen
  \bibfield  {author} {\bibinfo {author} {\bibfnamefont {M.}~\bibnamefont
  {Dakna}}, \bibinfo {author} {\bibfnamefont {T.}~\bibnamefont {Anhut}},
  \bibinfo {author} {\bibfnamefont {T.}~\bibnamefont {Opatrn\'y}}, \bibinfo
  {author} {\bibfnamefont {L.}~\bibnamefont {Kn\"oll}}, \ and\ \bibinfo
  {author} {\bibfnamefont {D.~G.}\ \bibnamefont {Welsch}},\ }\bibfield  {title}
  {\emph {\bibinfo {title} {Generating schr\"odinger-cat-like states by means
  of conditional measurements on a beam splitter},\ }}\href {\doibase
  10.1103/PhysRevA.55.3184} {\bibfield  {journal} {\bibinfo  {journal} {Phys.
  Rev. A}\ }\textbf {\bibinfo {volume} {55}},\ \bibinfo {pages} {3184}
  (\bibinfo {year} {1997})}\BibitemShut {NoStop}%
\bibitem [{\citenamefont {Dakna}\ \emph {et~al.}(1999)\citenamefont {Dakna},
  \citenamefont {Clausen}, \citenamefont {Kn\"oll},\ and\ \citenamefont
  {Welsch}}]{Dakna1999PRA}%
  \BibitemOpen
  \bibfield  {author} {\bibinfo {author} {\bibfnamefont {M.}~\bibnamefont
  {Dakna}}, \bibinfo {author} {\bibfnamefont {J.}~\bibnamefont {Clausen}},
  \bibinfo {author} {\bibfnamefont {L.}~\bibnamefont {Kn\"oll}}, \ and\
  \bibinfo {author} {\bibfnamefont {D.~G.}\ \bibnamefont {Welsch}},\ }\bibfield
   {title} {\emph {\bibinfo {title} {Generation of arbitrary quantum states of
  traveling fields},\ }}\href {\doibase 10.1103/PhysRevA.59.1658} {\bibfield
  {journal} {\bibinfo  {journal} {Phys. Rev. A}\ }\textbf {\bibinfo {volume}
  {59}},\ \bibinfo {pages} {1658} (\bibinfo {year} {1999})}\BibitemShut
  {NoStop}%
\bibitem [{\citenamefont {Wenger}\ \emph {et~al.}(2004)\citenamefont {Wenger},
  \citenamefont {Tualle-Brouri},\ and\ \citenamefont
  {Grangier}}]{Wenger2004PRL}%
  \BibitemOpen
  \bibfield  {author} {\bibinfo {author} {\bibfnamefont {J.}~\bibnamefont
  {Wenger}}, \bibinfo {author} {\bibfnamefont {R.}~\bibnamefont
  {Tualle-Brouri}}, \ and\ \bibinfo {author} {\bibfnamefont {P.}~\bibnamefont
  {Grangier}},\ }\bibfield  {title} {\emph {\bibinfo {title} {Non-gaussian
  statistics from individual pulses of squeezed light},\ }}\href {\doibase
  10.1103/PhysRevLett.92.153601} {\bibfield  {journal} {\bibinfo  {journal}
  {Phys. Rev. Lett.}\ }\textbf {\bibinfo {volume} {92}},\ \bibinfo {pages}
  {153601} (\bibinfo {year} {2004})}\BibitemShut {NoStop}%
\bibitem [{\citenamefont {Fiur\'a\ifmmode~\check{s}\else \v{s}\fi{}ek}\ \emph
  {et~al.}(2005)\citenamefont {Fiur\'a\ifmmode~\check{s}\else \v{s}\fi{}ek},
  \citenamefont {Garc\'{\i}a-Patr\'on},\ and\ \citenamefont
  {Cerf}}]{Fiurafmmodeheckslsesiek2005PRA}%
  \BibitemOpen
  \bibfield  {author} {\bibinfo {author} {\bibfnamefont {J.}~\bibnamefont
  {Fiur\'a\ifmmode~\check{s}\else \v{s}\fi{}ek}}, \bibinfo {author}
  {\bibfnamefont {R.}~\bibnamefont {Garc\'{\i}a-Patr\'on}}, \ and\ \bibinfo
  {author} {\bibfnamefont {N.~J.}\ \bibnamefont {Cerf}},\ }\bibfield  {title}
  {\emph {\bibinfo {title} {Conditional generation of arbitrary single-mode
  quantum states of light by repeated photon subtractions},\ }}\href {\doibase
  10.1103/PhysRevA.72.033822} {\bibfield  {journal} {\bibinfo  {journal} {Phys.
  Rev. A}\ }\textbf {\bibinfo {volume} {72}},\ \bibinfo {pages} {033822}
  (\bibinfo {year} {2005})}\BibitemShut {NoStop}%
\bibitem [{\citenamefont {Kitagawa}\ \emph {et~al.}(2006)\citenamefont
  {Kitagawa}, \citenamefont {Takeoka}, \citenamefont {Sasaki},\ and\
  \citenamefont {Chefles}}]{Kitagawa2006PRA}%
  \BibitemOpen
  \bibfield  {author} {\bibinfo {author} {\bibfnamefont {A.}~\bibnamefont
  {Kitagawa}}, \bibinfo {author} {\bibfnamefont {M.}~\bibnamefont {Takeoka}},
  \bibinfo {author} {\bibfnamefont {M.}~\bibnamefont {Sasaki}}, \ and\ \bibinfo
  {author} {\bibfnamefont {A.}~\bibnamefont {Chefles}},\ }\bibfield  {title}
  {\emph {\bibinfo {title} {Entanglement evaluation of non-gaussian states
  generated by photon subtraction from squeezed states},\ }}\href {\doibase
  10.1103/PhysRevA.73.042310} {\bibfield  {journal} {\bibinfo  {journal} {Phys.
  Rev. A}\ }\textbf {\bibinfo {volume} {73}},\ \bibinfo {pages} {042310}
  (\bibinfo {year} {2006})}\BibitemShut {NoStop}%
\bibitem [{\citenamefont {Biswas}\ and\ \citenamefont
  {Agarwal}(2007)}]{Biswas2007PRA}%
  \BibitemOpen
  \bibfield  {author} {\bibinfo {author} {\bibfnamefont {A.}~\bibnamefont
  {Biswas}}\ and\ \bibinfo {author} {\bibfnamefont {G.~S.}\ \bibnamefont
  {Agarwal}},\ }\bibfield  {title} {\emph {\bibinfo {title} {Nonclassicality
  and decoherence of photon-subtracted squeezed states},\ }}\href {\doibase
  10.1103/PhysRevA.75.032104} {\bibfield  {journal} {\bibinfo  {journal} {Phys.
  Rev. A}\ }\textbf {\bibinfo {volume} {75}},\ \bibinfo {pages} {032104}
  (\bibinfo {year} {2007})}\BibitemShut {NoStop}%
\bibitem [{\citenamefont {Ourjoumtsev}\ \emph {et~al.}(2007)\citenamefont
  {Ourjoumtsev}, \citenamefont {Dantan}, \citenamefont {Tualle-Brouri},\ and\
  \citenamefont {Grangier}}]{Ourjoumtsev2007PRL}%
  \BibitemOpen
  \bibfield  {author} {\bibinfo {author} {\bibfnamefont {A.}~\bibnamefont
  {Ourjoumtsev}}, \bibinfo {author} {\bibfnamefont {A.}~\bibnamefont {Dantan}},
  \bibinfo {author} {\bibfnamefont {R.}~\bibnamefont {Tualle-Brouri}}, \ and\
  \bibinfo {author} {\bibfnamefont {P.}~\bibnamefont {Grangier}},\ }\bibfield
  {title} {\emph {\bibinfo {title} {Increasing entanglement between gaussian
  states by coherent photon subtraction},\ }}\href {\doibase
  10.1103/PhysRevLett.98.030502} {\bibfield  {journal} {\bibinfo  {journal}
  {Phys. Rev. Lett.}\ }\textbf {\bibinfo {volume} {98}},\ \bibinfo {pages}
  {030502} (\bibinfo {year} {2007})}\BibitemShut {NoStop}%
\bibitem [{\citenamefont {Kim}(2008)}]{Kim2008JPB}%
  \BibitemOpen
  \bibfield  {author} {\bibinfo {author} {\bibfnamefont {M.~S.}\ \bibnamefont
  {Kim}},\ }\bibfield  {title} {\emph {\bibinfo {title} {Recent developments in
  photon-level operations on travelling light fields},\ }}\href {\doibase
  10.1088/0953-4075/41/13/133001} {\bibfield  {journal} {\bibinfo  {journal}
  {J. Phys. B: At., Mol. Opt. Phys.}\ }\textbf {\bibinfo {volume} {41}},\
  \bibinfo {pages} {133001} (\bibinfo {year} {2008})}\BibitemShut {NoStop}%
\bibitem [{\citenamefont {Duan}\ \emph {et~al.}(2000)\citenamefont {Duan},
  \citenamefont {Giedke}, \citenamefont {Cirac},\ and\ \citenamefont
  {Zoller}}]{Duan2000PRL}%
  \BibitemOpen
  \bibfield  {author} {\bibinfo {author} {\bibfnamefont {L.~M.}\ \bibnamefont
  {Duan}}, \bibinfo {author} {\bibfnamefont {G.}~\bibnamefont {Giedke}},
  \bibinfo {author} {\bibfnamefont {J.~I.}\ \bibnamefont {Cirac}}, \ and\
  \bibinfo {author} {\bibfnamefont {P.}~\bibnamefont {Zoller}},\ }\bibfield
  {title} {\emph {\bibinfo {title} {Entanglement purification of gaussian
  continuous variable quantum states},\ }}\href {\doibase
  10.1103/PhysRevLett.84.4002} {\bibfield  {journal} {\bibinfo  {journal}
  {Phys. Rev. Lett.}\ }\textbf {\bibinfo {volume} {84}},\ \bibinfo {pages}
  {4002} (\bibinfo {year} {2000})}\BibitemShut {NoStop}%
\bibitem [{\citenamefont {Eisert}\ \emph {et~al.}(2004)\citenamefont {Eisert},
  \citenamefont {Browne}, \citenamefont {Scheel},\ and\ \citenamefont
  {Plenio}}]{Eisert2004AP}%
  \BibitemOpen
  \bibfield  {author} {\bibinfo {author} {\bibfnamefont {J.}~\bibnamefont
  {Eisert}}, \bibinfo {author} {\bibfnamefont {D.}~\bibnamefont {Browne}},
  \bibinfo {author} {\bibfnamefont {S.}~\bibnamefont {Scheel}}, \ and\ \bibinfo
  {author} {\bibfnamefont {M.}~\bibnamefont {Plenio}},\ }\bibfield  {title}
  {\emph {\bibinfo {title} {Distillation of continuous-variable entanglement
  with optical means},\ }}\href {\doibase
  https://doi.org/10.1016/j.aop.2003.12.008} {\bibfield  {journal} {\bibinfo
  {journal} {Ann. Phys.}\ }\textbf {\bibinfo {volume} {311}},\ \bibinfo {pages}
  {431 } (\bibinfo {year} {2004})}\BibitemShut {NoStop}%
\bibitem [{\citenamefont {Takahashi}\ \emph {et~al.}(2010)\citenamefont
  {Takahashi}, \citenamefont {Neergaard-Nielsen}, \citenamefont {Takeuchi},
  \citenamefont {Takeoka}, \citenamefont {Hayasaka}, \citenamefont {Furusawa},\
  and\ \citenamefont {Sasaki}}]{Takahashi2010nphoton}%
  \BibitemOpen
  \bibfield  {author} {\bibinfo {author} {\bibfnamefont {H.}~\bibnamefont
  {Takahashi}}, \bibinfo {author} {\bibfnamefont {J.~S.}\ \bibnamefont
  {Neergaard-Nielsen}}, \bibinfo {author} {\bibfnamefont {M.}~\bibnamefont
  {Takeuchi}}, \bibinfo {author} {\bibfnamefont {M.}~\bibnamefont {Takeoka}},
  \bibinfo {author} {\bibfnamefont {K.}~\bibnamefont {Hayasaka}}, \bibinfo
  {author} {\bibfnamefont {A.}~\bibnamefont {Furusawa}}, \ and\ \bibinfo
  {author} {\bibfnamefont {M.}~\bibnamefont {Sasaki}},\ }\bibfield  {title}
  {\emph {\bibinfo {title} {Entanglement distillation from gaussian input
  states},\ }}\href {\doibase 10.1038/nphoton.2010.1} {\bibfield  {journal}
  {\bibinfo  {journal} {Nature Photonics}\ }\textbf {\bibinfo {volume} {4}},\
  \bibinfo {pages} {178} (\bibinfo {year} {2010})}\BibitemShut {NoStop}%
\bibitem [{\citenamefont {Opatrn\'y}\ \emph {et~al.}(2000)\citenamefont
  {Opatrn\'y}, \citenamefont {Kurizki},\ and\ \citenamefont
  {Welsch}}]{Opatrny2000PRA}%
  \BibitemOpen
  \bibfield  {author} {\bibinfo {author} {\bibfnamefont {T.}~\bibnamefont
  {Opatrn\'y}}, \bibinfo {author} {\bibfnamefont {G.}~\bibnamefont {Kurizki}},
  \ and\ \bibinfo {author} {\bibfnamefont {D.~G.}\ \bibnamefont {Welsch}},\
  }\bibfield  {title} {\emph {\bibinfo {title} {Improvement on teleportation of
  continuous variables by photon subtraction via conditional measurement},\
  }}\href {\doibase 10.1103/PhysRevA.61.032302} {\bibfield  {journal} {\bibinfo
   {journal} {Phys. Rev. A}\ }\textbf {\bibinfo {volume} {61}},\ \bibinfo
  {pages} {032302} (\bibinfo {year} {2000})}\BibitemShut {NoStop}%
\bibitem [{\citenamefont {Cochrane}\ \emph {et~al.}(2002)\citenamefont
  {Cochrane}, \citenamefont {Ralph},\ and\ \citenamefont
  {Milburn}}]{Cochrane2002PRA}%
  \BibitemOpen
  \bibfield  {author} {\bibinfo {author} {\bibfnamefont {P.~T.}\ \bibnamefont
  {Cochrane}}, \bibinfo {author} {\bibfnamefont {T.~C.}\ \bibnamefont {Ralph}},
  \ and\ \bibinfo {author} {\bibfnamefont {G.~J.}\ \bibnamefont {Milburn}},\
  }\bibfield  {title} {\emph {\bibinfo {title} {Teleportation improvement by
  conditional measurements on the two-mode squeezed vacuum},\ }}\href {\doibase
  10.1103/PhysRevA.65.062306} {\bibfield  {journal} {\bibinfo  {journal} {Phys.
  Rev. A}\ }\textbf {\bibinfo {volume} {65}},\ \bibinfo {pages} {062306}
  (\bibinfo {year} {2002})}\BibitemShut {NoStop}%
\bibitem [{\citenamefont {Olivares}\ \emph {et~al.}(2003)\citenamefont
  {Olivares}, \citenamefont {Paris},\ and\ \citenamefont
  {Bonifacio}}]{Olivares2003PRA}%
  \BibitemOpen
  \bibfield  {author} {\bibinfo {author} {\bibfnamefont {S.}~\bibnamefont
  {Olivares}}, \bibinfo {author} {\bibfnamefont {M.~G.~A.}\ \bibnamefont
  {Paris}}, \ and\ \bibinfo {author} {\bibfnamefont {R.}~\bibnamefont
  {Bonifacio}},\ }\bibfield  {title} {\emph {\bibinfo {title} {Teleportation
  improvement by inconclusive photon subtraction},\ }}\href {\doibase
  10.1103/PhysRevA.67.032314} {\bibfield  {journal} {\bibinfo  {journal} {Phys.
  Rev. A}\ }\textbf {\bibinfo {volume} {67}},\ \bibinfo {pages} {032314}
  (\bibinfo {year} {2003})}\BibitemShut {NoStop}%
\bibitem [{\citenamefont {Dell'Anno}\ \emph {et~al.}(2007)\citenamefont
  {Dell'Anno}, \citenamefont {De~Siena}, \citenamefont {Albano},\ and\
  \citenamefont {Illuminati}}]{DellAnno2007PRA}%
  \BibitemOpen
  \bibfield  {author} {\bibinfo {author} {\bibfnamefont {F.}~\bibnamefont
  {Dell'Anno}}, \bibinfo {author} {\bibfnamefont {S.}~\bibnamefont {De~Siena}},
  \bibinfo {author} {\bibfnamefont {L.}~\bibnamefont {Albano}}, \ and\ \bibinfo
  {author} {\bibfnamefont {F.}~\bibnamefont {Illuminati}},\ }\bibfield  {title}
  {\emph {\bibinfo {title} {Continuous-variable quantum teleportation with
  non-gaussian resources},\ }}\href {\doibase 10.1103/PhysRevA.76.022301}
  {\bibfield  {journal} {\bibinfo  {journal} {Phys. Rev. A}\ }\textbf {\bibinfo
  {volume} {76}},\ \bibinfo {pages} {022301} (\bibinfo {year}
  {2007})}\BibitemShut {NoStop}%
\bibitem [{\citenamefont {Yang}\ and\ \citenamefont {Li}(2009)}]{Yang2009PRA}%
  \BibitemOpen
  \bibfield  {author} {\bibinfo {author} {\bibfnamefont {Y.}~\bibnamefont
  {Yang}}\ and\ \bibinfo {author} {\bibfnamefont {F.-L.}\ \bibnamefont {Li}},\
  }\bibfield  {title} {\emph {\bibinfo {title} {Entanglement properties of
  non-gaussian resources generated via photon subtraction and addition and
  continuous-variable quantum-teleportation improvement},\ }}\href {\doibase
  10.1103/PhysRevA.80.022315} {\bibfield  {journal} {\bibinfo  {journal} {Phys.
  Rev. A}\ }\textbf {\bibinfo {volume} {80}},\ \bibinfo {pages} {022315}
  (\bibinfo {year} {2009})}\BibitemShut {NoStop}%
\bibitem [{\citenamefont {Dell'Anno}\ \emph {et~al.}(2010)\citenamefont
  {Dell'Anno}, \citenamefont {De~Siena},\ and\ \citenamefont
  {Illuminati}}]{DellAnno2010PRA}%
  \BibitemOpen
  \bibfield  {author} {\bibinfo {author} {\bibfnamefont {F.}~\bibnamefont
  {Dell'Anno}}, \bibinfo {author} {\bibfnamefont {S.}~\bibnamefont {De~Siena}},
  \ and\ \bibinfo {author} {\bibfnamefont {F.}~\bibnamefont {Illuminati}},\
  }\bibfield  {title} {\emph {\bibinfo {title} {Realistic continuous-variable
  quantum teleportation with non-gaussian resources},\ }}\href {\doibase
  10.1103/PhysRevA.81.012333} {\bibfield  {journal} {\bibinfo  {journal} {Phys.
  Rev. A}\ }\textbf {\bibinfo {volume} {81}},\ \bibinfo {pages} {012333}
  (\bibinfo {year} {2010})}\BibitemShut {NoStop}%
\bibitem [{\citenamefont {Birrittella}\ and\ \citenamefont
  {Gerry}(2014)}]{Birrittella2014JOSAB}%
  \BibitemOpen
  \bibfield  {author} {\bibinfo {author} {\bibfnamefont {R.}~\bibnamefont
  {Birrittella}}\ and\ \bibinfo {author} {\bibfnamefont {C.~C.}\ \bibnamefont
  {Gerry}},\ }\bibfield  {title} {\emph {\bibinfo {title} {Quantum optical
  interferometry via the mixing of coherent and photon-subtracted squeezed
  vacuum states of light},\ }}\href {\doibase 10.1364/JOSAB.31.000586}
  {\bibfield  {journal} {\bibinfo  {journal} {J. Opt. Soc. Am. B}\ }\textbf
  {\bibinfo {volume} {31}},\ \bibinfo {pages} {586} (\bibinfo {year}
  {2014})}\BibitemShut {NoStop}%
\bibitem [{\citenamefont {Braun}\ \emph {et~al.}(2014)\citenamefont {Braun},
  \citenamefont {Jian}, \citenamefont {Pinel},\ and\ \citenamefont
  {Treps}}]{Braun2014PRA}%
  \BibitemOpen
  \bibfield  {author} {\bibinfo {author} {\bibfnamefont {D.}~\bibnamefont
  {Braun}}, \bibinfo {author} {\bibfnamefont {P.}~\bibnamefont {Jian}},
  \bibinfo {author} {\bibfnamefont {O.}~\bibnamefont {Pinel}}, \ and\ \bibinfo
  {author} {\bibfnamefont {N.}~\bibnamefont {Treps}},\ }\bibfield  {title}
  {\emph {\bibinfo {title} {Precision measurements with photon-subtracted or
  photon-added gaussian states},\ }}\href {\doibase 10.1103/PhysRevA.90.013821}
  {\bibfield  {journal} {\bibinfo  {journal} {Phys. Rev. A}\ }\textbf {\bibinfo
  {volume} {90}},\ \bibinfo {pages} {013821} (\bibinfo {year}
  {2014})}\BibitemShut {NoStop}%
\bibitem [{\citenamefont {Wang}\ \emph {et~al.}(2019)\citenamefont {Wang},
  \citenamefont {Xu}, \citenamefont {Xu},\ and\ \citenamefont
  {Zhang}}]{Wang2019OC}%
  \BibitemOpen
  \bibfield  {author} {\bibinfo {author} {\bibfnamefont {S.}~\bibnamefont
  {Wang}}, \bibinfo {author} {\bibfnamefont {X.~X.}\ \bibnamefont {Xu}},
  \bibinfo {author} {\bibfnamefont {Y.~J.}\ \bibnamefont {Xu}}, \ and\ \bibinfo
  {author} {\bibfnamefont {L.~J.}\ \bibnamefont {Zhang}},\ }\bibfield  {title}
  {\emph {\bibinfo {title} {Quantum interferometry via a coherent state mixed
  with a photon-added squeezed vacuum state},\ }}\href {\doibase
  https://doi.org/10.1016/j.optcom.2019.03.068} {\bibfield  {journal} {\bibinfo
   {journal} {Opt. Commum.}\ }\textbf {\bibinfo {volume} {444}},\ \bibinfo
  {pages} {102 } (\bibinfo {year} {2019})}\BibitemShut {NoStop}%
\bibitem [{\citenamefont {Zhong}\ \emph {et~al.}(2020)\citenamefont {Zhong},
  \citenamefont {Wang}, \citenamefont {Zhou}, \citenamefont {Xu},\ and\
  \citenamefont {Sheng}}]{Zhong2020SC}%
  \BibitemOpen
  \bibfield  {author} {\bibinfo {author} {\bibfnamefont {W.}~\bibnamefont
  {Zhong}}, \bibinfo {author} {\bibfnamefont {F.}~\bibnamefont {Wang}},
  \bibinfo {author} {\bibfnamefont {L.}~\bibnamefont {Zhou}}, \bibinfo {author}
  {\bibfnamefont {P.}~\bibnamefont {Xu}}, \ and\ \bibinfo {author}
  {\bibfnamefont {Y.~B.}\ \bibnamefont {Sheng}},\ }\bibfield  {title} {\emph
  {\bibinfo {title} {Quantum enhanced-interferometry with asymmetric beam
  splitters},\ }}\href {https://doi.org/10.1038/nphoton.2010.268} {\bibfield
  {journal} {\bibinfo  {journal} {Sci. China Phys. Mech. Astron.}\ }\textbf
  {\bibinfo {volume} {63}},\ \bibinfo {pages} {260312} (\bibinfo {year}
  {2020})}\BibitemShut {NoStop}%
\bibitem [{\citenamefont {Holland}\ and\ \citenamefont
  {Burnett}(1993)}]{Holland1993PRL}%
  \BibitemOpen
  \bibfield  {author} {\bibinfo {author} {\bibfnamefont {M.~J.}\ \bibnamefont
  {Holland}}\ and\ \bibinfo {author} {\bibfnamefont {K.}~\bibnamefont
  {Burnett}},\ }\bibfield  {title} {\emph {\bibinfo {title} {Interferometric
  detection of optical phase shifts at the heisenberg limit},\ }}\href
  {\doibase 10.1103/PhysRevLett.71.1355} {\bibfield  {journal} {\bibinfo
  {journal} {Phys. Rev. Lett.}\ }\textbf {\bibinfo {volume} {71}},\ \bibinfo
  {pages} {1355} (\bibinfo {year} {1993})}\BibitemShut {NoStop}%
\bibitem [{\citenamefont {Glauber}(1963)}]{Glauber1963PR}%
  \BibitemOpen
  \bibfield  {author} {\bibinfo {author} {\bibfnamefont {R.~J.}\ \bibnamefont
  {Glauber}},\ }\bibfield  {title} {\emph {\bibinfo {title} {The quantum theory
  of optical coherence},\ }}\href {\doibase 10.1103/PhysRev.130.2529}
  {\bibfield  {journal} {\bibinfo  {journal} {Phys. Rev.}\ }\textbf {\bibinfo
  {volume} {130}},\ \bibinfo {pages} {2529} (\bibinfo {year}
  {1963})}\BibitemShut {NoStop}%
\bibitem [{\citenamefont {Brown}\ and\ \citenamefont
  {Twiss}(1956)}]{BROWN1956nature}%
  \BibitemOpen
  \bibfield  {author} {\bibinfo {author} {\bibfnamefont {R.~H.}\ \bibnamefont
  {Brown}}\ and\ \bibinfo {author} {\bibfnamefont {R.~Q.}\ \bibnamefont
  {Twiss}},\ }\bibfield  {title} {\emph {\bibinfo {title} {Correlation between
  photons in two coherent beams of light},\ }}\href {\doibase 10.1038/177027a0}
  {\bibfield  {journal} {\bibinfo  {journal} {Nature}\ }\textbf {\bibinfo
  {volume} {177}},\ \bibinfo {pages} {27} (\bibinfo {year} {1956})}\BibitemShut
  {NoStop}%
\bibitem [{\citenamefont {Weedbrook}\ \emph {et~al.}(2016)\citenamefont
  {Weedbrook}, \citenamefont {Pirandola}, \citenamefont {Thompson},
  \citenamefont {Vedral},\ and\ \citenamefont {Gu}}]{Weedbrook2016NJP}%
  \BibitemOpen
  \bibfield  {author} {\bibinfo {author} {\bibfnamefont {C.}~\bibnamefont
  {Weedbrook}}, \bibinfo {author} {\bibfnamefont {S.}~\bibnamefont
  {Pirandola}}, \bibinfo {author} {\bibfnamefont {J.}~\bibnamefont {Thompson}},
  \bibinfo {author} {\bibfnamefont {V.}~\bibnamefont {Vedral}}, \ and\ \bibinfo
  {author} {\bibfnamefont {M.}~\bibnamefont {Gu}},\ }\bibfield  {title} {\emph
  {\bibinfo {title} {How discord underlies the noise resilience of quantum
  illumination},\ }}\href {\doibase 10.1088/1367-2630/18/4/043027} {\bibfield
  {journal} {\bibinfo  {journal} {New Journal of Physics}\ }\textbf {\bibinfo
  {volume} {18}},\ \bibinfo {pages} {043027} (\bibinfo {year}
  {2016})}\BibitemShut {NoStop}%
\bibitem [{\citenamefont {Bradshaw}\ \emph {et~al.}(2017)\citenamefont
  {Bradshaw}, \citenamefont {Assad}, \citenamefont {Haw}, \citenamefont {Tan},
  \citenamefont {Lam},\ and\ \citenamefont {Gu}}]{Bradshaw2017PRA}%
  \BibitemOpen
  \bibfield  {author} {\bibinfo {author} {\bibfnamefont {M.}~\bibnamefont
  {Bradshaw}}, \bibinfo {author} {\bibfnamefont {S.~M.}\ \bibnamefont {Assad}},
  \bibinfo {author} {\bibfnamefont {J.~Y.}\ \bibnamefont {Haw}}, \bibinfo
  {author} {\bibfnamefont {S.-H.}\ \bibnamefont {Tan}}, \bibinfo {author}
  {\bibfnamefont {P.~K.}\ \bibnamefont {Lam}}, \ and\ \bibinfo {author}
  {\bibfnamefont {M.}~\bibnamefont {Gu}},\ }\bibfield  {title} {\emph {\bibinfo
  {title} {Overarching framework between gaussian quantum discord and gaussian
  quantum illumination},\ }}\href {\doibase 10.1103/PhysRevA.95.022333}
  {\bibfield  {journal} {\bibinfo  {journal} {Phys. Rev. A}\ }\textbf {\bibinfo
  {volume} {95}},\ \bibinfo {pages} {022333} (\bibinfo {year}
  {2017})}\BibitemShut {NoStop}%
\bibitem [{\citenamefont {Kim}\ \emph {et~al.}(2023{\natexlab{b}})\citenamefont
  {Kim}, \citenamefont {Hwang}, \citenamefont {Jung},\ and\ \citenamefont
  {Park}}]{Kim2023QINP}%
  \BibitemOpen
  \bibfield  {author} {\bibinfo {author} {\bibfnamefont {M.}~\bibnamefont
  {Kim}}, \bibinfo {author} {\bibfnamefont {M.-R.}\ \bibnamefont {Hwang}},
  \bibinfo {author} {\bibfnamefont {E.}~\bibnamefont {Jung}}, \ and\ \bibinfo
  {author} {\bibfnamefont {D.}~\bibnamefont {Park}},\ }\bibfield  {title}
  {\emph {\bibinfo {title} {Is entanglement a unique resource in quantum
  illumination?}\ }}\href {\doibase 10.1007/s11128-023-03839-z} {\bibfield
  {journal} {\bibinfo  {journal} {Quantum Information Processing}\ }\textbf
  {\bibinfo {volume} {22}},\ \bibinfo {pages} {98} (\bibinfo {year}
  {2023}{\natexlab{b}})}\BibitemShut {NoStop}%
\end{thebibliography}%

\end{document}